%% file: TRK-20-002_temp.tex
\pdfoutput=1
\documentclass[11pt,twoside,a4paper,cmspaper,final,collab]{cms-tdr}
\def\svnVersion{84e631b}\def\svnDate{2025/08/25}\def\cmsCernNoTag{CERN-EP-2025-036}\def\cmsCernDate{\today}\def\cmsMessage{Published in the Journal of Instrumentation as \href{http://dx.doi.org/10.1088/1748-0221/20/08/P08027}{\doi{10.1088/1748-0221/20/08/P08027}.}}
\begin{document}\cmsNoteHeader{TRK-20-002}

\newcommand{\SST}{CMS silicon strip tracker}
\newcommand{\sovern}{\ensuremath{S\!/\!N}}
\newcommand{\ileak}{\ensuremath{I_{\text{leak}}}\xspace}
\newcommand{\neff}{\ensuremath{N_{\text{eff}}}\xspace}
\newcommand{\vdep}{\ensuremath{V_{\text{dep}}}\xspace}

\cmsNoteHeader{TRK-20-002}

\title{Operation and performance of the CMS silicon strip tracker with proton-proton collisions at the CERN LHC}

\date{\today}

\abstract{
Salient aspects of the commissioning, calibration, and performance of the CMS silicon strip tracker are discussed, drawing on experience during operation with proton-proton collisions delivered by the CERN LHC. The data were obtained with a variety of luminosities. The operating temperature of the strip tracker was changed several times during this period and results are shown as a function of temperature in several cases. Details of the system performance are presented, including occupancy, signal-to-noise ratio, Lorentz angle, and single-hit spatial resolution. Saturation effects in the APV25 readout chip preamplifier observed during early Run~2 are presented, showing the effect on various observables and the subsequent remedy. Studies of radiation effects on the strip tracker are presented both for the optical readout links and the silicon sensors. The observed effects are compared to simulation, where available, and they generally agree well with expectations.}

\hypersetup{%
pdfauthor={CMS Collaboration},%
pdftitle={Operation and performance of the CMS silicon strip tracker with proton-proton collisions at the CERN LHC},%
pdfsubject={CMS},%
pdfkeywords={CMS, strip tracker, detector performance}}%

\maketitle 

\tableofcontents

\section{Introduction}\label{sec:intro}

The silicon strip tracker~(SST) of the CMS
experiment~\cite{Chatrchyan:2008aa} at the CERN
LHC~\cite{Evans:2008zzb} is the world's largest silicon-based detector with an
active area of 200\unit{m$^2$}.
The SST detects charge deposits (hits) at discrete points along the
paths of charged particles arising from the collisions produced
by the LHC. These hits, together with those detected in the CMS pixel
detector~\cite{Chatrchyan:2008aa,Karimaki:368412,thetrackergroupofthecmscollaboration2020cms,CMSTrackerGroup:2020edz},
are used to reconstruct the trajectories of charged particles
traversing the detector. Because of the bending of the particle trajectories in the 3.8 tesla field of the CMS solenoid magnet, the transverse momenta of the particles are measured.
The SST was initially
proposed in 1997~\cite{Karimaki:368412} as a part of a larger tracker featuring micro-strip gas
chambers, but the central tracker was changed to an all-silicon design
in the year 2000~\cite{CMS:2000aa}. It was assembled and tested at
the Tracker Integration Facility at
CERN~\cite{Adam:1291192,Adam:2009ac}, installed in CMS in late 2007, and subsequently
commissioned in 2008~\cite{Chatrchyan:2009ad}.

\begin{figure}[bh]
\centering
\includegraphics[width=\textwidth]{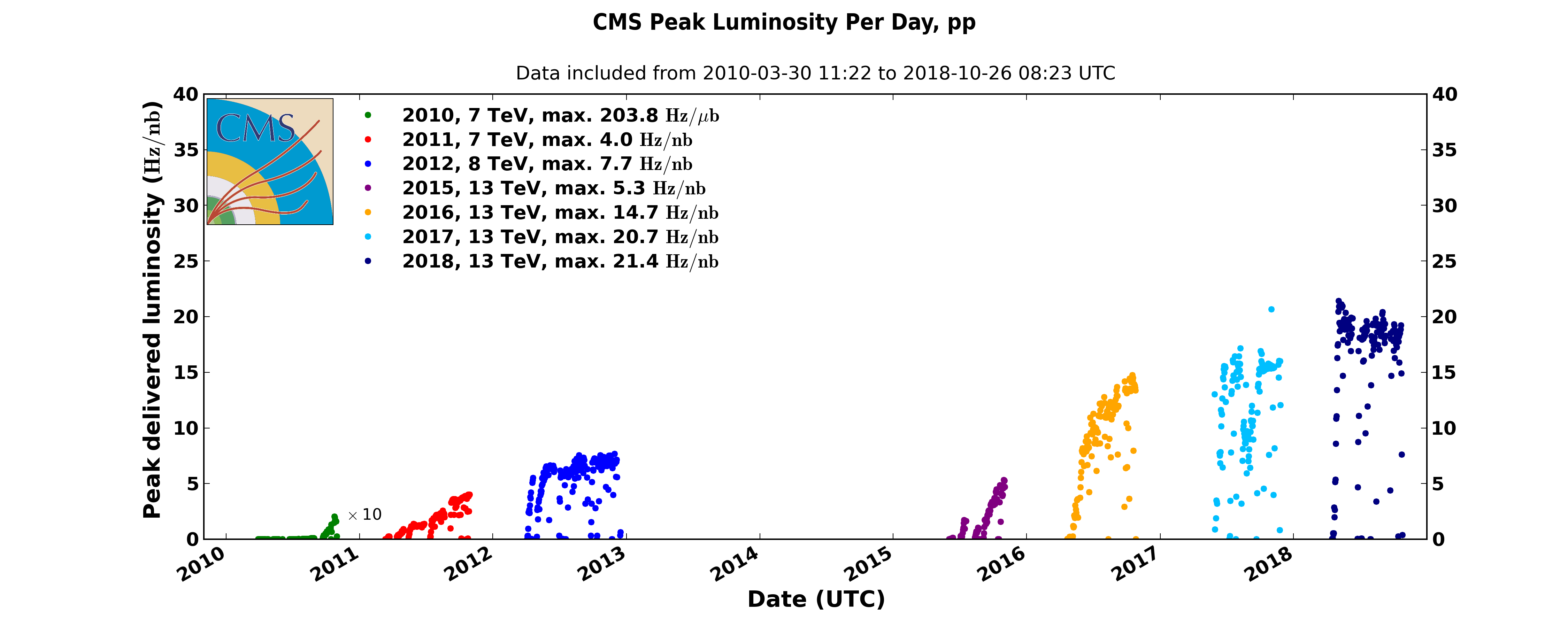}
\caption{ Peak luminosity delivered to CMS during stable
  $\Pp\Pp$ collisions for 2010--2012 and 2015--2018, as a function of time~\cite{CMSLumiPub}.
  The luminosity for the year 2010 is multiplied by a factor of 10.}
\label{fig:inst_lumi_evolution}
\end{figure}

The LHC started data operation with proton-proton~($\Pp\Pp$) collisions at 7\TeV in early 2010.
The center-of-mass energy was increased to 8\TeV in
2012. After the LHC Long Shutdown~1~(LS1), during
the years 2013--14, the LHC restarted at an increased center-of-mass
energy of 13\TeV. The data-taking
period 2010--2012 is commonly referred to as LHC Run~1, the years
2015--2018 as LHC Run~2. During Run~1 the bunch spacing in the LHC machine was 50\unit{ns}.
During Run~2 the bunch spacing was
reduced to its design value of 25\unit{ns}, after an initial period of about 6 weeks with 50\unit{ns} bunch spacing. The peak
instantaneous luminosity to which the SST has been exposed has changed
by orders of magnitude from a few
$10^{32}\percms$ during 2010 to up to a maximum of \mbox{$2.13\ten{34}\percms$} during
2018, more than twice the design luminosity, as shown in
Fig.~\ref{fig:inst_lumi_evolution}. The mean number of $\Pp\Pp$
collisions (pileup) in a single LHC bunch crossing was around 31
during the years 2017 and 2018, with the tails of the pileup
distribution extending to values around 70. The total integrated
luminosity of $\Pp\Pp$ collisions delivered to CMS in Runs~1 and 2 was 192.3\fbinv. At the time of publication, LHC Run~3 is ongoing at a center-of-mass energy of 13.6\TeV.
The SST was designed for a total integrated luminosity of 500\fbinv
and a lifetime of at least 10~years with no ability to access the detector for maintenance during this time.

This paper covers the years of data taking up
to the start of the LHC Long Shutdown~2~(LS2) in late 2018 and is
organized as follows. A brief overview of the SST and
the detector status at the start of LS2 is given in
Sections~\ref{sec:system_description} and \ref{sec:detector-status}. A description of the calibration
procedures and the variation of system properties such as optical link 
gain and noise is shown for different operating temperatures
in Section~\ref{sec:calibration}. In Section~\ref{sec:simulation}, the
ingredients necessary to simulate collision events in
the SST and model its performance are described. The performance of the SST with LHC collisions,
including quantities such as occupancy, hit efficiency, and the signal-to-noise
ratio, is shown in
Section~\ref{sec:collision_performance}. Section~\ref{sec:rad-mon}
is dedicated to the description of radiation effects in the SST, their
change with time and integrated luminosity, as well as studies of
the longevity of the existing tracker until the end of Run~3, when it is expected to be replaced.
A summary and an outlook for the future
of the SST are given in Section~\ref{sec:outlook}.
A glossary is supplied in Section~\ref{chap:glossary} for special terms and acronyms used in this paper.
A description of the track and vertex
reconstruction performance with the full CMS tracking system in Run~1 is given
in Ref.~\cite{Chatrchyan:2014fea}. Some results prepared for this paper have already been made public in Ref.~\cite{CMS:2023gfb}.

\section{Overview of the silicon strip tracker}\label{sec:system_description}
The SST is located at the center of the CMS experiment inside the
solenoidal magnet, which provides a homogeneous magnetic field of 3.8\unit{T}
parallel to the beam line. The inner bore of the SST houses the CMS
pixel detector.  The original CMS pixel detector~\cite{Karimaki:368412,
  Chatrchyan:2008aa} was operated during the years 2009 to 2016. It was
replaced by the Phase-1 pixel
detector~\cite{thetrackergroupofthecmscollaboration2020cms,CMSTrackerGroup:2020edz} in 2017.
The SST is surrounded by the electromagnetic and hadron
calorimeters, which are also placed inside the solenoid, and the steel flux-return
yoke of the magnet is instrumented with muon chambers.  A more
detailed description of the CMS apparatus is report in
Refs.~\cite{Chatrchyan:2008aa,CMS:2023gfb}.

The kinematic variables and spatial coordinates relevant for this paper
are described in the following. CMS uses a right-handed coordinate
system with the origin at the nominal interaction point inside the
experiment. The $x$ axis points towards the center of the LHC
ring, the $y$ axis points upwards (perpendicular to the LHC
plane), and the $z$ axis points
along the beam line in the direction of the counter-clockwise circulating beam. The
azimuthal angle $\phi$ is measured from the $x$ axis in the $x$-$y$
plane, and the radial coordinate in this plane is denoted by $r$. The
polar angle $\theta$ is measured from the $z$ axis. The pseudorapidity
is defined as $\eta = - \ln \tan\left(\theta/2\right)$. 

\begin{figure}[tbhp]
\centering
\includegraphics[width=0.75\textwidth]{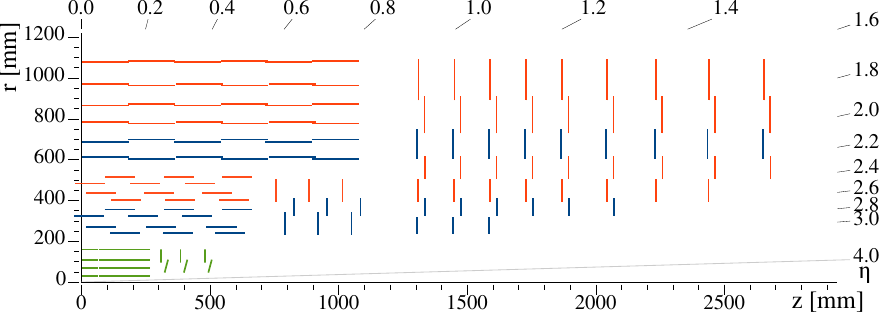}
  \caption{An $r$-$z$ view of one quarter of the CMS silicon strip
    tracker. Layers with stereo modules (details are given in the main text)
    are drawn as blue lines, layers with single modules as red lines. The Phase-1 pixel detector, installed in 2017, is shown in green. 
    }
  \label{fig:rz-tracker}
\end{figure}

The SST occupies a cylindrical volume of 6\unit{m} in length and 2.2\unit{m} in
diameter around the beam line. It is composed of 15\,148
individual silicon detection modules comprising 9.3 million readout strips. An $r$-$z$ view of the SST is
shown in Fig.~\ref{fig:rz-tracker}. 
The barrel section of the SST consists of the tracker inner barrel
(TIB) and the tracker outer barrel (TOB), composed of 4 and 6
concentric layers of silicon modules, respectively. The strips in the
TIB and TOB are oriented parallel to the beam line, enabling a precise
measurement of the $r$-$\phi$ coordinate of a charged-particle
track. The first two layers of both TIB and TOB are made of ``stereo
modules'', where two independent silicon modules are mounted back-to-back with
the sensor strips in the second module at a 100~mrad ``stereo'' angle to enable a three-dimensional point reconstruction. The
SST is complemented in the forward region by the two small tracker
inner disk (TID) subdetectors and the larger tracker endcaps (TECs). On each side
of the barrel region, the TID consists of three disks each with three rings
of modules. Each of the two TECs consist of 9 disks; the number of rings
decreases from 7 in the first three disks to 4 in disk 9 to
reduce the number of silicon modules while ensuring coverage for
$\vert\eta\vert < 2.5$.
The rings 1 and 2 of the TID and rings 1, 2, and 5 of the TEC also contain stereo modules. The modules in the TID
and the TEC are wedge-shaped with the strips pointing radially outwards from the nominal beam line to
enable a precise measurement in $\phi$.
Within each layer, modules are arranged so that there is a small
overlap between neighboring modules. This arrangement ensures
full acceptance and allows the use of tracks passing through overlapping
modules for alignment purposes and other studies that benefit from a
short extrapolation distance.
The SST can reconstruct tracks with transverse momenta upward of a few hundred \MeV
with full efficiency being reached from around 1\GeV.

\begin{figure}[th]
\centering
\includegraphics[width=0.75\textwidth]{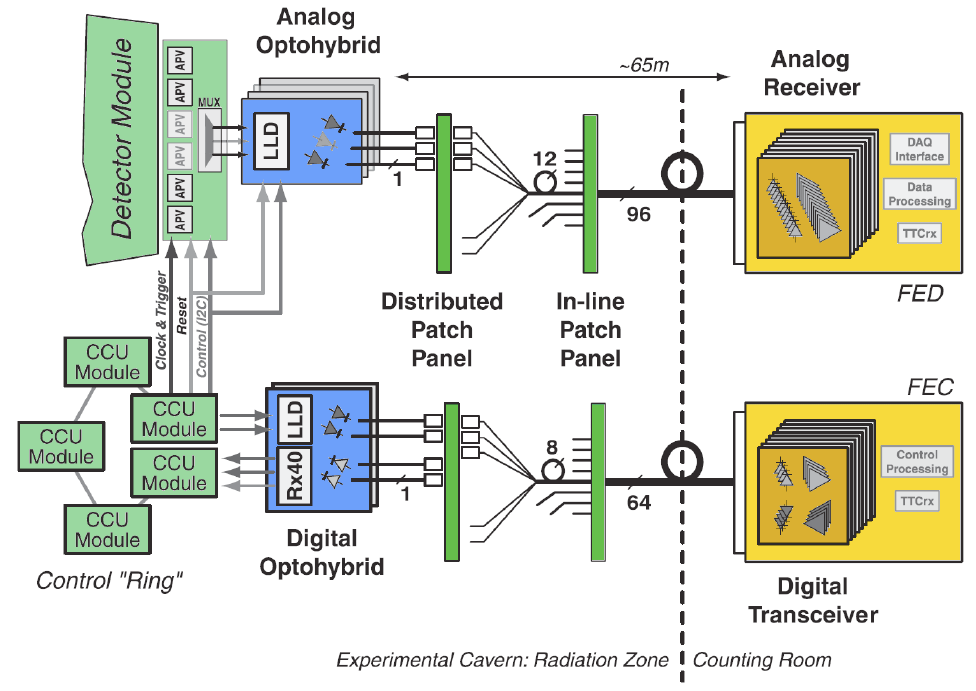}
  \caption{Overview of the control and readout scheme of the SST.    }
  \label{fig:tk_readout_control}
\end{figure}

The control and readout system of the SST is shown in
Fig.~\ref{fig:tk_readout_control}. The basic building blocks of the
detector are the detector modules. These contain the silicon sensors,
the APV25 front-end chips~\cite{Jones:432224}, and other auxiliary application-specific integrated circuits~(ASIC) for
readout and monitoring. The detector module readout is performed
via an optical link. The data from pairs of APV25 chips are time-multiplexed and
passed to a Linear Laser Driver~(LLD)~\cite{949887}, located on an Analog Optohybrid~(AOH)~\cite{Troska:2003yr}, which sends the data via an optical
fiber to the Front-End Driver~(FED) back-end boards~\cite{Coughlan:722058} located in the
service cavern. The modules are controlled 
by Communication and Control Units (CCUs)~\cite{Paillard:593914}, which are electrically daisy-chained into control rings. The control information is sent from
Front-End Controller~(FEC) boards~\cite{Gill:921198}, also located in the service cavern, via a
bi-directional digital optical link to a Digital Optohybrid~(DOH)~\cite{Troska:2003yr} that serves as
an entry to, and exit from, a control ring.

\subparagraph*{Silicon strip detector modules}~\\
The SST detector module consists of one or two silicon strip
sensors and a front-end hybrid (FEH) printed circuit board, which
houses the readout and auxiliary elec\-tronics, on a light-weight
carbon fiber frame. The silicon sensors are of the float-zone p-in-n type, with a
uniform n$^{++}$ back-side implant with phosphorous doping for the n-bulk material. The unprecedented scale of the CMS SST
prompted the move from 4\unit{inch} to 6\unit{inch} diameter wafers, and a $\langle100\rangle$ lattice
orientation was chosen~\cite{Borrello:687861} since this was shown to result in improved
radiation tolerance of the surface interface layer~\cite{Braibant:2002jy}. The spacing between the
individual p$^{+}$ implants of the readout strips (pitch) varies
between 80 and 205\mum, depending on the radial position in the
tracker, with the pitch mostly increasing with radius. The readout
strip is connected via a polysilicon bias resistor to a p$^+$ bias
ring, held at ground, which also defines the active area of the
sensor. A high voltage bias supply is connected to the
aluminum-covered, deep-diffused
n$^{++}$ layer of approximately 20--30\mum thickness that provides a robust barrier to charge injection.
In the inner parts of the tracker, silicon sensors of 320\mum thickness are
used, referred to as ``thin'' in the following. In the outer parts of the tracker (TOB and rings~5--7 of the
TECs), where occupancy and radiation exposure are lower, two silicon sensors are daisy-chained
to increase the cell size to about 20\unit{cm} in length. To maintain a
sufficiently high signal-to-noise ratio, 500\mum thick silicon
sensors, referred to as ``thick'' in the following, are used for those modules.

\begin{figure}[bh]
  \centering
  \includegraphics[width=0.75\textwidth]{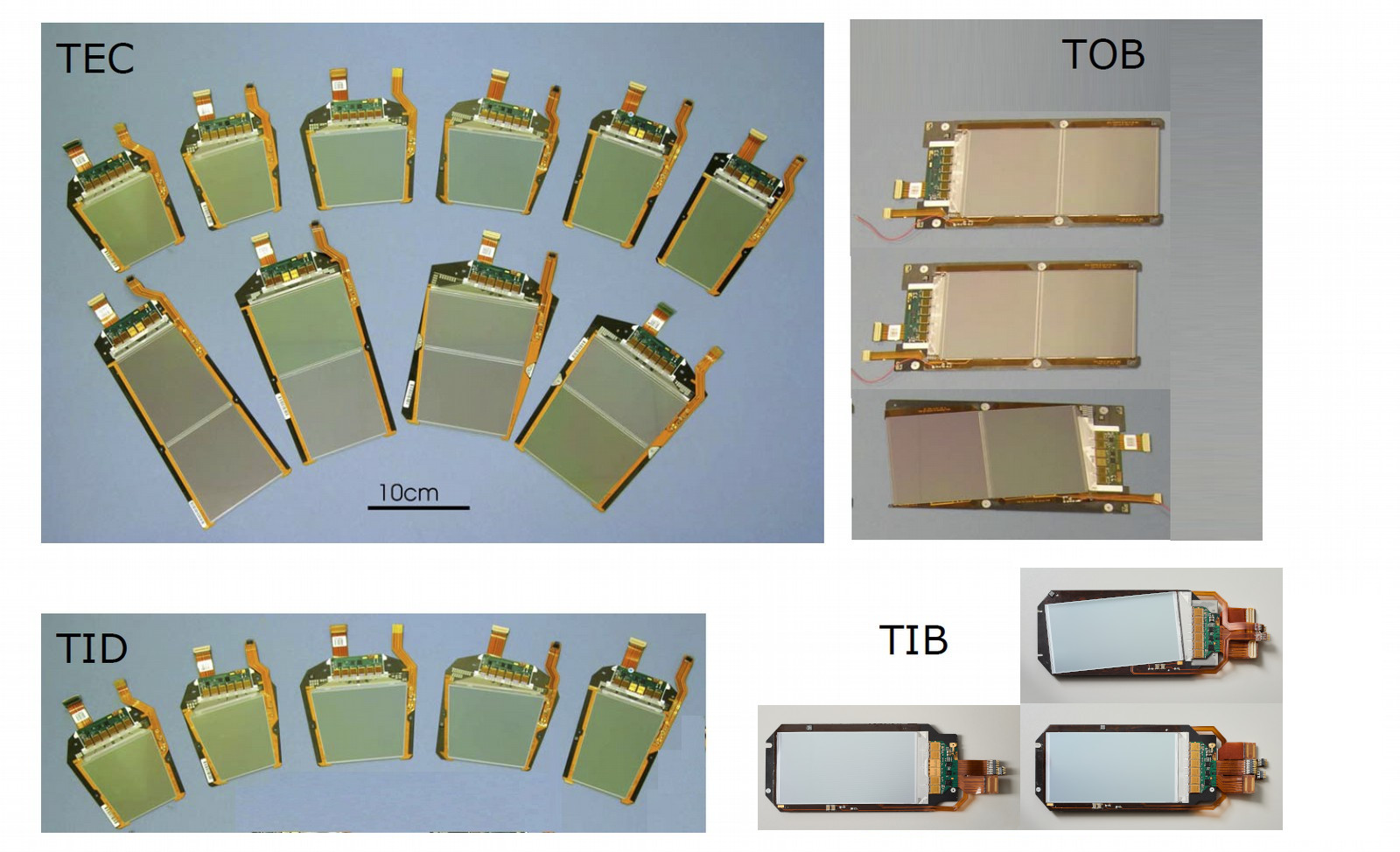}
  \caption{Module types of the SST.}
  \label{fig:picture_modules}
\end{figure}

The p$^{+}$ implants of the individual strips are covered by a thin silicon
oxide and nitride multilayer and overlaid by an aluminum strip. This
structure serves as a capacitor that AC-couples signals from
the sensor to the 128 individual inputs of an APV25 chip.
To enable the
connection of the different sensor pitches to the APV25 chip pitch, glass
pitch adapters are used. A detector module has either 512 or 768
strips and is read out by either four or six~APV25 chips. Modules with
six~APV25 chips are used to instrument TIB layers~1 and~2, TOB layers~5 and~6,
TID rings~1 and~2, and TEC rings~1,~2, and~5. All other
layers/rings have modules with four~APV25 chips. Pictures of all SST
module types can be seen in Fig.~\ref{fig:picture_modules}. 
A summary of the strip pitch for different detector locations, as well
as the number of APV25 chips per module is given in
Table~\ref{tab:module_summary}. More details about the physical
dimensions of the silicon sensors are reported
in Ref.~\cite{Borrello:687861}.

\begin{table}[th]
  \topcaption{Summary of number of APV25 chips per module and strip
    pitch (strip pitch range) for barrel (endcap) sensor geometries.}
  \label{tab:module_summary}
  \hspace{-0.5cm}
  \begin{tabular}{cc}
      \begin{tabular}{lcccc}
        \\
        Sub- & Layer & No. of & Pitch \\
        detector& & APV25s & [\!\mum\!\!]\\
        \hline
        TIB & 1, 2 & 6 & 80\\
        TIB & 3, 4 & 4 & 120\\
        TOB & 1--\,4 & 4 & 183\\
        TOB & 5, 6 & 6 & 122\\
        \hline\\
        \phantom{TOB}\\
        \phantom{TOB}\\
        \phantom{TOB}\\
      \end{tabular}
    &
      \begin{tabular}{lcccc}
        \\
        Sub- &  Ring & No. of & Pitch range\\
        detector& & APV25s & [\!\mum\!\!]\\
        \hline
        TID & 1 & 6 & 80.5--119\\
        TEC & 1 & 6 & \phantom{0}81--112\\
        TID/TEC & 2 & 6 &       113--143\\
        TID/TEC & 3 & 4 &       123--158\\
        TEC & 4 & 4 &           113--139\\
        TEC & 5 & 6 &           126--156\\
        TEC & 6 & 4 &           163--205\\
        TEC & 7 & 4 &           140--172\\
        \hline
      \end{tabular}
  \end{tabular}
\end{table}

\begin{figure}[th]
\centering
\includegraphics[width=\textwidth]{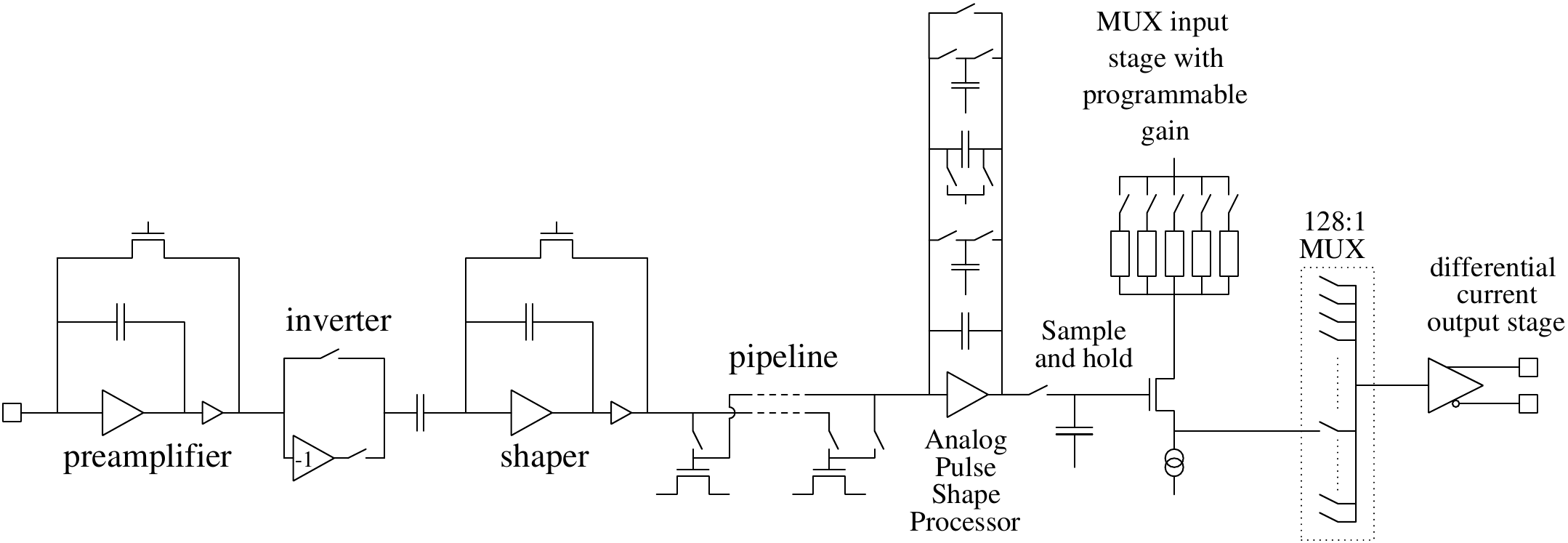}
  \caption{Functional schematic of a single channel of the APV25 chip.}
  \label{fig:apv25_single_channel}
\end{figure}

The APV25 chip is used to read out groups of 128 strips.
In the name, APV stands for ``analog pipeline voltage`` in
the memory pipeline, and the ``25`` in the name refers to the 250\unit{nm}
processing technology.
For each event, information from all strips is collected and
stored on the chip in an analog pipelined memory for potential readout.
A functional schematic of a single APV25 chip readout channel is shown in
Fig.~\ref{fig:apv25_single_channel}.
The APV25 chip samples at the LHC bunch crossing~(BX) frequency of 40\unit{MHz}. It has a preamplifier,
a unit-gain inverter stage that can be activated or
deactivated, a
shaper, a 192~BX deep analog memory pipeline, and a
\mbox{deconvolution} circuit. The pipeline enables buffering of the
information of a given event for a maximum of 4\mus. 

The APV25 chip has
two principal readout modes, called peak and deconvolution mode. In peak mode, a
single cell from the pipeline is read out for a given level-1 trigger
accept signal~(L1A). In deconvolution mode, an algorithm described in the following is used, as the shaper of the APV25 chip has a shaping time of
50\unit{ns}, which is too long to properly distinguish between hits resulting from collisions in consecutive
bunch crossings. The deconvolution algorithm~\cite{Gadomski:1992xu},
implemented in the analog pulse shape processor~(APSP) as a switched-capacitor network, effectively
shortens the pulse length to approximately 25\unit{ns}, thereby reducing substantially the contribution of hits from particles traversing in
adjacent (earlier or later) bunch crossings.
The algorithm performs a three-sample weighted sum to
determine the charge value to be sent out for a particular bunch
crossing. The shorter effective peaking time comes at the expense of increased noise, about
a factor of 1.5 higher compared with the peak mode. The peak mode is used during special commissioning periods and during runs
using cosmic triggers, whereas the deconvolution mode is used for collision
data taking.
The APV25 chip can reserve pipeline cells for readout of 31 events
in peak mode and 10 events in deconvolution mode. A special pipeline emulator board in the underground service
cavern close to the central trigger
system can block triggers using the CMS trigger throttling system~(TTS) 
to prevent buffer overflow in the chips.

\begin{figure}[t!h]
\centering
\includegraphics[width=0.75\textwidth]{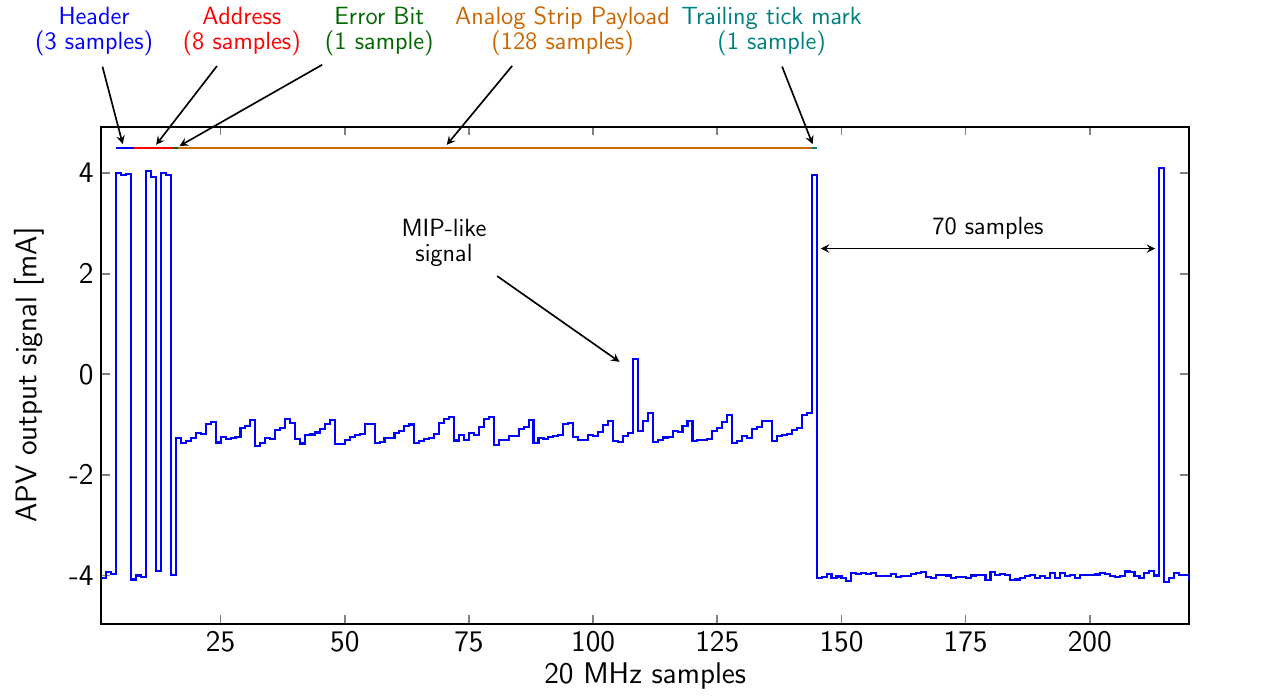}
\caption{Example of an APV25 chip output signal. A data frame consists of
  a 3-bit start-of-frame marker, an 8-bit pipeline address,
  an error bit, the 128 strip analog payload, and the trailing tick
  mark. In the absence of a level-1 trigger accept signal,
  the APV25 chip issues tick mark synchronization pulses every 70 clock
  cycles.}
\label{fig:apv_data_frame}
\end{figure}

Upon reception of an L1A, the APV25 chip outputs a sequence of analog signals corresponding to the charge present on each strip for the corresponding
event, at 20\unit{MHz} frequency.
An example of an APV25 chip data frame
is shown in Fig.~\ref{fig:apv_data_frame}.
Each APV25 chip first sends a digital header
with a start-of-frame marker, the address of the readout pipeline
cell and an error bit, followed by the analog strip payload. The payload consists of 
a series of 128 analog signals with an amplitude proportional to the charge on the individual strips.
A digital signal called a tick mark is issued as a synchronization
signal and end-of-frame marker. A tick mark is also issued every
70~clock cycles if no L1A is received by the APV25 chip. The signal
from two APV25 chips is time-multiplexed at 40\unit{MHz} via a dedicated ASIC
called the APVMUX (MUX in Fig.~\ref{fig:tk_readout_control}). The resulting data frame is then transferred electrically to the
AOH, where the electrical signal is transformed to an analog
optical signal by the LLD with four selectable gain stages based
on InGaAsP edge-emitting laser diodes. The signal from two APV25 chips
is sent via a single optical fiber at 40\unit{MHz} to one of the back-end readout boards. 

Each detector module
contains a Detector Control Unit~(DCU)~\cite{1323692} and a phase-locked loop (PLL)~\cite{Placidi:435986}
chip. The DCU reads the temperature and leakage current of the silicon
sensor as well as low voltages~(LV) and temperatures on the FEH.
The PLL chip can delay the clock and trigger signals
sent to the module to perform a precise adjustment of the readout
timing in steps of 1.04\unit{ns}.

\subparagraph*{Optical links and back-end readout}~\\
The readout signal from the LLD is transmitted via optical fibers out
of the underground experimental cavern to the
service cavern where it is received by the FEDs. A total of 440~FEDs are used to read out the SST.
Twelve readout fibers are bundled into a ribbon, and
a total of 8 ribbons are combined into multiribbon cables of 40--60\unit{m}
length, bridging the path between the caverns. The FED is a 9U VME board with 8 identical
inputs; each receives the signals from one ribbon or 12
individual fibers. A FED can thus read the signals from up to 192 APV25
chips; the actual number varies depending on the location of the modules
in the detector and other considerations such as load balancing. The
FED has a 10-bit analog-to-digital converter~(ADC) at the input. In
normal physics data taking, the FED runs in zero-suppressed~(ZS) mode. 
In this mode it performs pedestal and common mode subtraction
as well as cluster finding on the data, before sending the zero-suppressed data 
to the CMS central data acquisition~(DAQ). Candidate hits from charged particle tracks are identified
by comparing the strip signal to a corresponding noise level. Dedicated runs outside
of the physics data-taking periods are used to determine the pedestal and noise level of each
strip. Data from a single strip with a
signal above five times its noise level will be transmitted. Additionally, data from a strip
whose signal is higher than twice its noise level will be sent out if at
least one of its neighbors also has a signal exceeding two times its
noise level. Each signal is reduced to 8-bit precision before transmission
with the two most significant bits dropped. The highest two ADC values
have a special role: 254 implies a digitized strip signal between 254 and 1022 counts,
and 255 indicates that the actual value was 1023 counts,
\ie, saturating the ADC. The FED can
also send unsuppressed~(NZS for ``no zero suppression'') data 
that contain the full information from all connected strips in the
SST with 10-bit precision. The event size from the SST is 0.5--1.0\unit{MB}
in the ZS mode, depending on the occupancy, and about 14\unit{MB} in NZS data
taking, which is used only in special data-taking periods.

\subparagraph*{Detector control}~\\
The central feature of the SST control system is control rings
made of CCUs that are electrically daisy-chained into a ring-like architecture. A
DOH serves as the optical entry and exit point for each control ring. The FEC boards form the back-end
hardware for the detector control system. These 9U VME
boards are equipped with up to 8 mezzanine cards, called mFECs, each of
which communicates via a bi-directional optical link with one of the
356 control rings in the SST. The optical signal from the mFEC is
received by the DOH and transmitted electrically around the control
ring.

Each CCU is connected to 2--12 detector modules
and transmits clock and trigger 
signals to them via dedicated lines. In addition, configuration parameters are sent using the I$^{2}$C protocol~\cite{i2cref}, and 
slow-control data from the DCUs are read back. Each control ring consists of between 3 and 12
CCUs, and control rings vary in size between about 30 and 60 individual
modules. The control ring has a secondary path to allow 
bypassing individual CCUs in case of failure. A redundant
DOH is available in case the primary DOH fails.

\subparagraph*{Readout partitions}~\\
The strip tracker is organized into four readout ``partitions'', each of which
can be operated independently. This
means, for example, that for calibration runs the readout partitions can be
triggered independently and different calibration
procedures can be performed in parallel. The four readout partitions of the tracker
are formed by the TIB and the TID together, the TOB, and each of the
two endcaps~(TEC$\pm$).

\subparagraph*{Services}~\\
The SST silicon sensor modules require 1.25\unit{V} and 2.5\unit{V} low voltage (LV)
to be supplied for the front-end readout electronics, and up to 600\unit{V}
high voltage~(HV) for biasing the silicon strip
sensors~\cite{Paoletti:814088}. These are provided by a set of 962
CAEN A4601 power supply modules~(PSM). Each PSM consists of two independent and identical power supply units~(PSU). Each PSU has two LV channels
that can supply up to 6\unit{A} at 1.25\unit{V} and 13\unit{A} at 2.5\unit{V}, as well as two HV channels, each capable of supplying up
to 12\unit{mA} at 600\unit{V}. The SST was operated 
at a nominal HV of 300\unit{V}. Between
2 and 12 modules are supplied with LV and HV by a single PSU. Such 
a set of modules is called a ``power group''. The control rings are
supplied with 2.5\unit{V} LV by 110 CAEN 4602 power supply modules,
which can power four control rings each. The power supplies are
located in the experimental cavern and are distributed among 29 racks. They are
controlled by 29 branch controllers, which are housed inside 8
CAEN 2527 mainframes. The total power consumption at the end of Run~2 was
about 32\unit{kW} consumed in the front-end electronics and about 60\unit{kW}
delivered. The difference is dissipated along the 40--60\unit{m} long power cables. 

The SST is cooled by two cooling plants that circulate liquid
C$_6$F$_{14}$. Each plant feeds 90 independent sets of cooling loops. The total
cooling capacity is about 90\unit{kW}, which is sufficient
to remove the heat produced by the readout electronics and to keep the
silicon sensors at subzero temperatures over the entire lifetime of the detector.
The tracker support tube contains a thermal screen
system, which includes heating foils and cooling plates. Toward the
outside, the heating foils maintain a temperature of +18$^{\circ}$C{} to avoid possible
condensation and to prevent any thermal effect on the CMS
electromagnetic calorimeter. The cooling plates in the tracker support tube are fed by two
independent cooling plants with 8 loops each. These plants also circulate
liquid C$_6$F$_{14}$. The thermal screen cooling prevents heating
of the tracker from the outside and can be used as backup cooling
when the SST is powered off, independently of the two main cooling plants, to ensure that the
tracker volume can be kept cold at all times.

\section{Detector status}\label{sec:detector-status}

The SST was installed in CMS in December 2007, and, after commissioning, 98.6\% of the channels were fully operational~\cite{Chatrchyan:2009ad}. Over time, a number of
failures developed; the most significant are described below, representing the status of the end of the running year 2018.

A total of 3 out of the 356 control
rings can no longer be operated. This
amounts to an inactive channel fraction of 0.7\%. There is
one inactive control ring in TIB layer~1, one in TIB layer~2, and
one in TOB layer~4. The detector regions corresponding to the
defective control rings do not overlap in $\eta$ or $\phi$.
A total of 6 out of the 1924 power groups are inoperable and the
corresponding modules are excluded from data taking. Six out of the
3848 HV channels cannot be put under bias because of short circuits or
excessive currents, and again all modules connected are excluded from
data taking.
The above failures sum up to about 1.1\% of the modules in the SST.

\begin{figure}[bh]
\centering
\includegraphics[width=\textwidth]{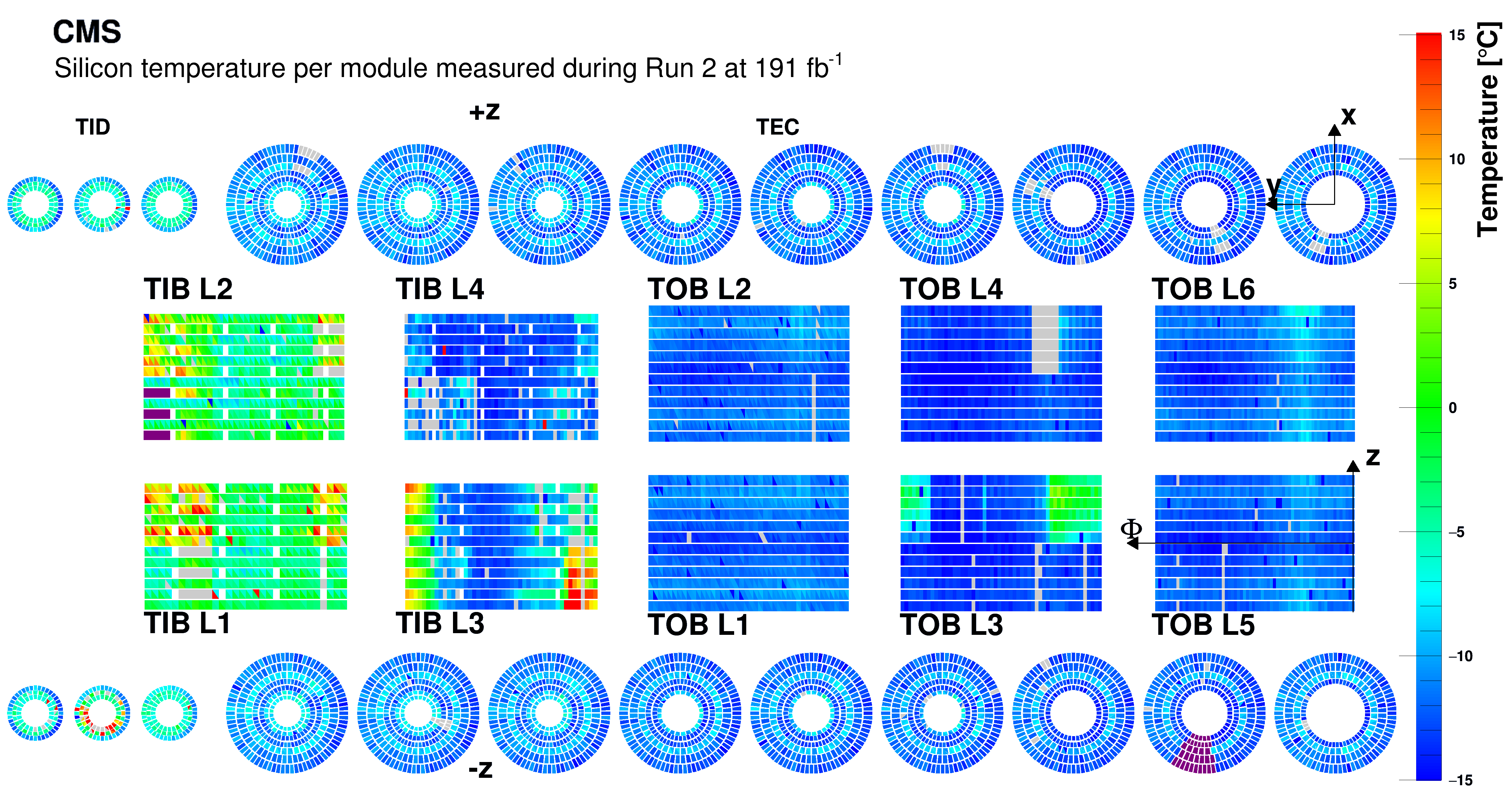}
\caption{Tracker map where each silicon module is represented by a
  rectangle in the barrel and a trapezoid in the endcap; in stereo
  modules each submodule constitutes one half of this area.
  In the TID and TEC the disks are shown with the distance from the
  interaction point increasing from left to right. The color
  scale represents the silicon sensor temperature measured by DCUs
  after 191\fbinv of integrated luminosity at a cooling plant set point
  of $-20^\circ$C. Modules in gray are excluded from the data acquisition.
  The large gray regions in TIB layers 1, 2 and
  TOB layer 4 are nonfunctioning control rings. The large
  purple regions in TIB layer 2 and TEC$-$ disk 8 are detector parts
  that are functional in the data acquisition, but have problems in
  the readout of slow control data via the DCUs.}
\label{fig:tkmap-tsil}
\end{figure}

In late 2009 the cooling circuits connected to one tracker cooling plant suffered from an overpressure
incident. Both the supply and the return valve of the 90 cooling loops
were closed while the coolant was still at +4$^\circ$C. The coolant
gradually warmed up to room temperature and the thermal expansion led
to high pressure inside the cooling lines. A few lines developed leaks
in the process and a total of 5 of the 180 cooling loops were inoperable by the end of
Run~2: three in TIB layer 3, one in TID$-$ disk 2, and one in TOB layer 3. The modules on these cooling pipes continue to be
operated but are cooled only from the neighboring detector parts via
the carbon fiber structure.
As a result of the over-pressure incident, some TIB cooling pipes, which have
an oval cross section to fit within space constraints, were deformed and came
into contact with the back side of the silicon sensor, creating a
short circuit on the HV line rendering all modules connected to the
same HV line unusable for data taking. These modules are
mostly located in TIB layers~3 or~4, and some are related to the faulty HV channels mentioned above. For some other modules the
cooling contact between the cooling line and the module has been
partially or fully detached leaving the modules with insufficient
cooling. These modules are mostly located in TIB layer~1 or~2. They
continue to be operable, but show high temperatures compared with
properly cooled detector parts. Layers and rings with stereo modules in addition generally show elevated temperatures compared to single-sided ones. This information is summarized in
Fig.~\ref{fig:tkmap-tsil}, which shows a tracker map where each silicon
module is represented by a rectangle in the barrel region and a trapezoid
in the endcap region, and one part of a stereo module constitutes one
half of this area. The large purple regions in TIB layer 2 and TEC$-$ 
disk 8 are detector parts that are functional in the data
acquisition, but where the readout of slow control data via the DCUs fails.
Owing to the multiple overlapping layers of the SST, there is no appreciable loss of
physics performance from all failures listed above. More details on the total number of
active channels during data taking can be found in
Section~\ref{sec:bad-components}.

The SST was operated at $+4^\circ$C coolant temperature during
Run~1. The reason to choose a temperature set point much higher than
foreseen was insufficient
humidity control in the tracker service channels and in the interface
region between the SST volume and the environmental seal towards the
outside (called the bulkhead). The SST volume itself had very low
humidity values at all times with dew points well below $-30^\circ$C.
In LS1 a number of measures were undertaken to enable operation at low
temperature. A dry gas delivery system with much larger capacity and
fine-grained distribution was installed, and vapor
insulation was improved with seamless sealing from the edge of the
solenoid magnet to the bulkhead. To avoid
condensation on the surface of the cooling ducts and the outside of the
tracker cold volume, additional heating elements were
added. Monitoring of humidity was also improved by installing a fine-grained system with both in situ
humidity sensors and a gas extraction system with remote humidity measurements. 

The cooling system also underwent a
major refurbishment including new and improved heat exchangers with better
regulation and instrumentation, and new housings with better insulation for the plant core.
The tracker now has the ability to operate safely with coolant temperature as low as
$-25^\circ$C until LS3---the foreseen end of life
for the SST---when the highest power demand is expected because of
accumulated radiation damage.
The evolution of the cooling plant temperature set points is summarized in
Table~\ref{tab:cp_set_history}.

\begin{table}
  \topcaption{Cooling plant set points of the SST during different
    operating periods and integrated luminosity acquired during each
    period.}
  \label{tab:cp_set_history}
  \centering
  \begin{tabular}{ccc}
    \vspace{-0.075cm} Operating & Cooling &  \\
    period    & set point [$^\circ$C] & $\mathcal{L_{\text{int}}}$ [\!\fbinv\!\!] \\
    \hline
    2009--2012 & $+$4  & 29.4    \\
    2015--2017 & $-$15 & 95.0 \\
    2018         & $-$20 & 67.9  \\
    \hline
  \end{tabular}  
\end{table}

The coolant temperature is of great importance because of the 
temperature dependence of the effects of radiation damage to the silicon sensors (Section~\ref{sec:rad-mon}). Charged
and neutral particles crossing the detector will cause damage to the
silicon lattice structure that results in the introduction of
additional energy levels in the silicon bandgap and resulting changes
in the macroscopic sensor properties. The most relevant changes for
the SST are an increase in the dark or leakage current of the silicon
diodes and a change in the full depletion voltage due to a change in
the effective doping concentration of the silicon bulk material. At
temperatures above about 0$^\circ$C, defects in the silicon lattice
will undergo slow migration processes that result in a
reconfiguration of the defects. This migration of defects is commonly
referred to as ``annealing''. For the leakage current, annealing
results in a reduction of the additional leakage current caused by
irradiation. For the full depletion voltage a short-term (days to a few
weeks) process called ``beneficial annealing'' causes a reversal of
the effective doping concentration change. This effect then saturates and at
longer timescales (weeks to months and beyond) a process called
``reverse annealing'' becomes dominant, which results in a further
degradation of the sensor material, on top of the initial damage
caused by irradiation. The annealing typically occurs over periods without irradiation, but
it will also occur while radiation damage is being incurred, provided
temperatures are sufficiently high.

\section{Detector commissioning and calibration}\label{sec:calibration}

The procedures for commissioning the SST are described in detail in
Refs.~\cite{Stringer:1291201,Chatrchyan:2009ad}. The most important aspects
of the procedures are reviewed below, and the results of commissioning at different temperatures are shown and compared with
expectations and with previous results where available.

An initial step, which only needs to be performed once, is to
validate the connection scheme between on-~and off-detector
components. For this, a unique pattern of high and low signals is
generated by each laser driver. This is received by the FED and allows
unique identification of the individual connections. 

Next, the following calibrations are performed to ensure the optimal performance of the SST:
\begin{itemize}
\item internal time alignment, 
\item tuning of laser driver gain and bias,
\item adjustment of FED frame-finding thresholds,
\item tuning of the APV25 chip baseline,
\item calibration of the APV25 chip pulse shape,
\item pedestal and noise measurement, and
\item synchronization to external triggers.
\end{itemize}

\begin{figure}[htb]
\centering
\includegraphics[width=0.45\textwidth]{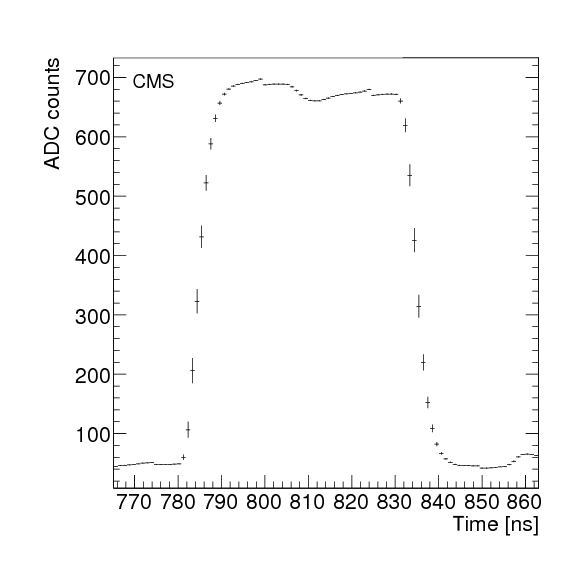}
\caption{High-resolution time domain capture of the tick marks from two
  APV25 chips (1~laser) from one module. The signals from the two APV25 chips are time multiplexed. The tick-height corresponds to an amplitude of about 800 mV.}
\label{fig:example-tick}
\end{figure}

Several commissioning steps make use of the periodically issued tick
mark of the APV25 chip. A high-resolution time domain picture of two 25\unit{ns} wide
time-multiplexed tick marks can be seen in Fig.~\ref{fig:example-tick}. The
tick mark is a digital~1, which corresponds to a $+4$\unit{mA} electrical
signal or $+400$\unit{mV} at the input of the AOH. The digital~0
corresponds to a level of $-4$\unit{mA} and $-400$\unit{mV}.

The SST is recalibrated several times per year to compensate for the
effects of radiation damage to the readout electronics and the silicon sensors, and to compensate
for potential drifts of calibration constants. In addition, changes of
the operating temperature, e.g., from $-15$ to
$-20^{\circ}$C between 2017 and 2018, necessitate a recalibration.

\subsection{Laser driver tuning}\label{sec:LLD_tuning}

\begin{figure}[t!bp]
\centering
\includegraphics[height=0.3\textwidth]{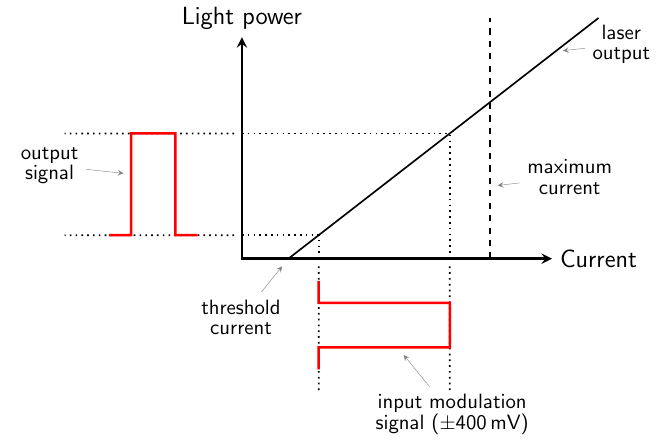}
\includegraphics[height=0.35\textwidth]{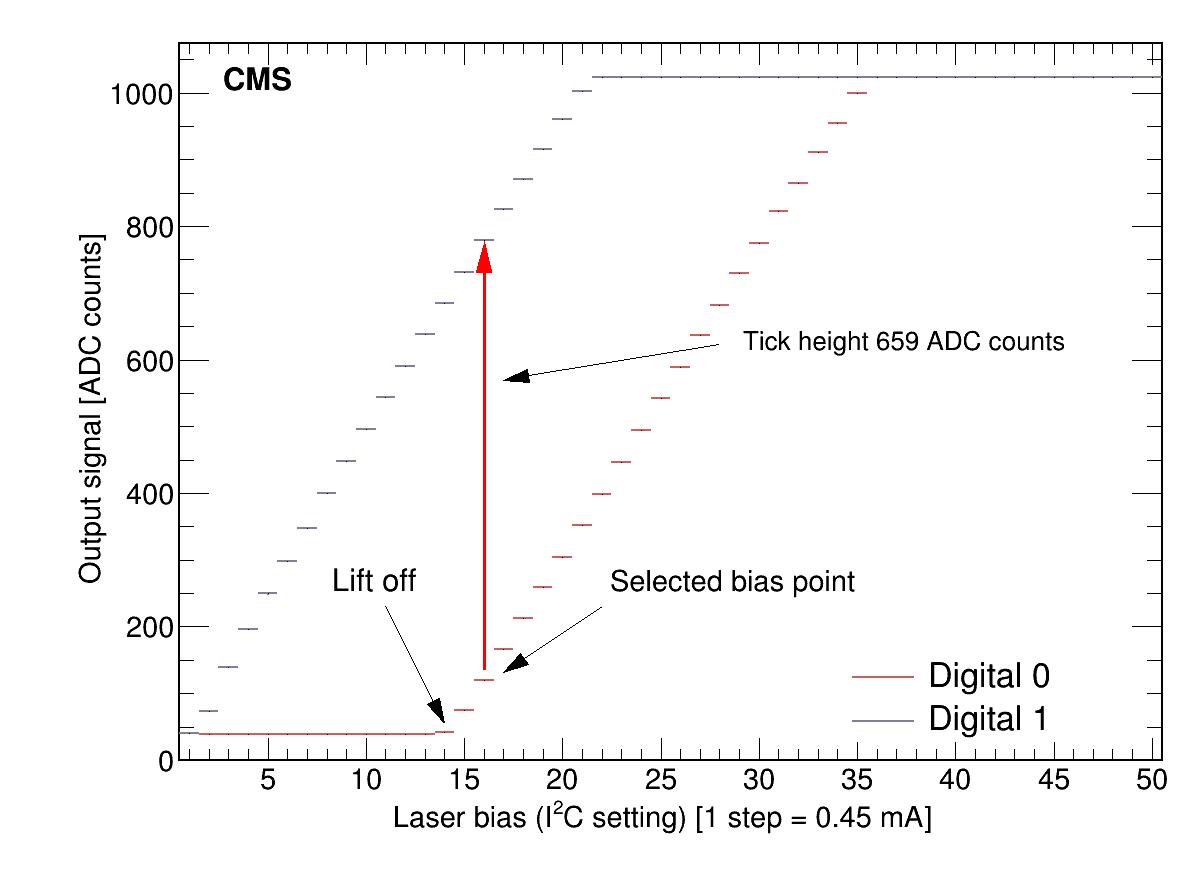}
\caption{Illustration of pulse modulation by the LLD
  (left) and visualization of the bias setting scan during the
  optical link setup run for one LLD gain setting (right). }
\label{fig:lld-pulse-modulation}
\end{figure}

Because of the analog nature of the data transmission to the
back-end electronics, the LLDs must be tuned in order to match the
expected range of charge deposits in the silicon sensors and the 10-bit dynamic range of the FED receivers.
This provides more precise charge measurements, leading to improvements in
spatial resolution from better determination of charge sharing between
strips, and improved
particle identification via the specific energy loss ($\ddinline Ex$).
The LLD has four switchable gain
stages to adjust the amplitude of the output signal for differences in
the optical link gain, caused by, for example, differences in the
laser-to-fiber alignment, or by sample variations of the
components. Gain setting 0 has the lowest gain and thus proportionally lowest light power output; gain setting 3 has the highest.
Figure~\ref{fig:lld-pulse-modulation}~(left) shows how an
input modulation signal is transformed to an output signal with a
certain gain when the laser is pre-biased at its working point.

The optical link setup run (called a gain scan or opto scan)
consists of a nested loop over the four gain stages of the laser driver
and the laser bias setting (0--22.5\unit{mA}). For each configuration
of laser gain and bias setting scanned in the loop, the signal
received by the FED is analyzed when sending a digital~0 and a
digital~1 (tick~mark), as shown in Fig.~\ref{fig:lld-pulse-modulation}~(right).
The laser bias is chosen to ensure that the lowest point
of the 4\unit{mA} differential APV25 chip output (digital~0) still produces a
signal within the dynamic range of the FED receiver.

The gain setting is chosen to correspond to the value of the tick mark height
closest to 690 ADC counts. The FED receiver has a response of about
1~ADC count/mV, so the tick mark height target corresponds to a link
gain of 0.863~V/V when comparing the voltage levels at the LLD input
and the FED receiver.  This central gain value means that charges of up to
2--3 times that of a minimum-ionizing particle (MIP) are
resolved in 8 bits of dynamic range for thin sensors and up to 
1.5--2 times a MIP for thick sensors.
Larger charge deposits will be flagged using the overflow bits
(Section~\ref{sec:system_description}).

\begin{figure}[t!bp]
\centering
\includegraphics[width=\textwidth]{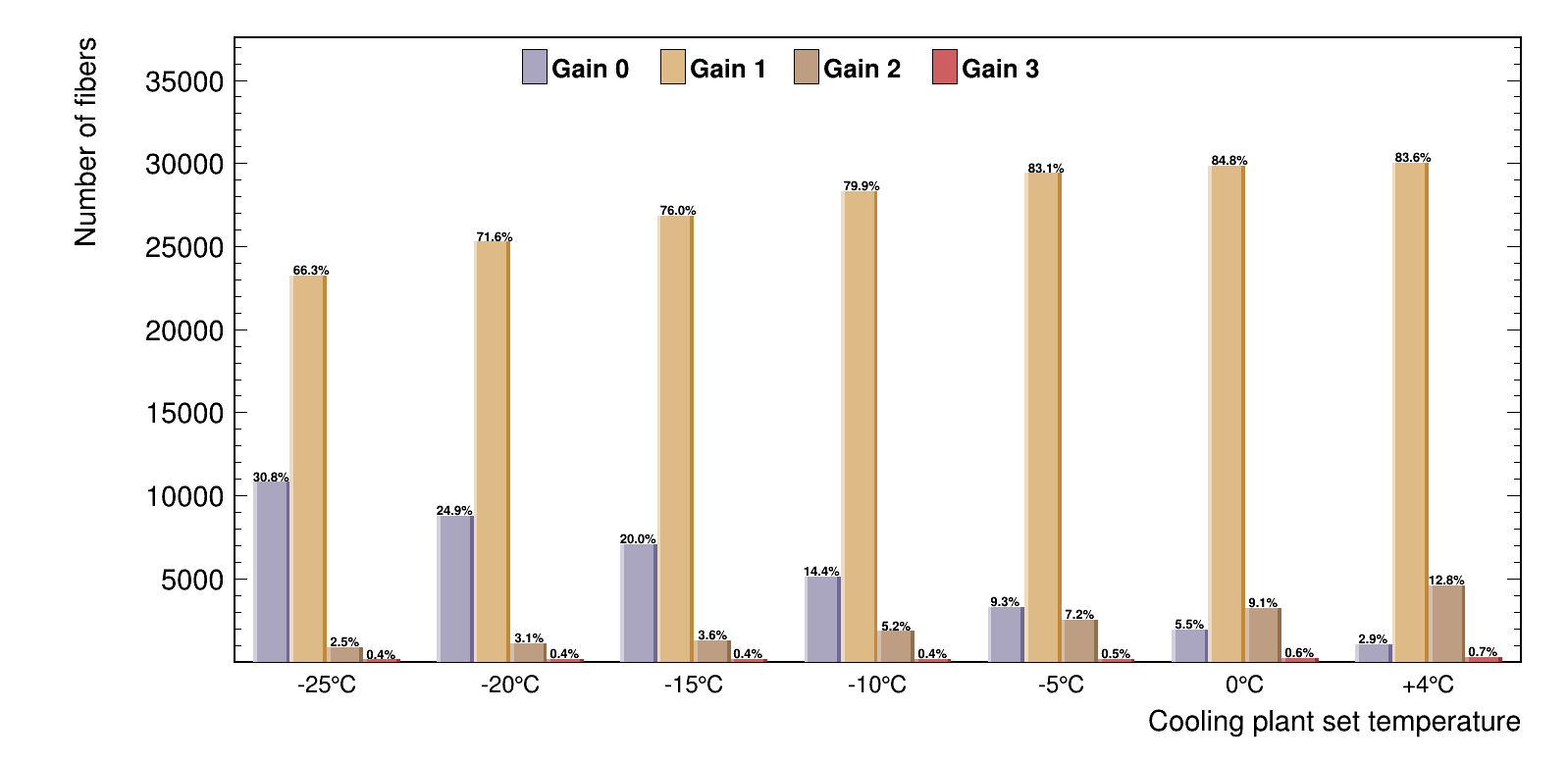} 
\caption{Chosen gain settings at different operating temperatures. The
  expected migration to lower gain settings with decreasing
  temperature is seen. }
\label{fig:gain-settings}
\end{figure}

\begin{figure}[t!bp]
\centering
\includegraphics[width=0.45\textwidth]{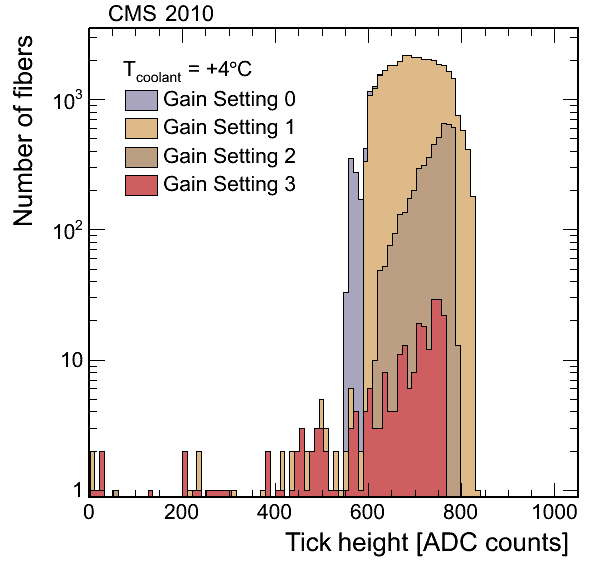} 
\includegraphics[width=0.45\textwidth]{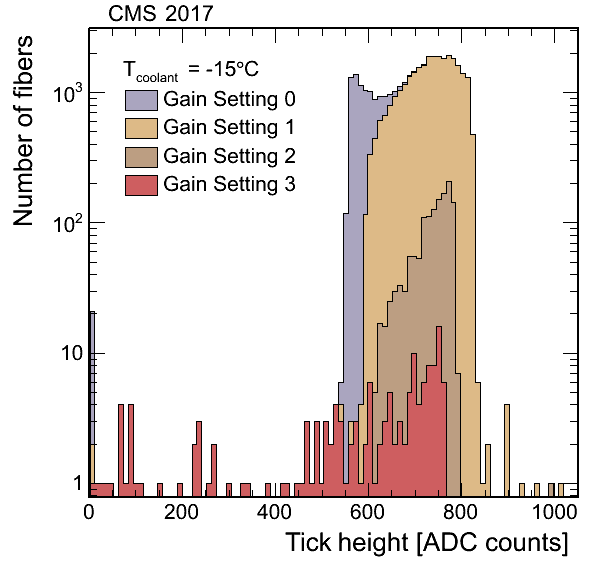} 
\caption{Example distributions of tick heights for each of the four laser driver
  gain settings for data taken in 2010 at $+4^\circ$C{} coolant
  temperature (left) and in 2017 at $-15^\circ$C{} coolant
  temperature (right). Distributions from different gain settings are
  stacked. }
\label{fig:tickheight-distribution}
\end{figure}

Figure~\ref{fig:gain-settings} shows the change in distribution of the chosen
gain settings when decreasing the cooling plant set point from $+4$ to
$-25^\circ$C in steps of 4~or~5$^\circ$C. The gain of the LLD is
expected to increase with decreasing temperature by about
0.8\% per $^\circ$C~\cite{Dris:1000407}. As a result of this, the number
of optical links with lower gain settings increases as
temperature decreases. In Fig.~\ref{fig:tickheight-distribution} two examples of
the resulting height of the tick mark as seen by the FED are
shown.
The spread of the distributions is about 270~ADC counts~(from 552
to 824~ADC counts). The vast majority of the links
fall into this range with sharp edges of the distributions on both
sides caused by LLDs switching into lower or higher gain settings, if available.
Tick heights below about 550~ADC counts correspond to links
with malfunctioning components. Such malfunctions can arise from temporary or permanent issues
with the programming of the LLDs, from damaged fibers, or from other
problems. Links with tick heights near zero are mostly caused by
nonfunctioning or (temporarily or permanently) unpowered APV25 chips.
Effects of radiation damage on the optical links will be discussed in
Section~\ref{sec:rad-mon}.

\subsection{Noise measurement}\label{sec:noise_performance}

The noise performance of the strip tracker is of crucial importance because
the noise is used both in the online zero-suppression of the data in
the tracker FEDs and in the offline identification of clusters originating 
from traversing particles. The noise is measured
from runs taken in the NZS readout mode in periods with no beam using low frequency 
triggers. These are called pedestal runs.
The pedestal of a strip is calculated as the mean of the
ADC values over several thousand events. All channels of an APV25 chip can
experience coherent event-to-event fluctuations or ``common mode'' shifts.
The common mode shift for an APV25 chip in a given event in the analysis of pedestal runs is
calculated as the mean of all strip ADC values after pedestal
subtraction. The square root of the variance of the
resulting quantity over many events is the ``common mode subtracted
  noise'' of a given strip. In the event processing inside the FEDs when running with zero-suppression, this common mode
value, in this case taken as the median of all strip ADC values after pedestal subtraction, is subtracted from the pedestal-subtracted signal of each strip
in the event before cluster finding.

\begin{figure}[th]
\centering
\includegraphics[width=0.8\textwidth]{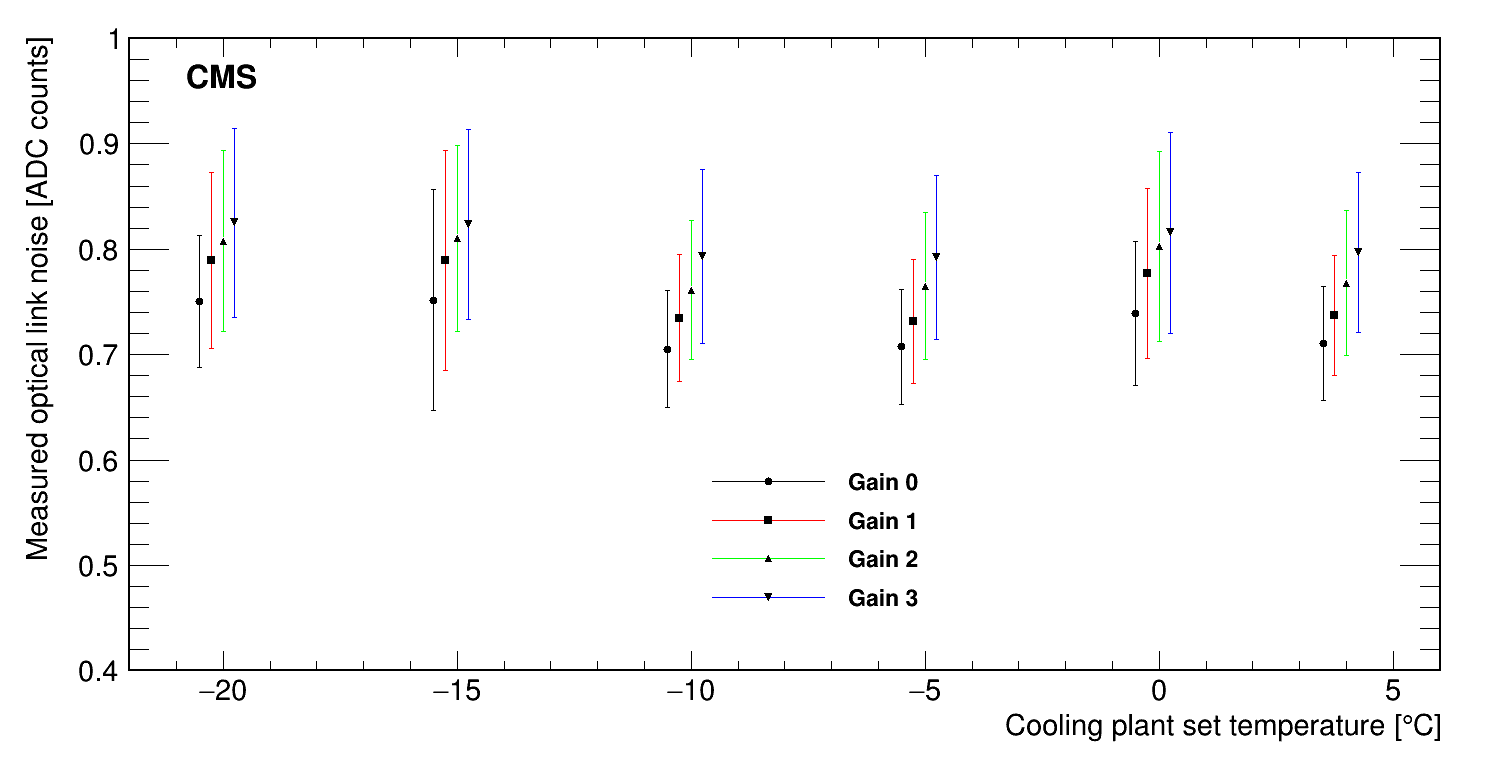}
\caption{Noise measured in the optical link chain during optical link
  setup runs at temperatures between $+4^\circ$C{} and $-20^\circ$C{}
  coolant temperature for the four laser driver gain settings. The
  error bars show the root-mean-square (RMS) of the individual noise distributions and
  are a measure of the spread of the noise among all optical links in
  the SST. Points for different gain settings at the same temperature
  are slightly displaced for visibility.}
\label{fig:link_noise}
\end{figure}

The noise measured with this procedure is the combination of many
sources including the silicon sensor, the APV25 chip, the LLD, and the FED receiver.
The combined noise from the LLD and the FED
receiver (referred to as ``link noise'') is measured in gain scan runs
from the fluctuations on the link output when the APV25 chip output is
biased at $-$4\unit{mA}, i.e., the digital~0. Measurements were performed at
operating temperatures between $+4$ and $-20^\circ$C{} and
are summarized in Fig.~\ref{fig:link_noise}, where the mean link noise
is shown for the four different gain stages of the LLD. The noise
increases slightly for higher gain settings, as expected, but is
quite stable as a function of temperature. Under all conditions the
measured link noise is below 1~ADC count, compared with a total
noise of 3--8~ADC counts measured during pedestal runs, showing that
the link noise is not the dominant noise source.

To obtain the noise in units of electrons, the readout
gain needs to be taken into account.
The tick mark at the input corresponds to a signal
of about 8 MIPs or 175\,000 electrons in 320\mum of silicon. Using
this information, the equivalent noise charge (ENC) in units of the electron charge $e$ is extracted from
the noise measurement in ADC counts. The tick mark height is measured
in a separate run. Runs used for the tick mark scaling are required to
be taken within 48~hours of the corresponding pedestal run and under
otherwise identical conditions. APV25 chips with very high ($>$15 ADC counts) and low 
($<$2 ADC counts) average noise are not used in
the analysis. Runs in which more than 100 of the about 72\,000 APV25 chips show very high
or low noise are rejected completely.

\begin{figure}[htb]
\centering
\includegraphics[width=0.65\textwidth]{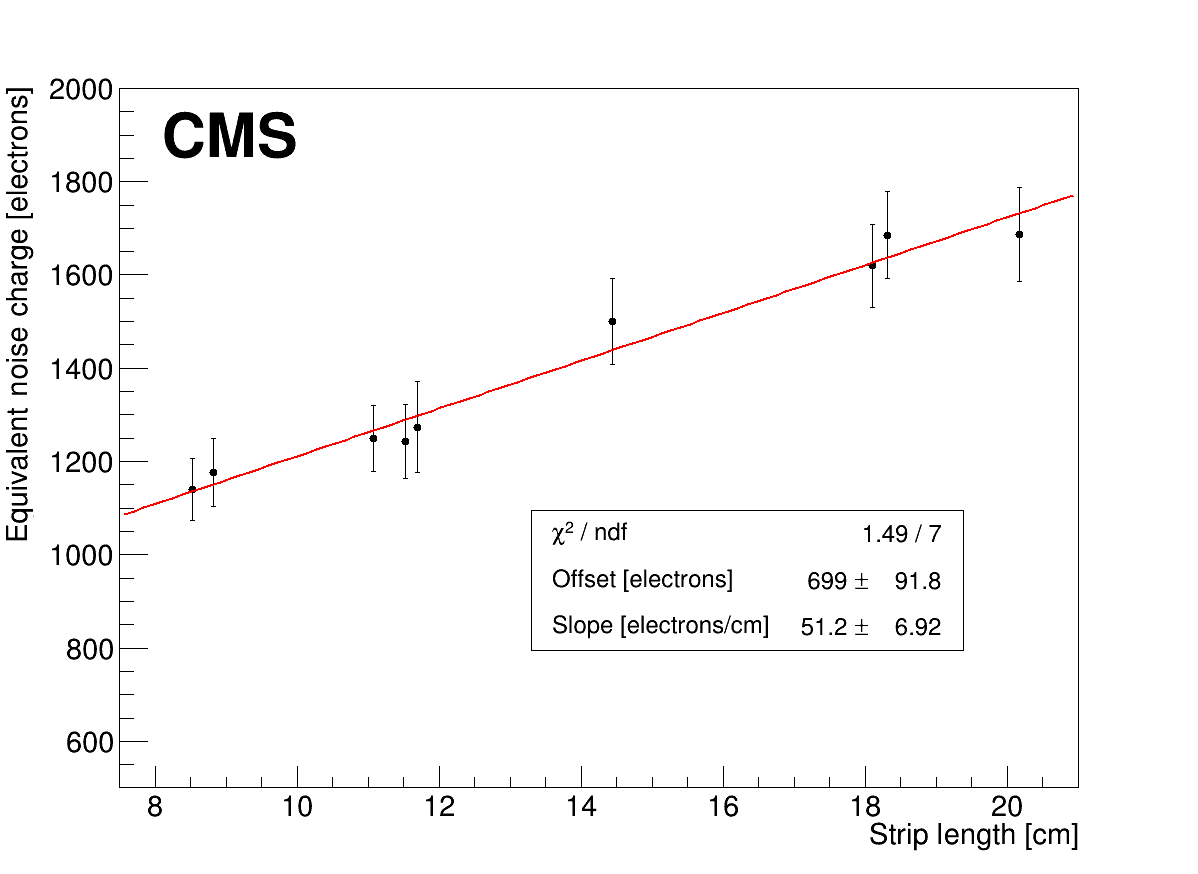}
\caption{Equivalent noise charge as a function of the strip length.}
\label{fig:noise_vs_strip_length}
\end{figure}

The capacitive noise of a silicon sensor increases with the strip length.
As a consequence the total noise is also expected to increase since other noise components do not change for modules with different strip length.
An example of this behavior from pedestal and tick mark
runs taken at $-15^{\circ}$C with the APV25 chips in deconvolution mode is presented in
Fig.~\ref{fig:noise_vs_strip_length}.

\begin{figure}
  \centering
  \includegraphics[width=0.49\textwidth]{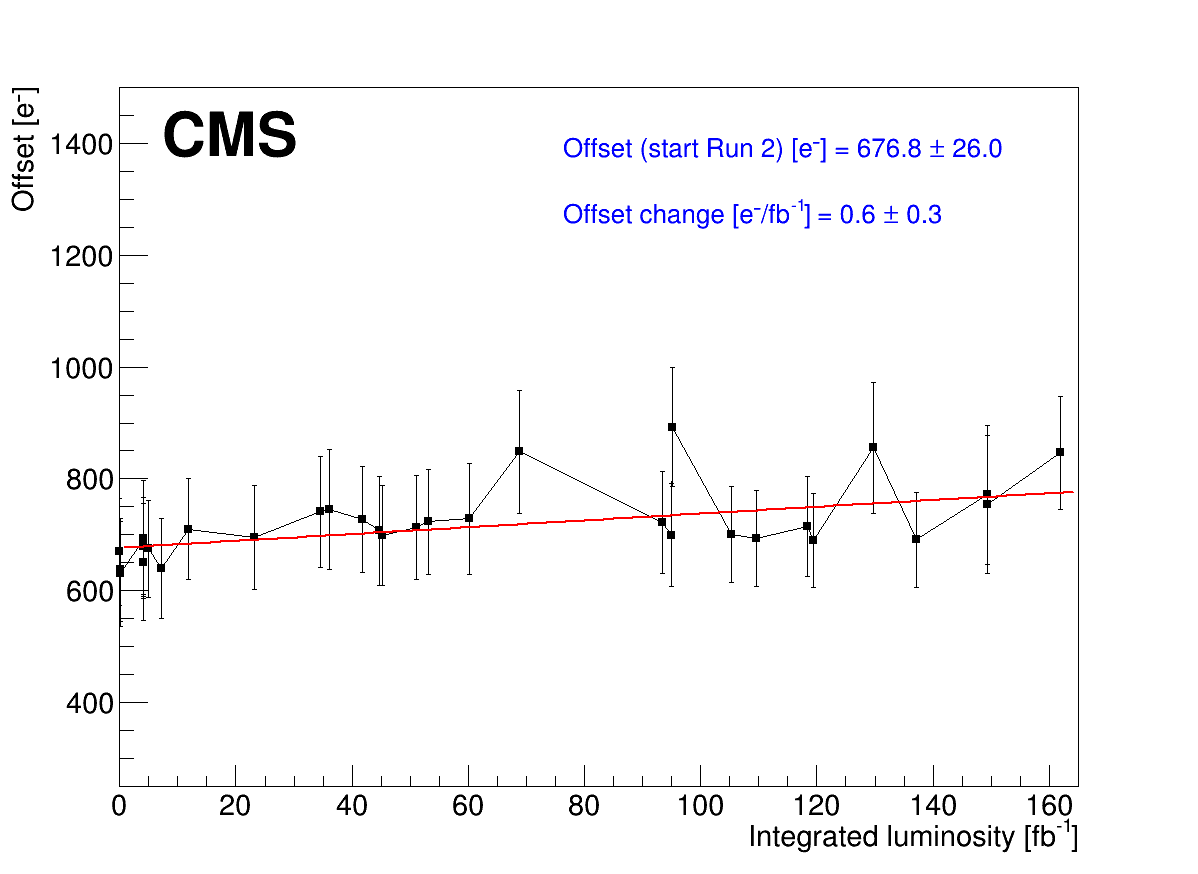}
  \includegraphics[width=0.49\textwidth]{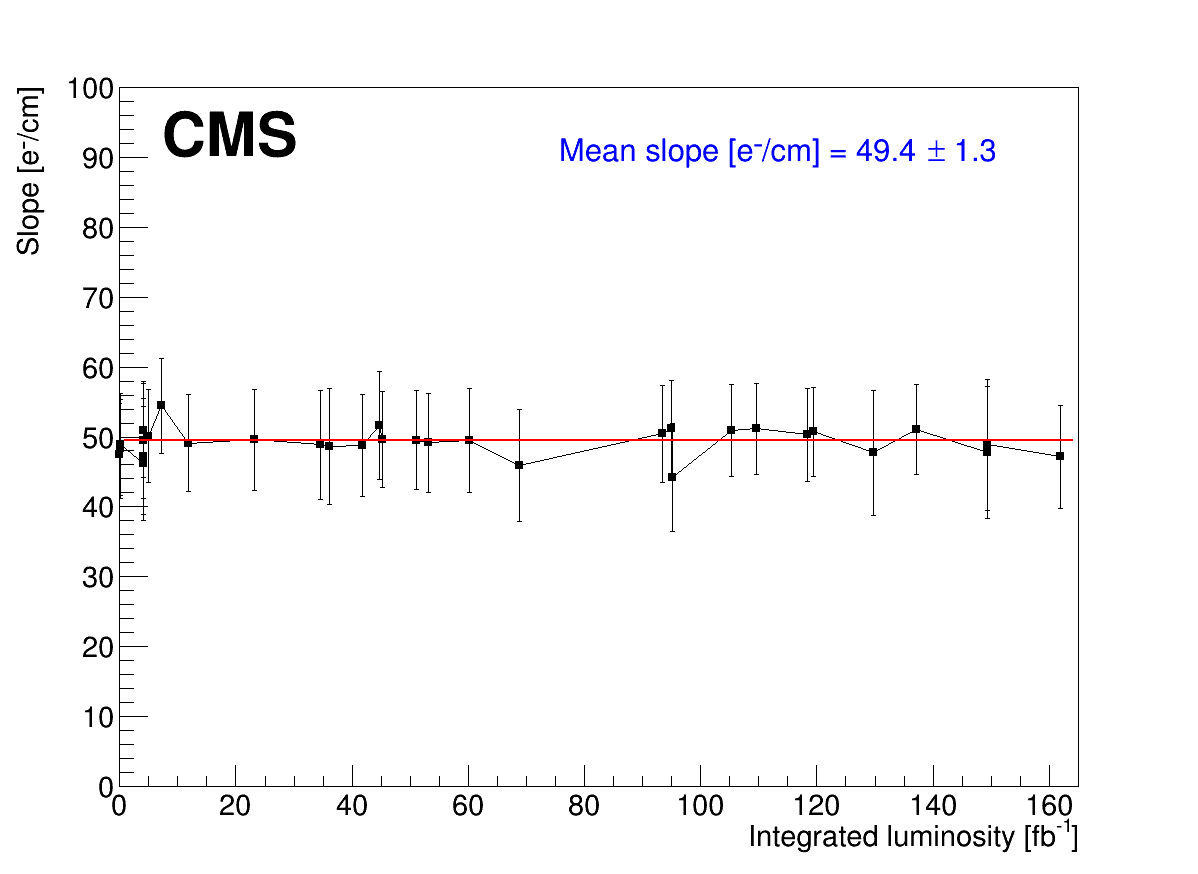}
  \caption{Offsets (left) and slopes (right) as a function of the
    integrated luminosity, derived from straight-line fits to the
    equivalent noise charge as a function of the strip
    length. }
  \label{fig:scaling_summary}
\end{figure}

The expected linear scaling behavior is observed. A straight-line fit
to the data yields the following scaling behavior for the equivalent noise charge in deconvolution mode (ENC$_{\text{deconvolution}}$):
\begin{equation}
        \text{ENC}_{\text{deconvolution}} = \text{ENC}_{\text{offset}} + \text{ENC}_{\text{slope}} L = ( 699\pm92 )\,e + (51.2\pm6.9)\,\frac{e}{\text{cm}}\,L
\end{equation}
where ENC$_{\text{offset}}$ and ENC$_{\text{slope}}$ are the offset and slope of the fit, respectively, and $L$ is the strip length in cm. The same analysis is
performed for all runs available from Run~2. The fit is only performed
if data are available from all readout partitions. The
results are summarized in Fig.~\ref{fig:scaling_summary}, where the
slopes and offsets are reported as a function of the integrated
luminosity, $\mathcal{L}_{\text{int}}$.
The offset shows a slight increase with the accumulated integrated
luminosity. A linear fit to the data yields
\begin{equation}
  \text{ENC}_{\text{offset}} ( \mathcal{L}_{\text{int}} ) = \left( 677 \pm 26 \right)\,e + ( 0.6 \pm 0.3 )\, \frac{e}{\fbinv}\,\mathcal{L}_{\text{int}} .
\end{equation}
The ENC$_{\text{slope}}$ is constant as a function of the
accumulated integrated luminosity. The central value is 
\begin{equation}
  \text{ENC}_{\text{slope}} = \left(49.4 \pm 1.3\right) e/\text{cm}.
\end{equation}
When fitting with a first-order polynomial, the result for
the increase with integrated luminosity is compatible with zero.
In~Ref.~\cite{Chatrchyan:2008aa} a compatible scaling behavior was observed
at low operating temperatures. 

\subsection{Time alignment and trigger synchronization}\label{sec:synchronization_to_lhc}

A precise time alignment of the detector readout to the LHC collisions is
required to obtain an optimal 
signal-to-noise ratio, while minimizing the contributions
from particles in adjacent bunch crossings. The time
alignment of all APV25 chips in the tracker is performed in multiple steps, several of which
take place before collision data taking. Initially 
the known length of the readout fibers from the modules to the back-end
electronics is taken into account. After this, the time alignment of the tick marks of the
individual APV25 chips is performed. The readout sampling is adjusted
to account for the time of flight of the particles through the
detector, assuming that these move at the speed of light along straight trajectories starting at the nominal collision point. The
required time adjustment ranges from about 1\unit{ns} for the central region
of the inner barrel to about 9\unit{ns} in the outer rings of the last disk
of the tracker endcaps. The internal latency of each APV25 chip
is adjusted to ensure that, for a given L1A, the
information from the appropriate BX is read out from the APV25 chip pipeline
buffer. This uses data triggered on cosmic muons, and
is done before the start of beam
operation~\cite{Stringer:1291201}. These adjustments already
give good time alignment for recording $\Pp\Pp$ collision data. In addition, the time
alignment of each module is verified with $\Pp\Pp$ collisions.
For each module in the SST, a
random time shift in a window of $\pm$10\unit{ns} around the current
setting is applied, \ie, in a window with the width of about one LHC BX.
By shifting only up to a maximum of $\pm 10$\unit{ns}
it is expected that most detector modules are sufficiently close to
their original working point to still efficiently detect particle hits.
By shifting modules randomly, the likelihood of having
inefficient modules in consecutive tracking layers is minimized.
A certain number of tracks is collected with these settings. Only tracks with a transverse momentum above 1\GeV are used for the final analysis.
The timing of each module
is then shifted by $+$1\unit{ns} relative to its current (random) setting. Modules
which reach a shift of $+$10\unit{ns} compared with their initial setting are next
set to $-$10\unit{ns} relative to the initial timing. This procedure is
repeated 21 times to ensure that all modules have taken data with
timing settings in the window [$-$10\unit{ns}, $+$10\unit{ns}\!] with respect to their
initial working point in steps of 1\unit{ns}. For each of these delay
steps, the leading strip charge of clusters associated to reconstructed particle tracks
is studied to find the timing point that maximizes this charge. 
Simulations show that maximizing the entire cluster charge would lead 
to higher contributions from collisions from adjacent bunch crossings, due to inter-strip cross talk~\cite{Delaere:1061284}.
Modules can be synchronized with a precision of around 1\unit{ns} with this method.
Most modules require little or no
adjustment after the time-of-flight adjustment.

\begin{figure}[tb]
\centering
\includegraphics[height=0.39\textwidth]{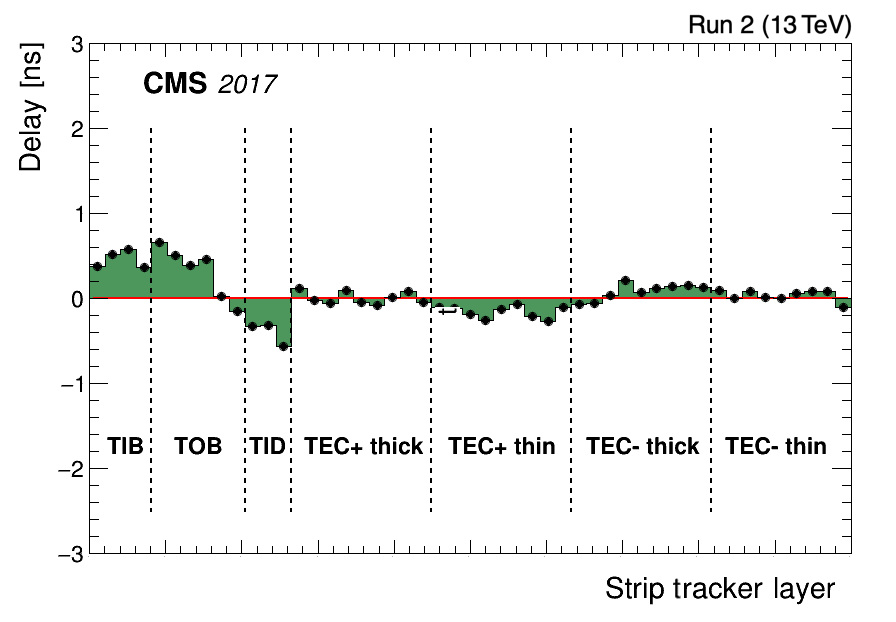}
\includegraphics[height=0.39\textwidth]{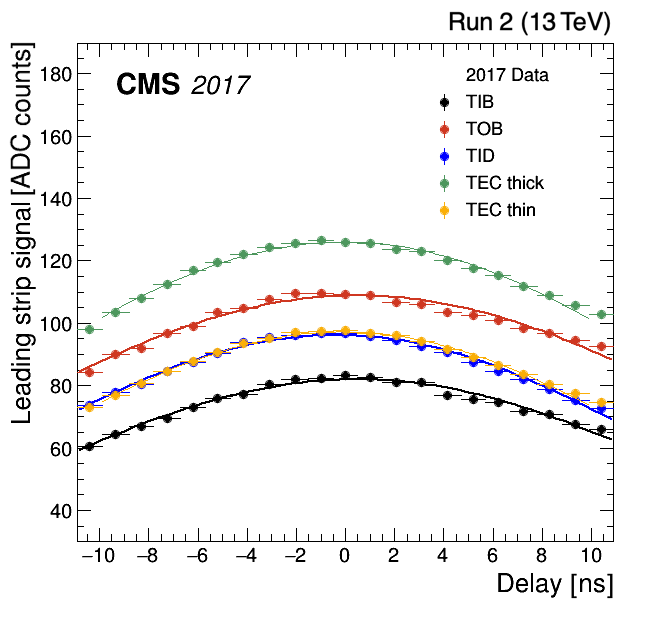}
\caption{Left: average delay adjustment relative to the original sampling
  point for each layer of the tracker. Right:
  leading strip charge in a cluster as function of the change in sampling
  point for different parts of the strip tracker. The data are fit
  with a Gaussian function to determine the position of the maximum. The
  smallest possible delay adjustment is 1\unit{ns}.}
\label{fig:delay_partition_layer}
\end{figure}

Figure~\ref{fig:delay_partition_layer} summarizes the adjustments
required for each layer and disk of the SST. The endcaps are split
according to the thickness of the
sensors. Figure~\ref{fig:delay_partition_layer} (left) shows the net
delay required on average in each layer; Fig.~\ref{fig:delay_partition_layer} (right) shows the
delay curves in which the signal decreases when moving away from the
optimal working point. Each data distribution is fit with a Gaussian function,
which approximates the pulse shape of the APV25 chip in deconvolution mode well in
the central part of the distribution.

\section{Simulation}\label{sec:simulation}
An accurate simulation is required to design and optimize a detector and to better understand its operation. Ultimately, those simulations are used as a fundamental tool to analyze and interpret the recorded data. The simulation chain starts with the generation of the
$\Pp\Pp$ collision products, then propagates the particles
inside the CMS detector through the passive components and
sensitive elements, and finally models the response of the readout electronics.
The reconstruction that follows is identical for data and simulated events.

A dedicated software package~\cite{simulation} based on the \GEANTfour
toolkit~\cite{Agostinelli:2002hh} has been developed by the CMS Collaboration. Each particle
is propagated through the detector volume. The particle entry and
exit points are recorded for each sensitive element, together with
the energy deposited. The granularity of the information saved matches
the design of the detector and its sensitive elements, which are strips in the case of
the SST.

To properly reproduce the interactions of the particles,
taking into account, \eg, nuclear interactions, photon conversions, and
electron bremsstrahlung, an accurate description of the passive
and active detector components was prepared following the
completion of the construction and integration of the SST in CMS in
2007. Individual components had been weighed during construction, and the results were then compared with the estimated weights of the corresponding components in the simulation. The sum of these weights for each partition and for the entire SST is given in Table~\ref{tab:weight}, showing agreement within 5\%. The TOB was never a standalone entity so could not be weighed independently. 

{\renewcommand{\arraystretch}{1.4}
\begin{table}
  \topcaption{Simulated and measured weights of the SST and of three of its partitions.}
  \label{tab:weight}
  \centering
  \begin{tabular}{ l c c c}
 & Measured weight [kg] & Simulated weight [kg] & Difference [\%]\\
\hline
SST & $3990_{-130}^{+90}$ & $4037$ & $-1$\\
TIB/TID & $450_{-12}^{+20}$ & $427$ & $+5$\\
TEC$+$ & $704.3$ & $691.7$ & $+1.8$\\
TEC$-$ & $700.2$ & $691.7$ & $+1.2$\\
     \hline
  \end{tabular}
\end{table}
}

\begin{figure}[tb]
\centering
\includegraphics[width=.45\textwidth]{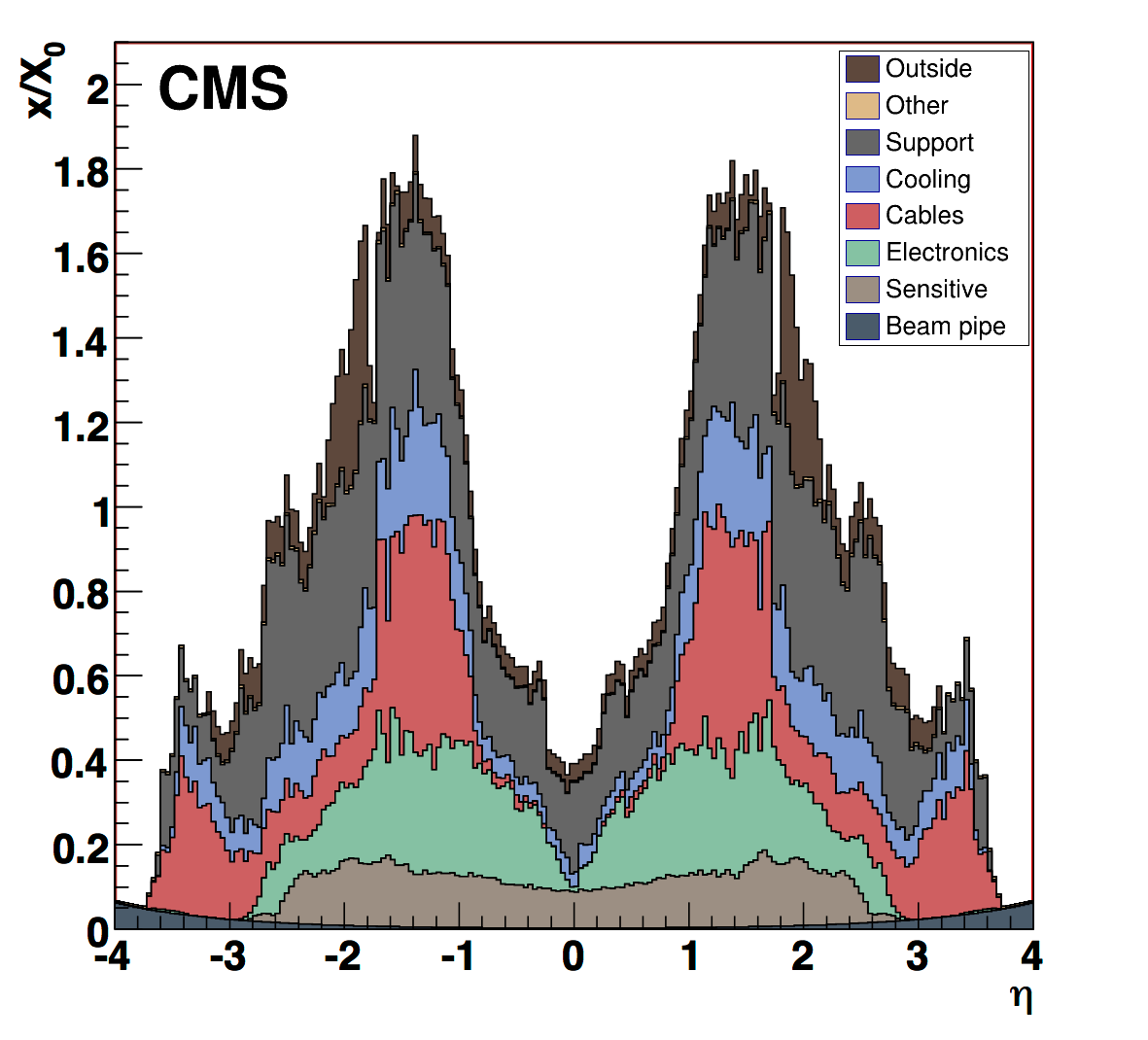}
\includegraphics[width=.45\textwidth]{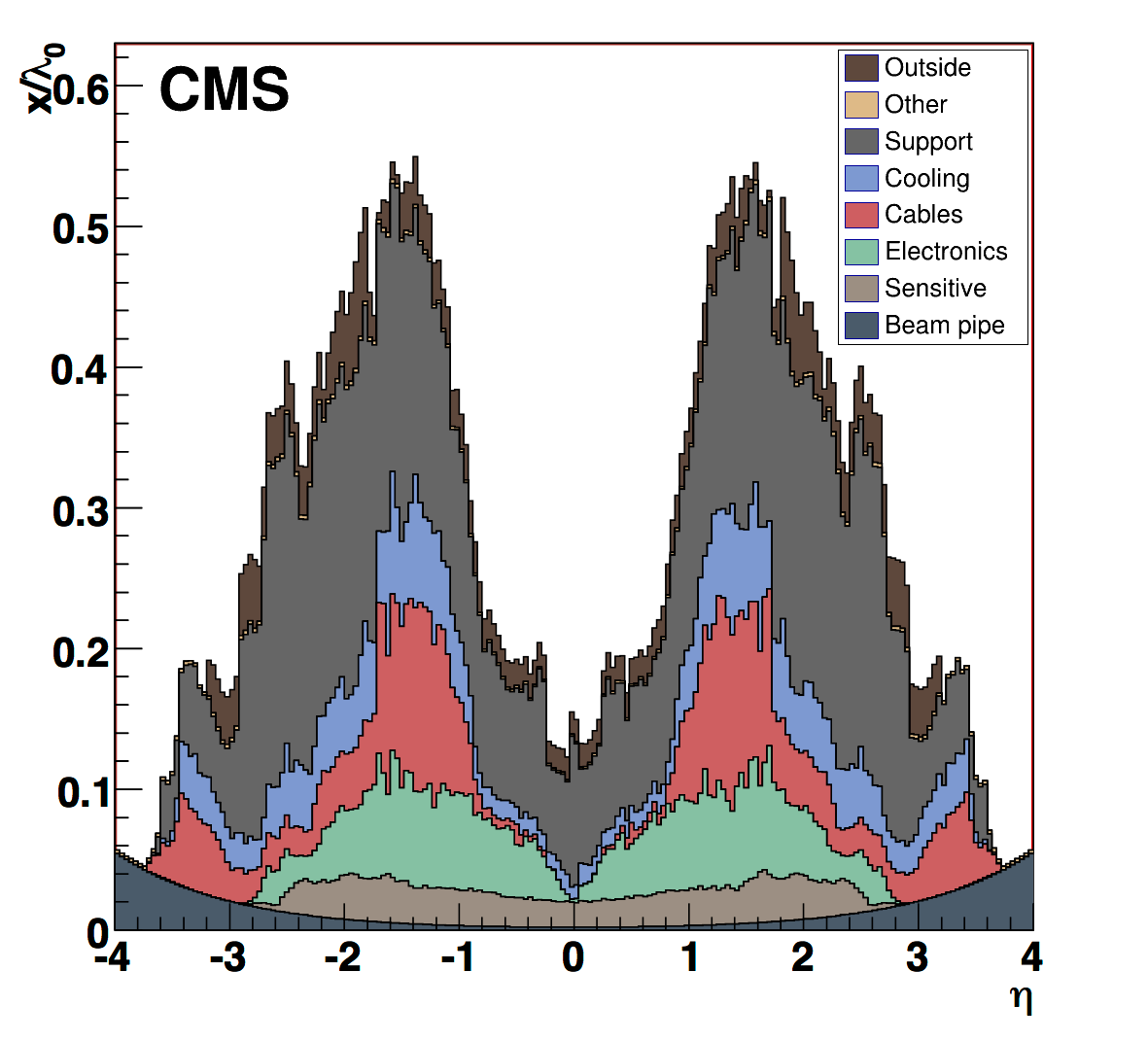}
\caption{Material budget in units of radiation length (left) and
  interaction length (right) as a function of $\eta$ in the SST
  simulation, shown for the different material categories: beam pipe,
  silicon sensitive volumes, electronics, cables, cooling pipes and
  fluid, support mechanics and outside structures (support tube, thermal
  screen and bulkheads).}
\label{fig:matBud}
\end{figure}

The SST material budget given in units of radiation length ($X_0$) and interaction length ($\lambda_0$) is shown in Fig.~\ref{fig:matBud} as a
function of $\eta$, with different contributions stacked. The sensitive elements contribute only about 9\% of the
total. The dominant contribution comes from the support structures
(${\approx}36\%$) and services (power cables, cooling pipes). In the
barrel-to-endcap transition regions, $1.2<\abs{\eta}<1.8$, the material
budget reaches 1.8\,$X_0$, due to the routing of
the cables and cooling pipes from the inner regions. In the central part of the tracker
($\abs{\eta}<0.8$) the material budget remains below 0.8\,$X_0$.

After the \GEANTfour energy deposition step in the sensitive material described above, the SST simulation proceeds
as follows. First, the behavior of charges in the silicon is
simulated: the energy deposited by \GEANTfour is distributed along the path of
  the particle in the silicon. Then each individual subdeposit is migrated to the surface of the
  sensor, taking into account local effects in the silicon sensor,
  \eg, the drift due to the magnetic and
  electric fields, together with the effect of thermal diffusion. The result is a charge on the surface, with a position and a width. The charge contributions are merged for each strip. The resulting
signal is then processed to include the effects of the electronics and digitization 
to derive the so-called digis, which are the detector
event data as output by the electronics.  
By design, the digis have the same data format
as data from the detector.

The simulation implements the following steps:
\begin{itemize}
\item modeling of the signal pulse shape, in peak or deconvolution mode,
\item addition of noise (and the pedestal value in NZS readout mode),
\item addition of in-time and out-of-time pileup contributions by
  adding signals from 12 bunches before and 3 after the actual
  collision bunch, using simulated minimum bias
  collisions,
\item modeling of the time of the signal hits: the time of each simulated hit is shifted taking into account the bunch of its origin, and the shifted time for each hit is considered when generating the
  pulse shapes. The signals from out-of-time particles are scaled down
  in size according to the time response of the electronics,
\item application of conversion factors (gains) for electronic channels,
\item addition of inter-strip cross talk.
\end{itemize}

\begin{figure}[bh]
\centering
\includegraphics[width=.45\textwidth]{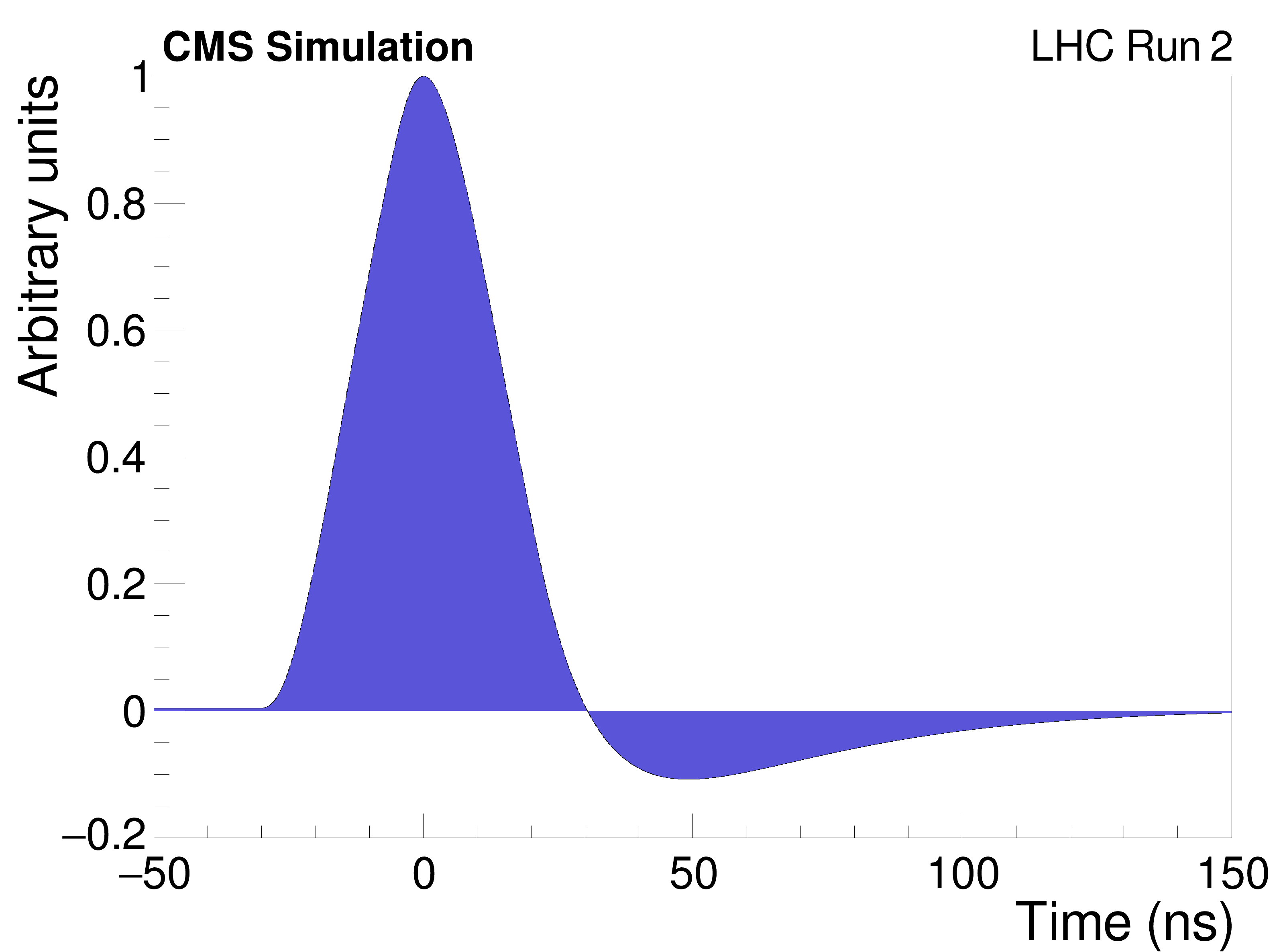}
\caption{Simulated APV25 pulse shape (deconvolution mode) in the CMS simulation software. The peak is centered at 0, corresponding to a perfect timing alignment with respect to LHC collisions.}
\label{fig:pulseshape}
\end{figure}

In deconvolution
mode the signal is processed with an algorithm to reduce its duration to a
25\unit{ns} time window. 
The deconvolution pulse shape is parametrized in the simulation based on
the results obtained from Ref.~\cite{Delaere:1061284}, with the shape
shown in Fig.~\ref{fig:pulseshape}.

Inter-strip cross talk is caused by inter-strip capacitance. The total charge deposited by a particle is
visible in three neighboring strips. In 2018, during a planned CMS maintenance with 0~magnetic field, the cross talk was
measured in NZS data using cosmic
ray muons~\cite{Jansova:these}. Table~\ref{tab:xtalk} summarizes the measurements obtained
during this dedicated campaign. The majority of the charge (between 75 and 86\%) is collected in the leading strip and its two neighbors (6--10\%). 

\begin{table}[tb]
  \topcaption{Cross talk measured in the barrel in 2018, obtained with
    cosmic ray data taken without magnetic field.  Results are
    summarized as the average fraction $f_0$ of observed charge for
    the leading strip, the fraction $f_1$ on each of the two
    neighboring strips, and the fraction $f_2$ on each of the strips
    next to these neighboring strips.}
\centering
\begin{tabular}{ l c c c }
 & $f_0$  & $f_1$ &  $f_2$ \\
\hline
TIB Layers 1, 2  & $0.836\pm0.009$  & $0.070\pm0.004$ & $0.012\pm0.002$ \\
TIB Layers 3, 4 & $0.862\pm0.008$  & $0.059\pm0.003$ & $0.010\pm0.002$ \\
TOB Layers 1--\,4  & $0.792\pm0.009$  & $0.083\pm0.003$ & $0.020\pm0.002$ \\
TOB Layers 5, 6  & $0.746\pm0.009$  & $0.100\pm0.003$ & $0.027\pm0.002$ \\
\hline
\end{tabular}
\label{tab:xtalk}
\end{table}

\begin{figure}[tb]
\centering
\includegraphics[width=0.4\textwidth]{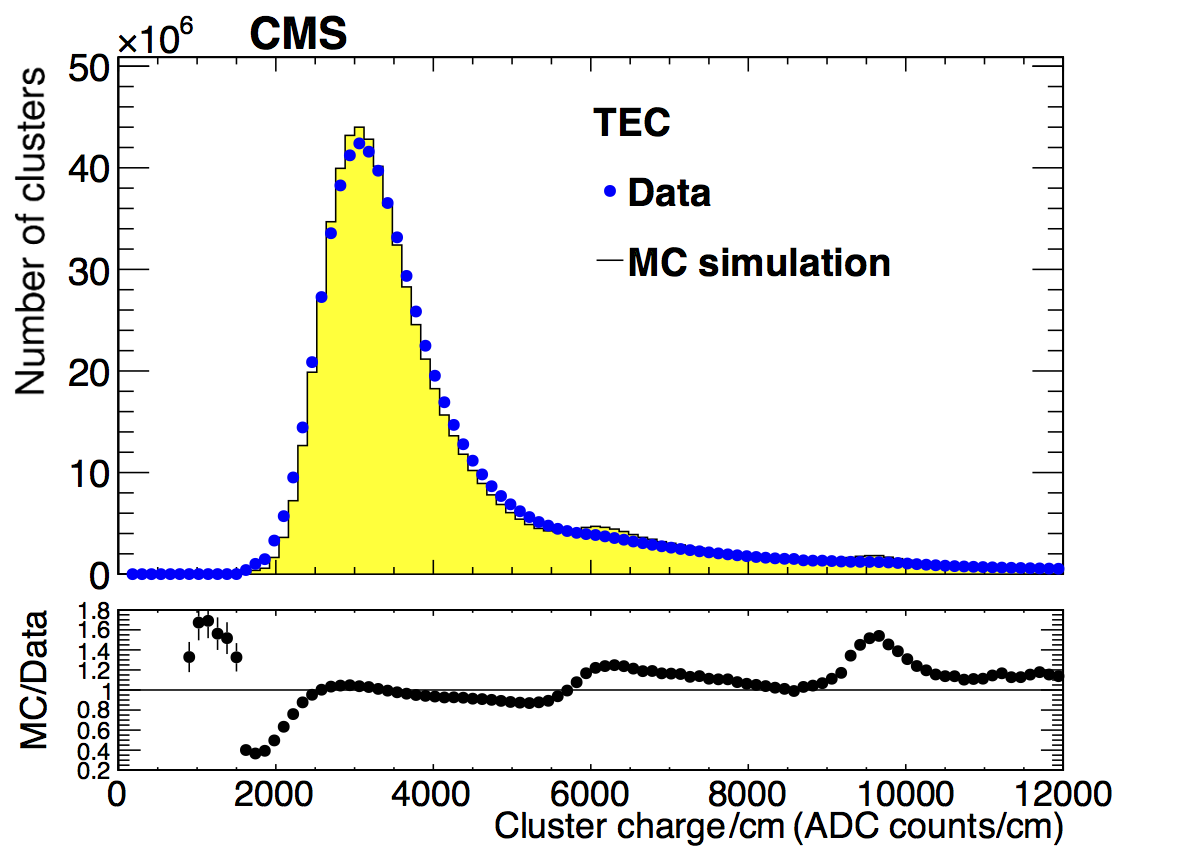}
\includegraphics[width=0.4\textwidth]{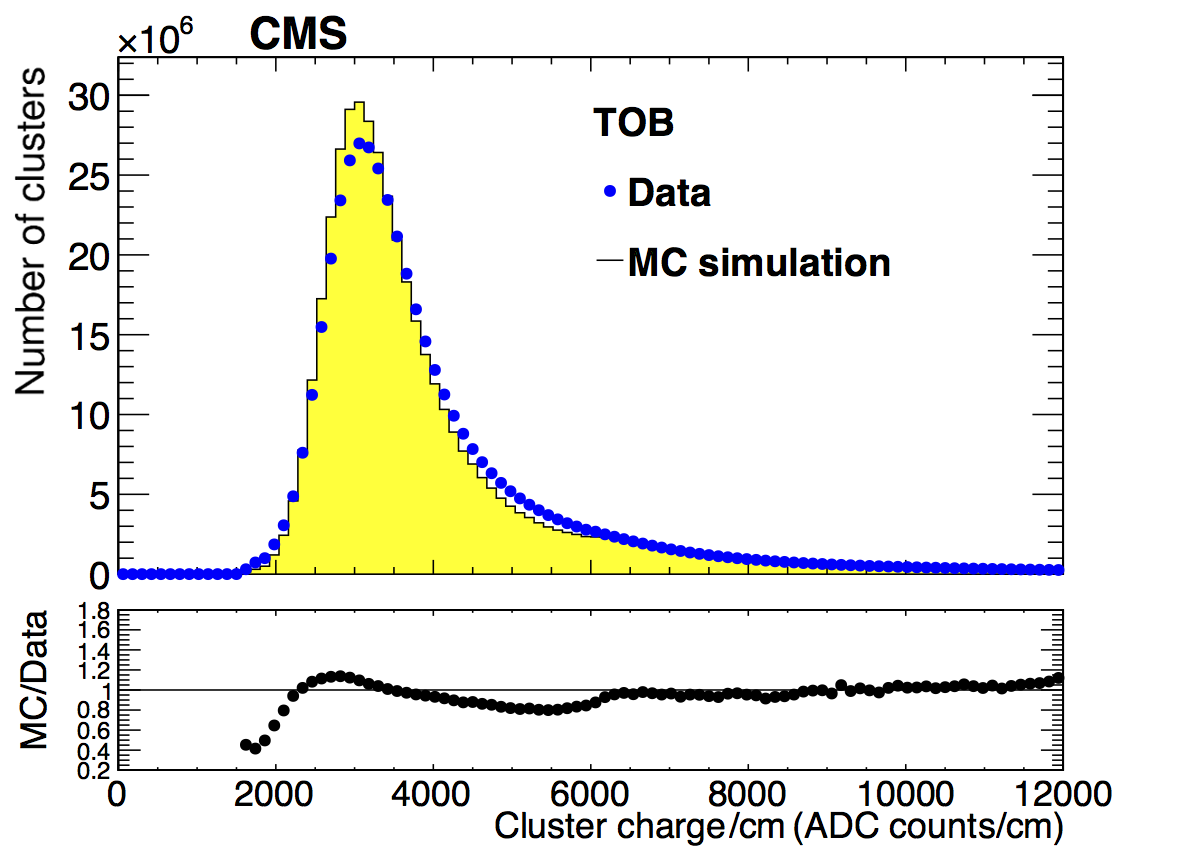}
\includegraphics[width=0.4\textwidth]{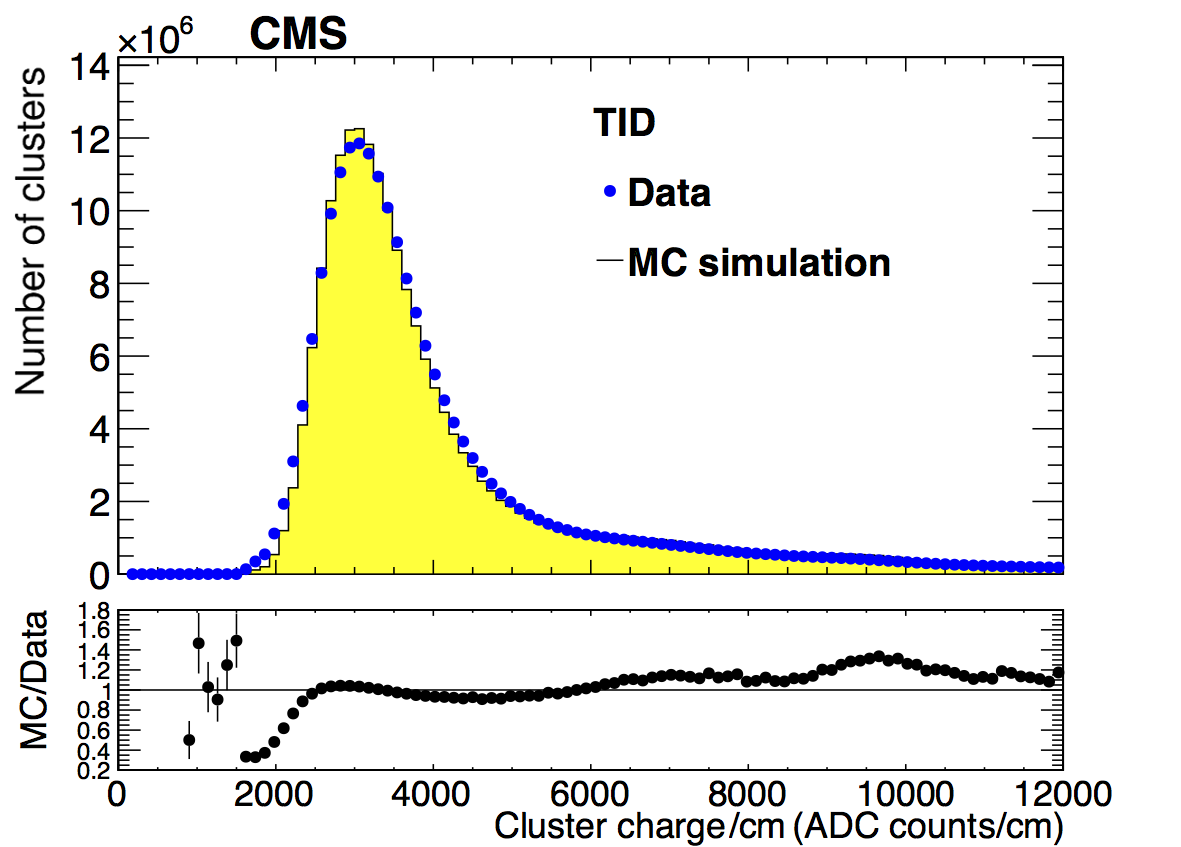}
\includegraphics[width=0.4\textwidth]{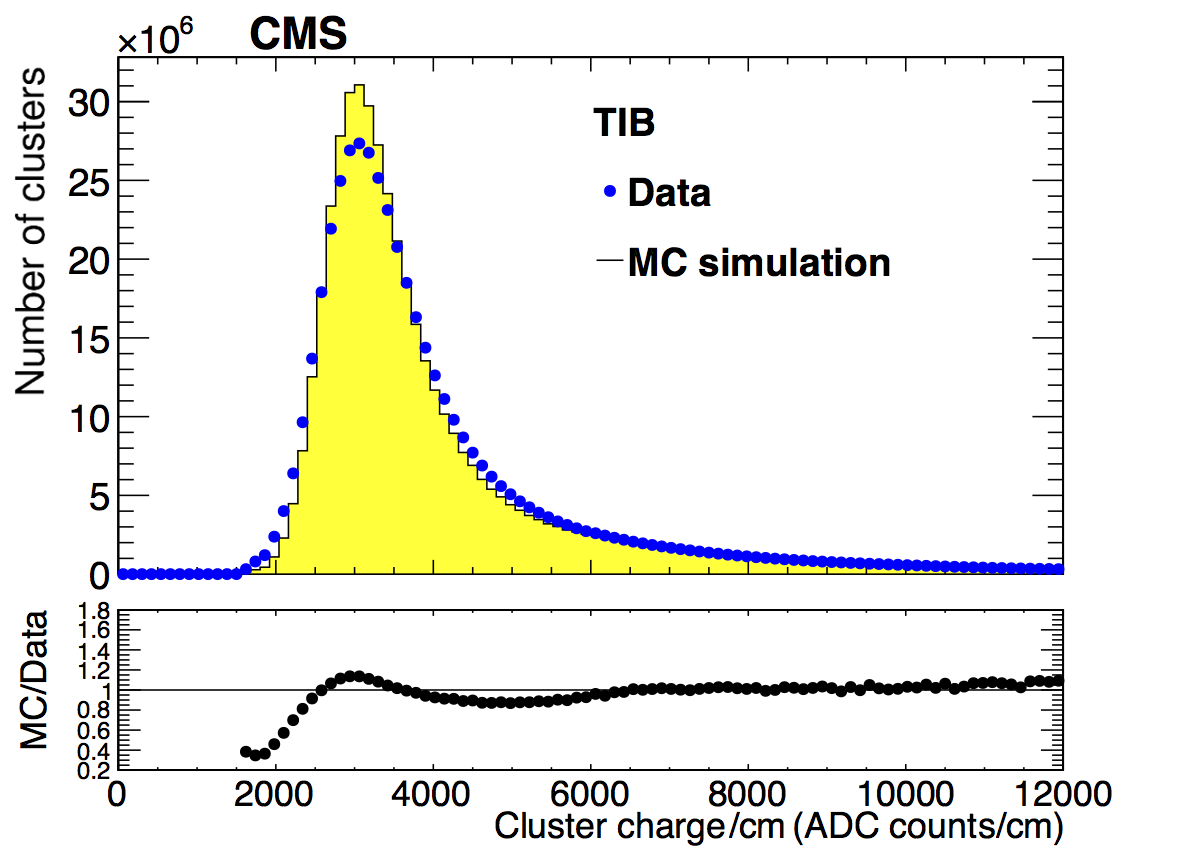}
\caption{Simulated and measured cluster charge normalized with the track path length for the different SST subdetectors: TEC (upper left), TOB (upper right), TID (lower left), and TIB (lower right). The measurements are shown by points, whereas the simulations are shown by yellow histograms. Lower panels show the ratios of the simulated predictions to data.}
\label{fig:clustercharge}
\end{figure}

One of the main figures of merit for the SST simulation is the
comparison between the simulated and measured cluster charge,
corresponding to the charge deposited by a particle crossing the
sensor. In Fig.~\ref{fig:clustercharge} the simulated cluster charge is in sufficient
agreement with the measured charge for each SST partition, as
can be seen.

\section{Detector performance with LHC collisions}\label{sec:collision_performance}

In this section, the performance of the SST during LHC
collision periods Run~1 (2009--2012) and Run~2 (2015--2018) is presented. 

The steps required to reconstruct charged-particle tracks from SST hit
data are summarized below. More details are reported
in~Ref.~\cite{Chatrchyan:2014fea}. A good spatial alignment of the
detector is a prerequisite for the precise reconstruction of particle
tracks. Details about the CMS strategy for the alignment of the SST
and the pixel detector are reported in Ref.~\cite{TRK-20-001}.

\paragraph*{Cluster finding}~\\
Clusters of strips are reconstructed starting from
strips with charges above threshold that are identified by the zero-suppression logic
implemented inside the FEDs. Further requirements are imposed for a strip to be included in a cluster. A three-threshold
algorithm is used. 
A strip is considered a seed for a cluster if its
charge is larger than three times its noise. The seed strip forms a
proto-cluster to which more strips can be added.
Strips adjacent to cluster candidates are
added to it if their charge is larger than twice their
noise. This procedure is repeated until no more strips are found
for addition to the cluster. No holes are allowed in a cluster unless a bad strip,
as defined in Section~\ref{sec:bad-components}, is encountered in which case the search
is continued. If a second consecutive bad strip is encountered the search
is terminated. 
A cluster candidate is retained if the summed
signal of all strips in the cluster candidate is larger than five
times the cluster noise, defined as
\begin{equation}\label{eq:cluster_noise}
  \sigma_{\text{cluster}} = \sqrt{\sum_{i=1}^{n_{\text{strips}}} \sigma_{\mathrm{i}}^{2} },
\end{equation}
where $\sigma_{\mathrm{i}}$ is the noise of an individual strip
and the sum runs over all $n_{\text{strips}}$ in the cluster candidate.

\begin{figure}[t!b]
    \centering
    \includegraphics[width=0.5\textwidth]{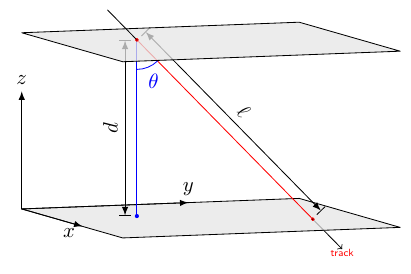}
    \caption{Path length $\ell$ of a particle crossing a detector of thickness $d$ at an angle $\theta$.}
    \label{fig:sketch_scaling}
  \end{figure}

\paragraph*{Path length correction}~\\
  For several of the measurements presented in this paper, the cluster charge is
  corrected for the length $\ell$ of the particle trajectory inside the
  active silicon volume (Fig.~\ref{fig:sketch_scaling}). The charge of a cluster from a
  track that passed through the sensor with thickness $d$ at an angle
  $\theta$ with respect to perpendicular incidence on the sensor is
  scaled as $\cos \theta = d/\ell$ so that $S_{\text{corr}} = S_{\text{raw}}\cos\left(\theta\right)$,
  where $S_{\text{raw}}$ is the cluster charge before any correction
  is applied and $S_{\text{corr}}$ is the corrected cluster charge.
  With this correction cluster charges are normalized to the one
  expected under vertical incidence.

\begin{figure}
  \centering
  \includegraphics[width=0.5\textwidth]{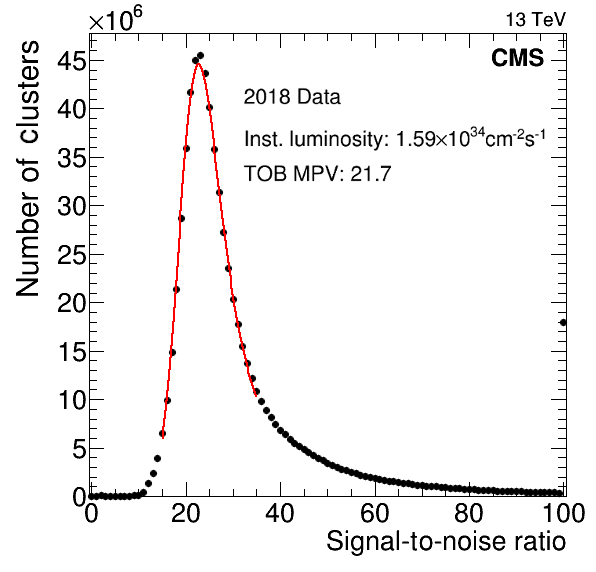}
  \caption{Example of a signal-to-noise distribution from the TOB recorded during 2018 at an instantaneous luminosity of 1.59\ten{34}\percms. The central part of the distribution is fitted with a Landau convoluted with a Gaussian distribution. The bin at 100 serves as the overflow bin.}
  \label{fig:sovern_example}
\end{figure}

\paragraph*{Cluster signal-to-noise ratio}~\\
A frequently used quantity is the signal-to-noise ratio~(\sovern) of
a cluster. In this paper we define this as the corrected cluster
charge $S_{\text{corr}}$, defined above, divided by the RMS
noise of the individual strips in the cluster. This can be
expressed as $S_{\text{corr}}/\left( \sigma_{\text{cluster}}/\sqrt{n_{\text{strips}}}\right)$ with the
cluster noise definition from Eq.~(\ref{eq:cluster_noise}).
The distribution of the energy loss in
silicon follows a Landau distribution~\cite{PDG2024}. An example of the resulting \sovern{}
distribution from a $\Pp\Pp$ collision run is shown in
Fig.~\ref{fig:sovern_example}. The central part of the distribution is
fitted with a Landau convolved with a Gaussian distribution, and the most probable value~(MPV) of the Landau distribution is
taken as a measure of the \sovern.

\paragraph*{Track finding} ~\\
  Tracks are reconstructed from hits in the pixel detector and the
  SST. Multiple iterations of track finding are performed. Early
  iterations reconstruct prompt tracks originating from the
  interaction region. Later iterations reconstruct
  topologies such as displaced tracks. Clusters already associated with high quality
  particle tracks are not reused in later iterations (excluding regional iterations near high transverse momentum jets or muons). Track seeds are
  constructed preferentially from hits in the pixel detector
  (triplets for the original, quadruplets for the upgraded pixel detector).
  Seeding in subsequent iterations relies on triplets or doublets of pixel hits or a mix of hits in the
  pixel and innermost layers of the SST, taking into account
  constraints from the size of the interaction region. Later iterations also use combinations of TOB and TEC layers for track seeding. The final track
  fit is performed using a Kalman filter algorithm. Clusters
  associated with a reconstructed track are called on-track clusters.

\subsection{Detector occupancy}\label{sec:occupancy}
\begin{figure}[tbp]
\centering
\includegraphics[width=0.7\textwidth]{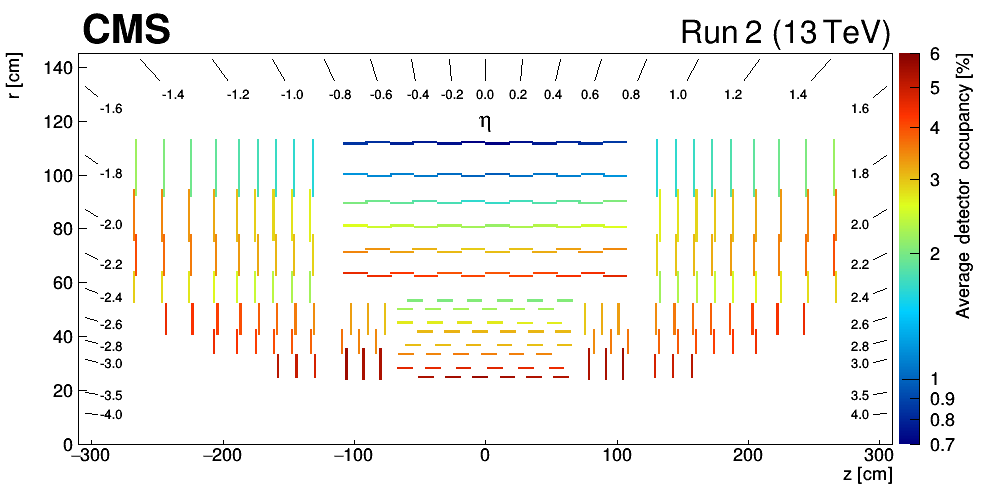}
\caption{View of the SST mean strip occupancy in the $r$-$z$ plane recorded at the beginning of a typical LHC fill where the pileup was at its maximum (of the order of 55).
} 
\label{fig:occupancy}
\end{figure}

The mean detector occupancy, defined as the average fraction of
detector cells that are traversed by one or more particles per event, is an important quantity to consider when designing 
detectors, in particular when defining the granularity of the sensors. The SST was designed to ensure the occupancy is at most a few percent. The occupancy depends on the pitch, 
the strip length, and the particle flux in a given location within the detector. It is measured via the strip occupancy, defined as the ratio between the 
total number of strips in the reconstructed
clusters associated with a particle crossing a sensor and the total
number of strips in the detector.  Figure~\ref{fig:occupancy} shows the average strip occupancy at the beginning of an LHC fill for a mean number of $\Pp\Pp$ interactions of about 55 per bunch crossing, which can be translated into an instantaneous luminosity of $2\times10^{34}\percms$. Even at this high value of pileup, more than twice the design value, the occupancy is at most a few percent, in line with the design values defined in Ref.~\cite{CMS:2000aa}.

The lowest occupancy (0.7\%) occurs in the last layer of the barrel (TOB) whereas the highest occupancy (5.6\%) occurs in the innermost ring of the TID. Readout deadtime due to the FED buffer occupancy, leading to a loss of data-taking efficiency, is expected at a level-1 trigger rate of 100\unit{kHz} only for occupancies above 8\%~\cite{Foudas:900442}, a value that is not reached in the SST. 

The occupancy is monitored daily, because a deviation from typical values can be a sign of a detector issue, \eg, an increase of the noise, or a localized detector failure.

\subsection{Bad component identification}\label{sec:bad-components}
Failure of components of the SST can occur for various reasons, ranging from radiation damage to thermal and electrical issues. 
Although some external failures, \eg, of the power supplies or the cooling plants, can be repaired on a short time scale, interventions within the detector itself are practically impossible. The detector is monitored continuously so that any failures can be rapidly recognized, and if possible, repaired. Likewise miscalibration or other operational issues can be identified and corrected. When noisy or inefficient electronic channels are identified, the information is propagated to the offline data processing so these 
channels can be removed before reconstructing the tracks. 
Since the detector has considerable redundancy, the omission of such channels does not degrade the track reconstruction performance in almost all cases.

During data taking, the identification of bad components in the SST is performed automatically in the 48~hours that CMS maintains between data taking and the start 
of offline reconstruction, during which an express processing of the CMS data is performed. An automatic analysis identifies defects at the level of the APV25 each time 150\,000 clusters are reconstructed within the SST during this express processing. During standard LHC operation, this occurs about once per day. An iterative comparison is made between the median occupancy of each APV25 chip and the mean of the medians of all APV25 chips in a given layer and the same $z$ region in the barrel, or in a ring and the same $r$ region in the endcap. At each iteration, 
APV25 chips with a median occupancy more than 3 standard deviations from the mean of the respective comparison sample of APV25 chips are removed from the next iteration. At the end of this process, all APV25 chips with outlier occupancies are removed from the offline reconstruction. A last iteration targets issues at the level of single strips. For the APV25 chips not identified as problematic in this process, the mean occupancy of each hit strip is estimated assuming a Poisson probability distribution. Single strips for which this probability is too low ($\leq 10^{-7}$) or that are outliers with respect to the mean occupancy in the considered chip are also ignored in the offline reconstruction.\\

\begin{figure}[th]
\centering
\includegraphics[width=1.1\textwidth]{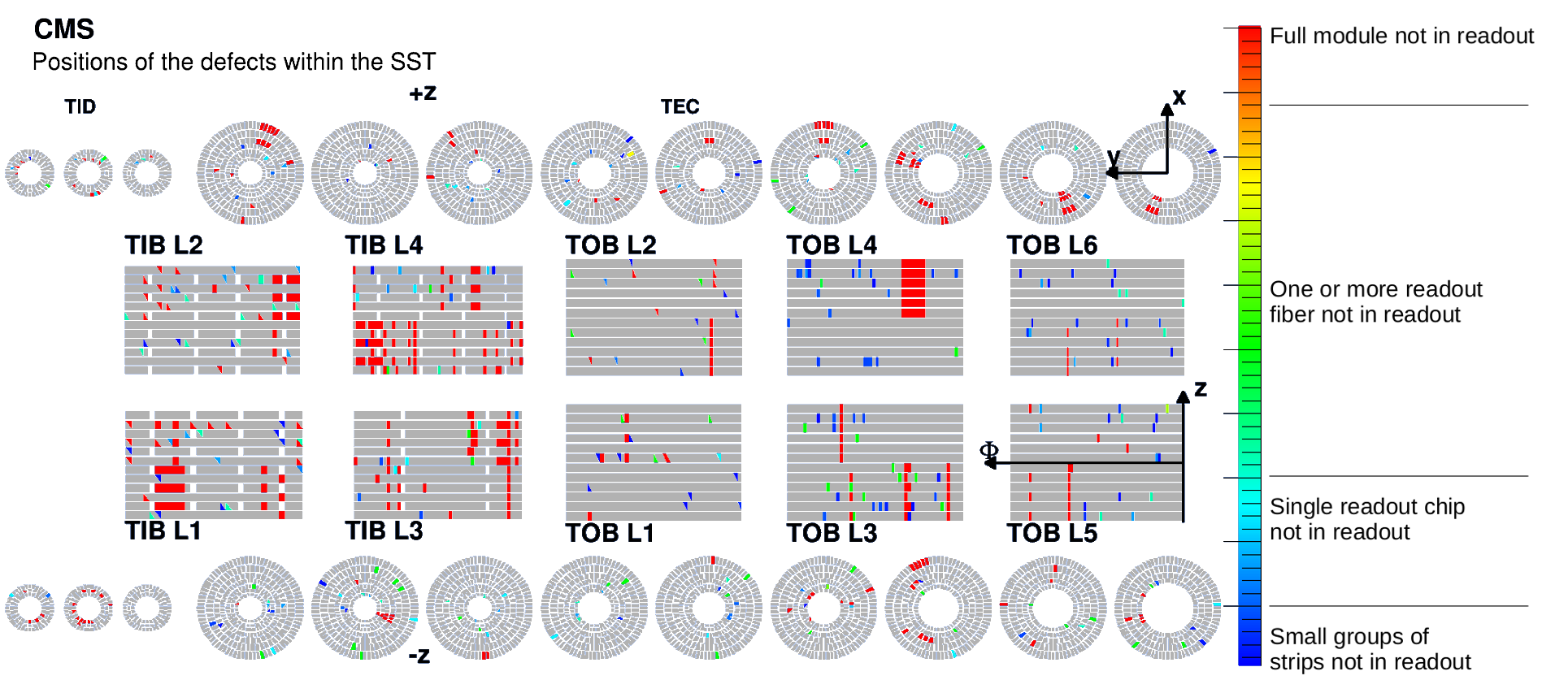}
\caption{Positions of the defects within the SST at the module level at the end of 2017. Disabled modules are shown in red. Those with at least one fiber not in the readout appear in green. The range blue to green indicates damage varying from single strips (dark blue) to all strips of a single chip. The total number of defects represents less than 5\% of the total number of the SST channels.}
\label{fig:trackermapbadcomp}
\end{figure}

Figure~\ref{fig:trackermapbadcomp} summarizes the channel status for a representative run at the end of 2017. Fully operational modules appear as gray, whereas modules with issues are colored. Disabled modules are also shown in red (these are the permanent defects described in Section~\ref{sec:detector-status}). Broken optical fibers affect a small number of modules (colored green). Modules colored blue suffer from other problems, ranging from a single dead strip to an entire defective APV25 chip.

\begin{figure}[tbp]
\centering
\includegraphics[width=0.65\textwidth]{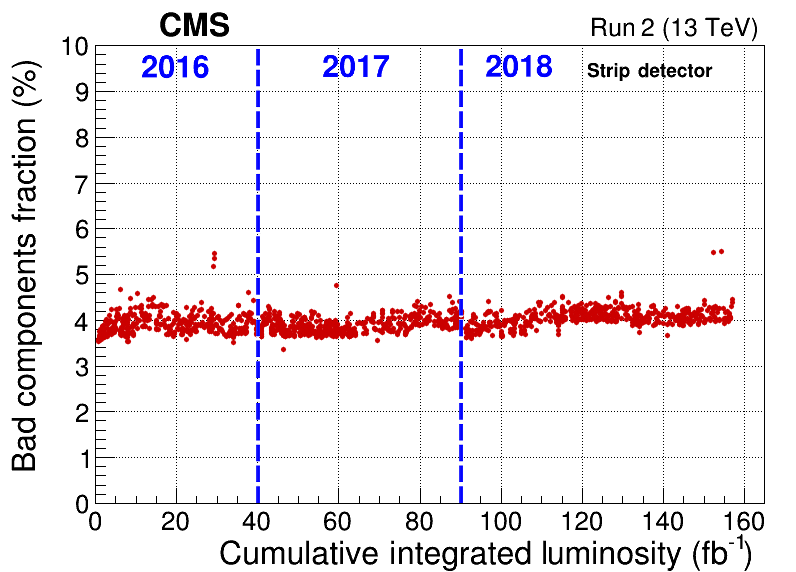}
\caption{Fraction of bad channels as a function of the delivered LHC integrated luminosity. The vertical lines indicate the start of each calendar year.}
\label{fig:badchantrend}
\end{figure}

Overall, the operational fraction of the strip tracker at the end of LHC Run~2 was 95.5\%. Table~\ref{tab:alive-fraction} shows the fraction of live channels for each SST partition. As shown in Fig.~\ref{fig:badchantrend}, the number of defects within the SST remained stable during Run~2 with an average fraction of bad channels just above $4\%$. The fluctuations visible in the live-channel fraction are mostly due to transient power supply issues or temporary failures of
component configuration. 
Both these causes generally affect only a small part of the detector, and only for relatively short times (typically from a few hours to a few days).

\begin{table}[tb]
\topcaption{Fraction of live channels in the different readout
  partitions of the SST at the end of data taking in 2018.}
\centering
\begin{tabular}{lp{1cm}l}
Partition & & Percentage of live channels\\
\hline
TIB/TID & & 91.5 \%\\
TOB & & 96.7 \%\\
TEC$+$ & & 96.9 \%\\
TEC$-$ & & 97.4 \%\\
\hline
SST & & 95.5 \%\\
\end{tabular}
\label{tab:alive-fraction}
\end{table}

\subsection{The APV25 preamplifier saturation}\label{sec:apv_saturation}

In late 2015 and early 2016 the strip tracker experienced a decrease in
signal-to-noise ratio, also associated with a loss of hits on tracks, with
the effect becoming more pronounced with increasing instantaneous
luminosity. It was eventually found that the problem arose from saturation
in the preamplifier of the APV25 chip at high
occupancies.
The capacitor in the feedback loop (Fig.~\ref{fig:apv25_single_channel}) is charged by the signal current,
and then discharges
through a resistance formed by the channel of a
field-effect transistor (FET) in parallel with the capacitor, over a time which is long compared with the 25\unit{ns}
bunch crossing. Although the effect of high
hit rates was studied during prototyping, no effects were
observed during CMS data taking until the instantaneous luminosity regularly
exceeded $3\times10^{33}\percms.$, which means more than 20 interactions per bunch crossing.

In the absence of a build up of charge on the feedback capacitor from previous particle crossings, the
response of the preamplifier is linear up to about 3~MIPs but for higher charges
the response is no longer linear~\cite{Jones:432224}. The maximum
signal size accommodated in the system is $\approx 7$~MIPs, limited by the
magnitude of the output voltage of the APV25 and the ADC inside the FED, whereas the preamplifier
alone maintains good linearity to about 30~MIPs. 

The significant reduction of the operating temperature for Run~2 resulted
in a large and unexpected increase of the preamplifier discharge time
compared to Run~1 when the SST was operated at $+4^{\circ}$C. 
The slow discharge speed together with the high-occupancy conditions
resulted in a gradual build up of charge in the preamplifier and consequently a
very nonlinear and reduced response of the amplifier to newly deposited
charges. Once the problem was understood, it was resolved by adjusting the
preamplifier feedback voltage bias (abbreviated as VFP for voltage feedback preamplifier), which controls the feedback FET
channel resistance, to reduce the preamplifier discharge time constant
($\tau$). 

\begin{figure}[t!b]
  \centering
  \includegraphics[width=0.6\textwidth]{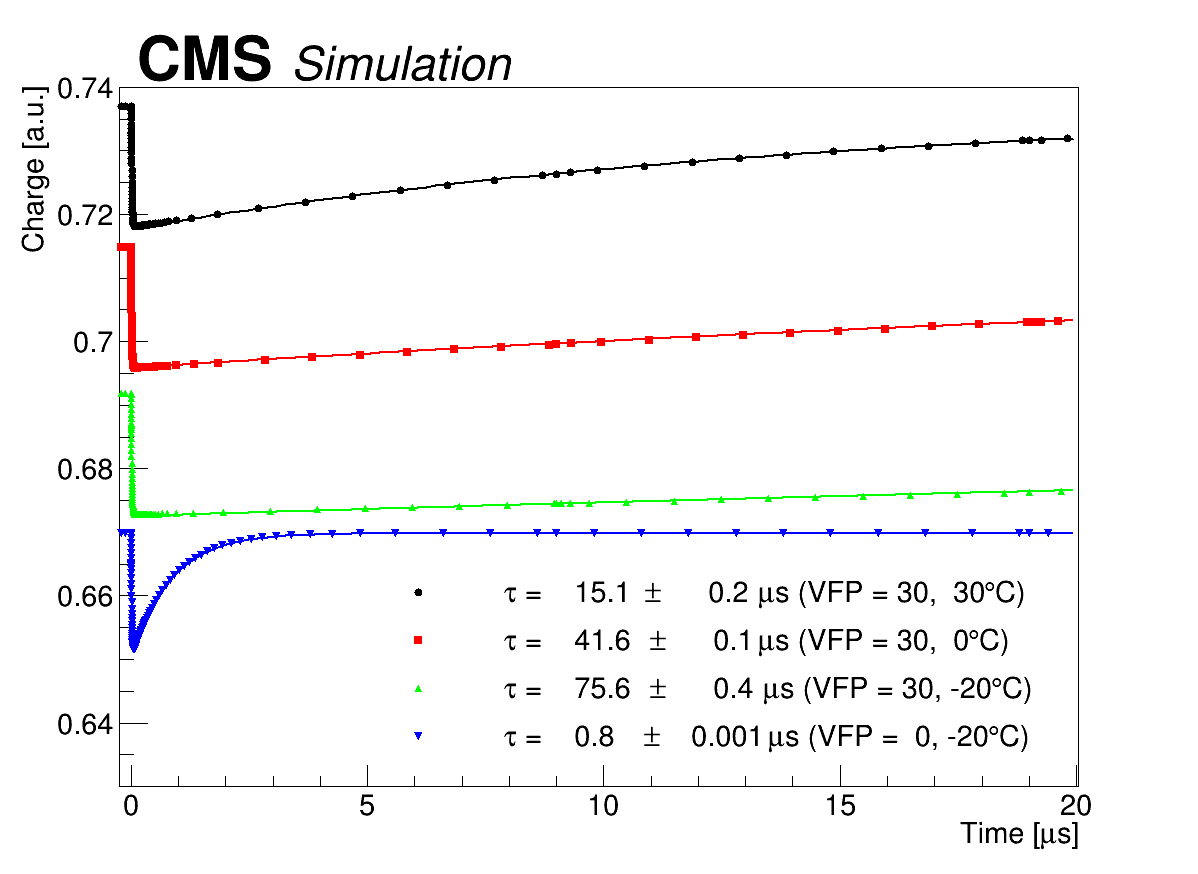}
  \caption{Simulated discharge behavior of the APV25 preamplifier for
    different temperatures with preamplifier feedback voltage bias
    (VFP) at 30 (Run~1 and 2015--2016) and at 0 (second half of 2016--2018).
    An exponential decay function with a discharge time
    constant $\tau$ is fitted to each set of simulated data. The
    resulting values for $\tau$ are displayed in the legend.}
  \label{fig:preampsat-discharge-time}
\end{figure}

The simulated behavior of the preamplifier is shown in
Fig.~\ref{fig:preampsat-discharge-time}. The charge state of the
preamplifier is changed (in this case, reduced) by the arrival of a charge
from the sensor. The charge then drains away until the original
charge state is recovered. One can see that for \mbox{$\text{VFP}=30$}, the
value used in Run~1 and early in Run~2, the discharge time increases
strongly with lower temperatures. With $\text{VFP}=0$ the discharge
time is below 1\mus even at low temperature. Given that no adverse
effects were seen from this change, the setting was retained at $0$,
the lowest possible value, for the remainder of Run~2.
The effect of the preamplifier saturation in the early 2016 data
taking will be discussed in the following sections.

Since the inefficiency cannot be recovered in the data
affected by the saturation effects (about 20\fbinv{}), the effect has to be 
included in the simulation. A model of the preamplifier
saturation was developed, which is applied on top of the detector
simulation described in Section~\ref{sec:simulation}.
The charge state of an APV25 preamplifier depends on the specific
layer of the SST, on the $z$ position of the detector module within the layer,
and on the pileup.

\begin{figure}
  \centering
  \includegraphics[width=0.45\textwidth]{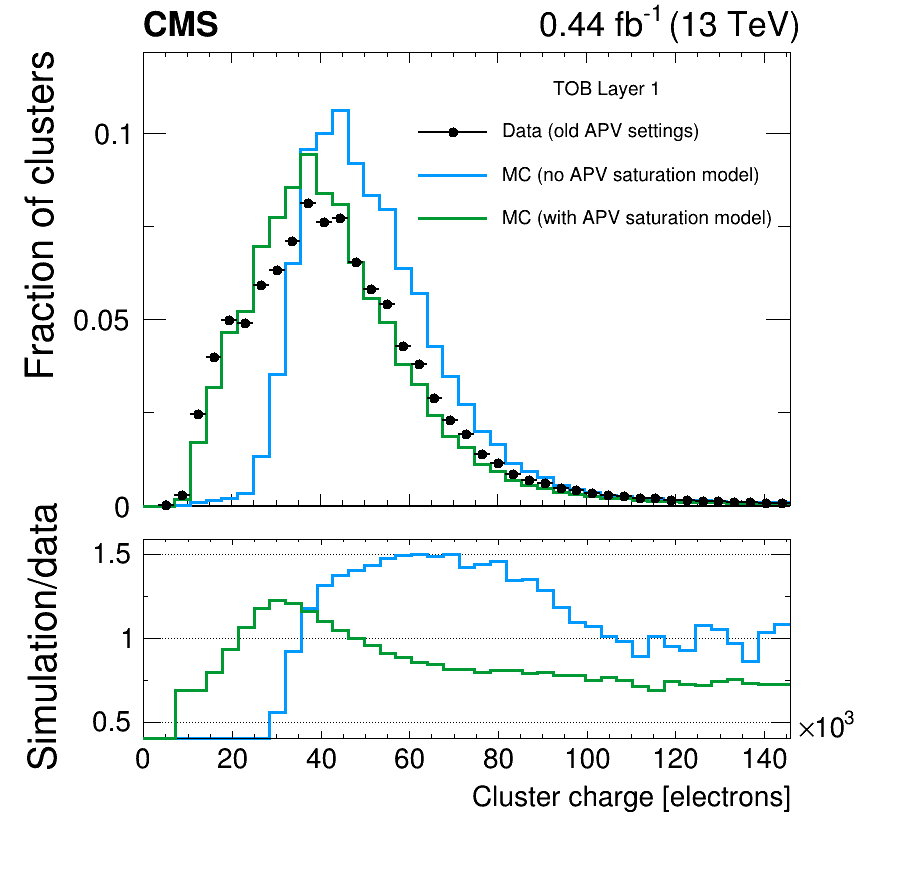}
  \includegraphics[width=0.45\textwidth]{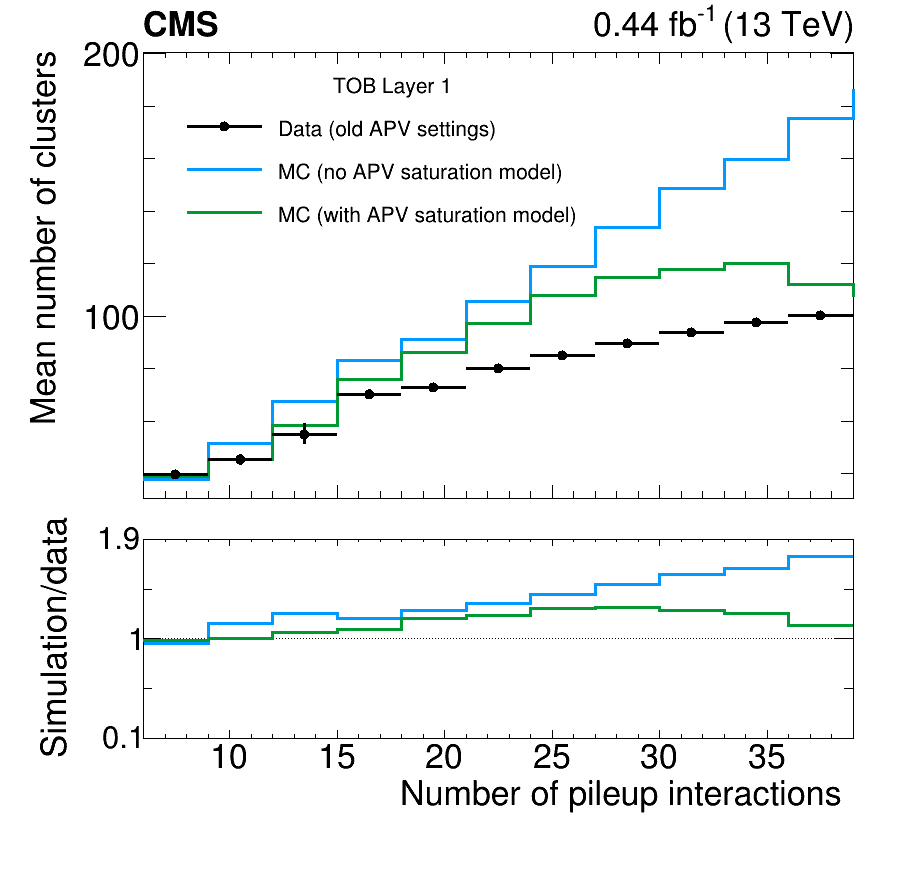}
  \caption{Left: cluster charge distribution for clusters on
    reconstructed particle tracks in the TOB layer~1. Right:
    multiplicity of clusters on reconstructed particle tracks in the
    TOB layer~1 as a function of the number of pileup interactions. The data from early 2016 (black dots) are compared with
    two different MC simulations; one (blue) does not contain any special treatment
    of the APV25 preamplifier saturation, the other
    (green) contains a modeling of the preamplifier saturation to
    account for the reduced charge response of the amplifier under
    high-occupancy conditions. Lower panels show the ratio of MC predictions to data.}
  \label{fig:APV_preamp_simu_cluster_charge}
\end{figure}

\begin{figure}
  \centering
  \includegraphics[width=0.45\textwidth]{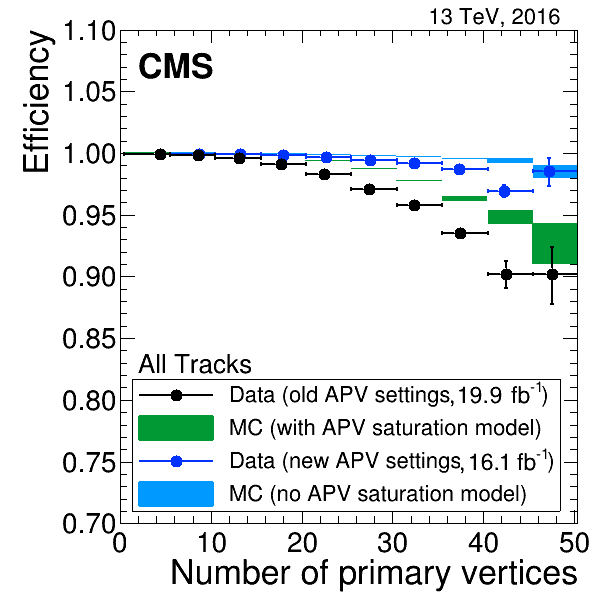}
  \caption{Tracking efficiency estimated using a tag-and-probe method
    as a function of the number of primary vertices for an inclusive
    muon track collection~\cite{CMS-DP-2020-035}. Data from early 2016
    (black dots) and late 2016 (blue dots) are compared with two
    different MC simulations: one (blue) does not contain any
    special treatment of the APV25 preamplifier saturation, the
    second (green) contains a modeling of the preamplifier
    saturation to account for the reduced charge response of the
    amplifier under high-occupancy conditions.}
  \label{fig:muon_eff_tp_run2legacy_including_vfp_simu}
\end{figure}

The effect of applying the model to simulated events can be seen in
Fig.~\ref{fig:APV_preamp_simu_cluster_charge}. 
For both the charge of on-track clusters (left) and the multiplicity of
clusters as function of pileup (right), the Monte Carlo~(MC) events including the APV25
preamplifier saturation describe the data significantly better than without it.
The effect on the muon reconstruction efficiency measured using a tag-and-probe
method~\cite{CMS:2010svw} 
is shown in
Fig.~\ref{fig:muon_eff_tp_run2legacy_including_vfp_simu} for an
inclusive muon track collection~\cite{CMS-DP-2020-035}. Two sets of data are shown: data
from early 2016 (black) affected by the
preamplifier saturation and data from late 2016 (blue) after the VFP parameter change.
They are compared with two MC simulations, one containing a modeling of the preamplifier saturation (green)
and one without it (blue). The sample with the preamplifier saturation model describes the data
from early 2016 significantly better than the sample without this model.
Similarly the data after the VFP change are better described by the MC without any preamplifier
saturation model, as expected.

\subsection{Signal evolution}\label{sec:signal_evolution}

\begin{figure}[htb]
  \centering
  \includegraphics[width=0.6\textwidth]{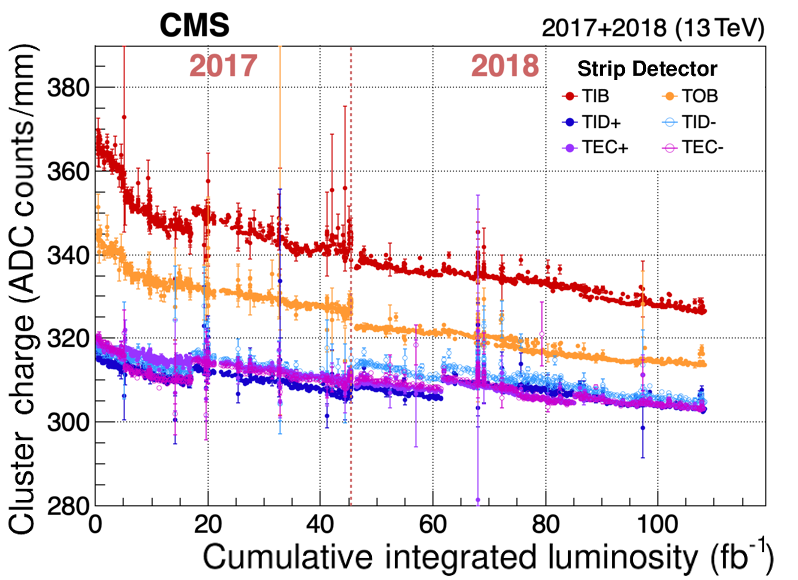}
  \caption{Evolution of the cluster charge normalized to unit length
    as a function of the integrated luminosity for different parts of
    the SST. The zero point of the plot is after the integrated luminosities of Run~1 and the years 2015 and 2016 for a total of 75.3\fbinv{}.}
  \label{fig:signal_evolution}
\end{figure}

After the identification and exclusion of
nonoperational components, the stability of the signal
induced by charged particles in the detector can be studied.  The
cluster charge per unit path length as a function of the integrated
luminosity is shown in Fig.~\ref{fig:signal_evolution} for 
2017 and 2018. Each point in the distribution is a run during an LHC fill with more
than 1200~proton bunches colliding in CMS. Only runs for which the data have been 
certified as good are included~\cite{Tuura_2010}.
A Landau function is fitted to the cluster signal distribution. The
MPV from the fit is plotted, with the error bar being the uncertainty
on the MPV from the fit.  Large error bars arise from runs with few
events.  Data from 2016 are excluded because of the APV25 preamplifier
saturation described above. The MPV decreases as a
function of time with some discontinuities. The overall decrease
is caused by the accumulating effects of radiation exposure. The discontinuities occur at
recalibrations, particularly of the optical
links~(Section~\ref{sec:LLD_tuning}).

\subsection{Signal-to-noise performance}\label{sec:sovern_performance}

In this section the signal-to-noise performance
of the SST is discussed. The \sovern{} value is very important for the SST,
as the FED zero suppression and the offline selection of clusters for
the track reconstruction both depend on it.

\begin{figure}[th]
\centering
\includegraphics[width=0.65\textwidth]{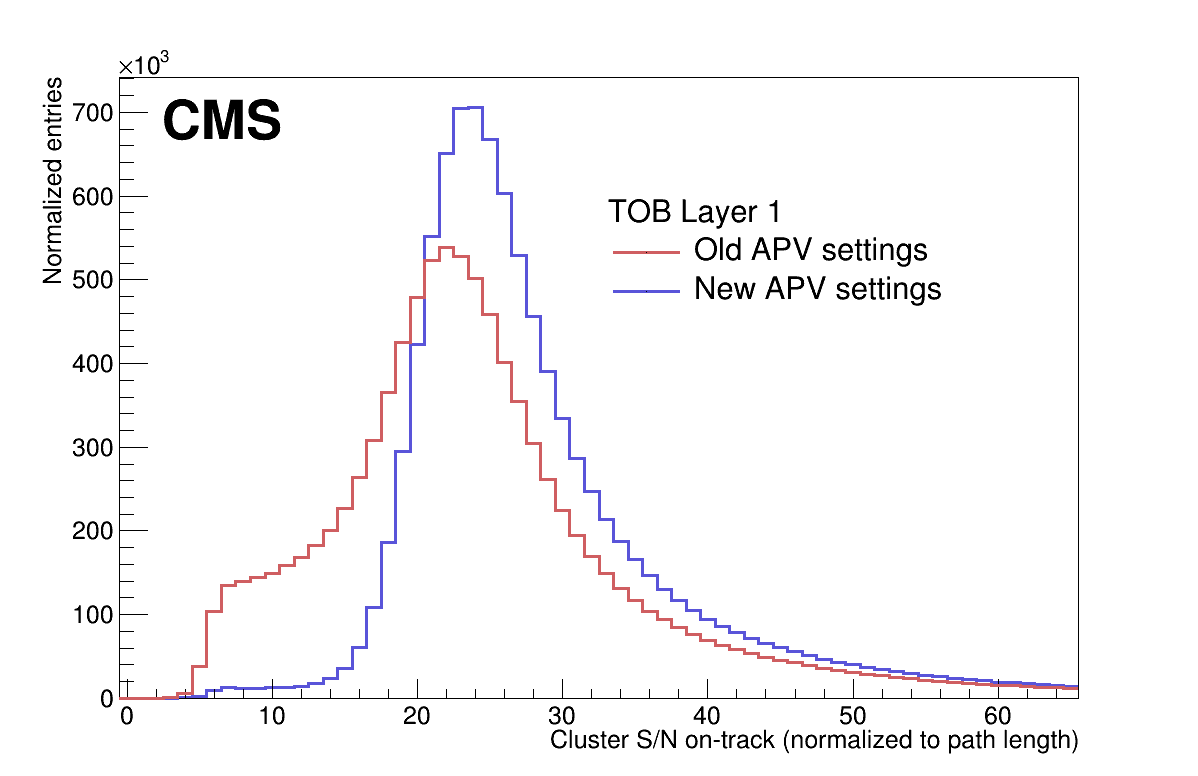}
\caption{Signal-to-noise ratio for clusters on reconstructed particle tracks
  in TOB layer 1 for two runs in 2016. The first run
  (red curve) is affected by saturation effects in the APV25
  preamplifier. In the second run (blue curve), the preamplifier voltage
  feedback (VFP) has been changed to shorten the discharge time of the
  preamplifier. Both curves are normalized to the same number of entries.}
\label{fig:preampsat-sovernshape}
\end{figure}

The impact of the APV25 preamplifier saturation in early 2016 on the \sovern{}
performance of the system is illustrated in
Fig.~\ref{fig:preampsat-sovernshape}, where the \sovern{} for hits on tracks
is shown for TOB layer~1, which was the most affected region of the
detector. With the old settings, the shape of the \sovern{} deviates from the
expected Landau-like shape. A downward shift of the MPV 
and an increased population in the low-end tail of the
distribution can be seen. The second run~(new APV settings) was taken after the change of the
VFP parameter.  Under very similar running conditions (peak
instantaneous luminosity around $1\times10^{34}\percms$) 
the \sovern{} distribution
after the VFP change is completely recovered and shows the expected
Landau-like shape, with only a very small population in the low-end
tail.

\begin{figure}[th]
\centering
\includegraphics[width=0.45\textwidth]{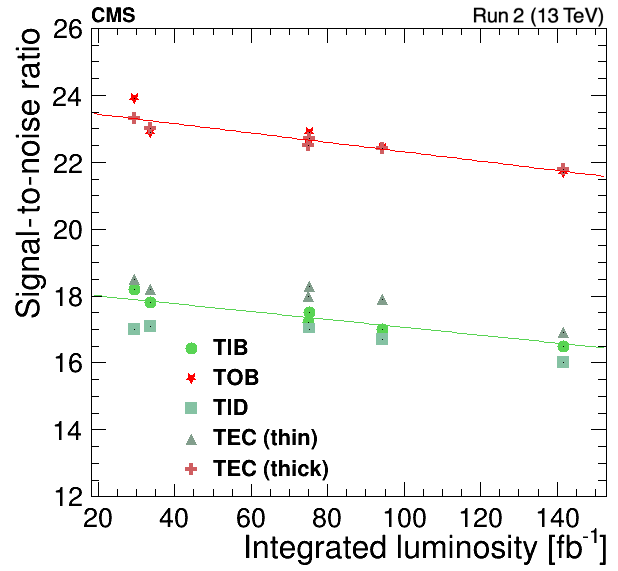}
\caption{Signal-to-noise ratio as a function of the integrated luminosity for modules
  from different parts of the tracker. Trend lines are obtained by
  separately fitting a linear function to the points for thin and
  thick sensors.}
\label{fig:sovern-trend-vs-fluence}
\end{figure}

The \sovern{} distributions for the different regions of the SST are
each fitted with a Landau convoluted with a Gaussian distribution. The MPV of the respective Landau
fit curve is quoted as the \sovern{} value for a specific detector
region.  The evolution of the \sovern{} as a function of the
integrated luminosity is summarized in
Fig.~\ref{fig:sovern-trend-vs-fluence}. In the TEC the \sovern{} is
plotted separately for thin and thick sensors. Trend lines are fitted
to the two populations of thin and thick sensors. The trend lines show
a decrease of 0.12/\fbinv for thin and 0.14/\fbinv for thick
sensors. Differences due to the different radial positions beyond the
separation of thin and thick sensors are neglected in this
plot. Extrapolating both curves to the design end of life of the SST,
\ie, to 500\fbinv of integrated luminosity, yields an expected
\sovern{} of 12.4 for thin and 16.7 for thick sensors, in good
agreement with the measurements in
Ref.~\cite{Chatrchyan:2008aa}. These predictions exceed the
detector design specification of an \sovern{} value of~10.

\subsection{Signal equalization using particles}\label{sec:g2_calibration}
The energy deposited by a charged particle crossing a sensor in the SST is reconstructed as a cluster of charge signals on individual strips. Residual nonuniformities at the level of $15\%$ in the signal response are expected to come from the LLD even after the signal 
equalization described in Section~\ref{sec:LLD_tuning}. These cannot be corrected by the calibrations performed with the tick height method described in Section~\ref{sec:noise_performance}.
Particle identification using energy loss in the sensors is sensitive to these inhomogeneities. Therefore, signals from MIPs crossing the 
SST are used to calibrate the detector for uniform response across the full SST as well as for individual modules~\cite{Quertenmont:2010ota}. This equalization compensates also for the signal loss due to the radiation inasmuch as it is not corrected through the change of LLD gain following an opto scan.

For each APV25, the calibration requires the distribution of the charge normalized to the path length for all 
clusters associated with MIPs reconstructed in the appropriate silicon module. 
The MPV is then extracted from a Landau fit to this distribution. The calibration constant (or gain factor~$G$) is determined by 
normalizing the MPV of the Landau distribution to the same value (300~ADC counts/mm), corresponding to the value 
expected for a MIP.

\begin{figure}[th]
\centering
\includegraphics[width=0.45\textwidth]{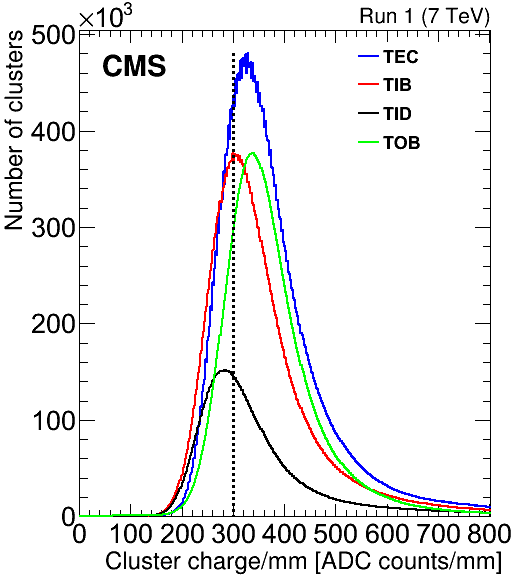}
\includegraphics[width=0.45\textwidth]{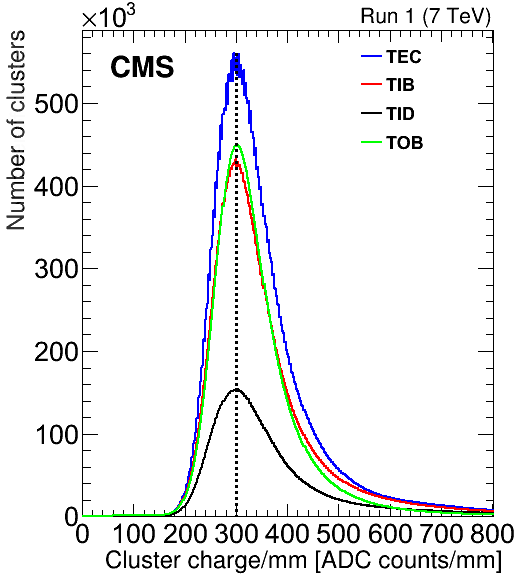}
\caption{The distribution of the charge normalized to the path length after the tick mark calibration before (left) and after (right) applying the signal equalization. 
The dashed line represents the calibration value (set to 300 ADC counts/mm).}
\label{fig:gaincalibrationresults}
\end{figure}

As an illustration of the method, the charge normalized to the path length is shown in Fig.~\ref{fig:gaincalibrationresults}, before (left) and after (right) 
applying the gain factor $G$ to each APV25 of the SST for data recorded during LHC Run~1. The result is a clear alignment of the MPVs over the different regions, corresponding to different positions,
sensor thicknesses, and radiation exposures of the modules. When performing the clustering of strips, this gain factor is used as a 
correction to the 
strip charge signal in order to guarantee stability and uniformity of the most probable value of the cluster charge over time.

\begin{figure}[th]
\centering
\includegraphics[width=0.6\textwidth]{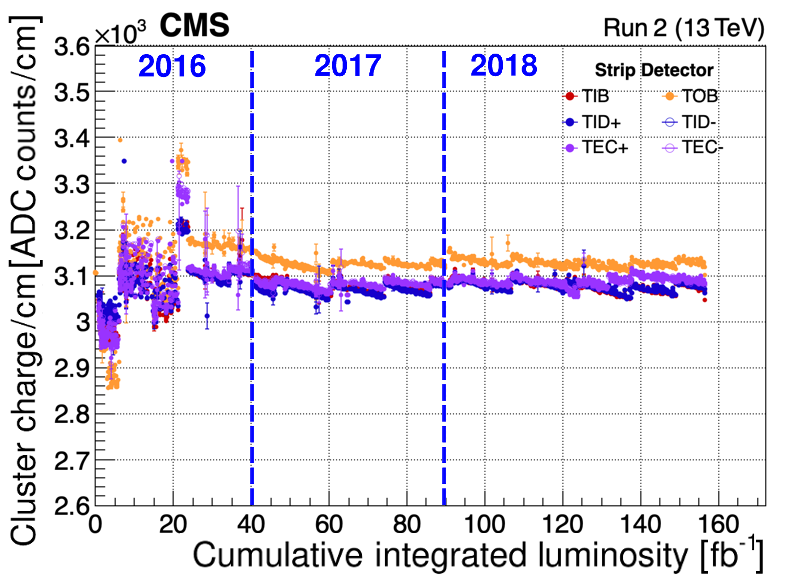}
\caption{Cluster charge normalized to the path length after the offline calibration for the SST barrel (TIB and TOB) and the plus and minus sides of the TID and TEC
as a function of the delivered integrated luminosity. The vertical lines indicate the start of each calendar year. The large fluctuations in 2016 are a consequence of the APV25 saturation issue, see text for details.}
\label{fig:gaincalibrationtrend}
\end{figure}

The cluster charge is monitored over time and is displayed in Fig.~\ref{fig:gaincalibrationtrend} for the different SST parts during LHC Run~2. For daily monitoring of the gain stability, a simpler fitting procedure than that presented in Section~\ref{sec:signal_evolution} is used over a restricted cluster charge range, which leads to a slightly biased value for the MPV, 310~ADC counts/mm instead of the actual 300~ADC counts/mm set by the calibration procedures just discussed. Despite the bias the overall stability is clear. 
This bias does not affect monitoring of the stability.

The large fluctuations of the cluster charge in the first $20\fbinv$ of the data taken in 2016 were caused by the saturation of the preamplifier of the APV25 chip (Section~\ref{sec:apv_saturation}). This effect had a strong instantaneous luminosity dependence and resulted in significant deviation from a Landau distribution, leading to these fluctuations in the calibration.
After the APV25 saturation issue was solved in mid 2016, the SST cluster charge remained stable. The discontinuities occur at intended updates of the LLD gain, pedestal, and noise values.

\subsection{Lorentz angle measurement}\label{sec:lorentz_angle}

\begin{figure}[tb]
\centering
\includegraphics[width=0.65\textwidth]{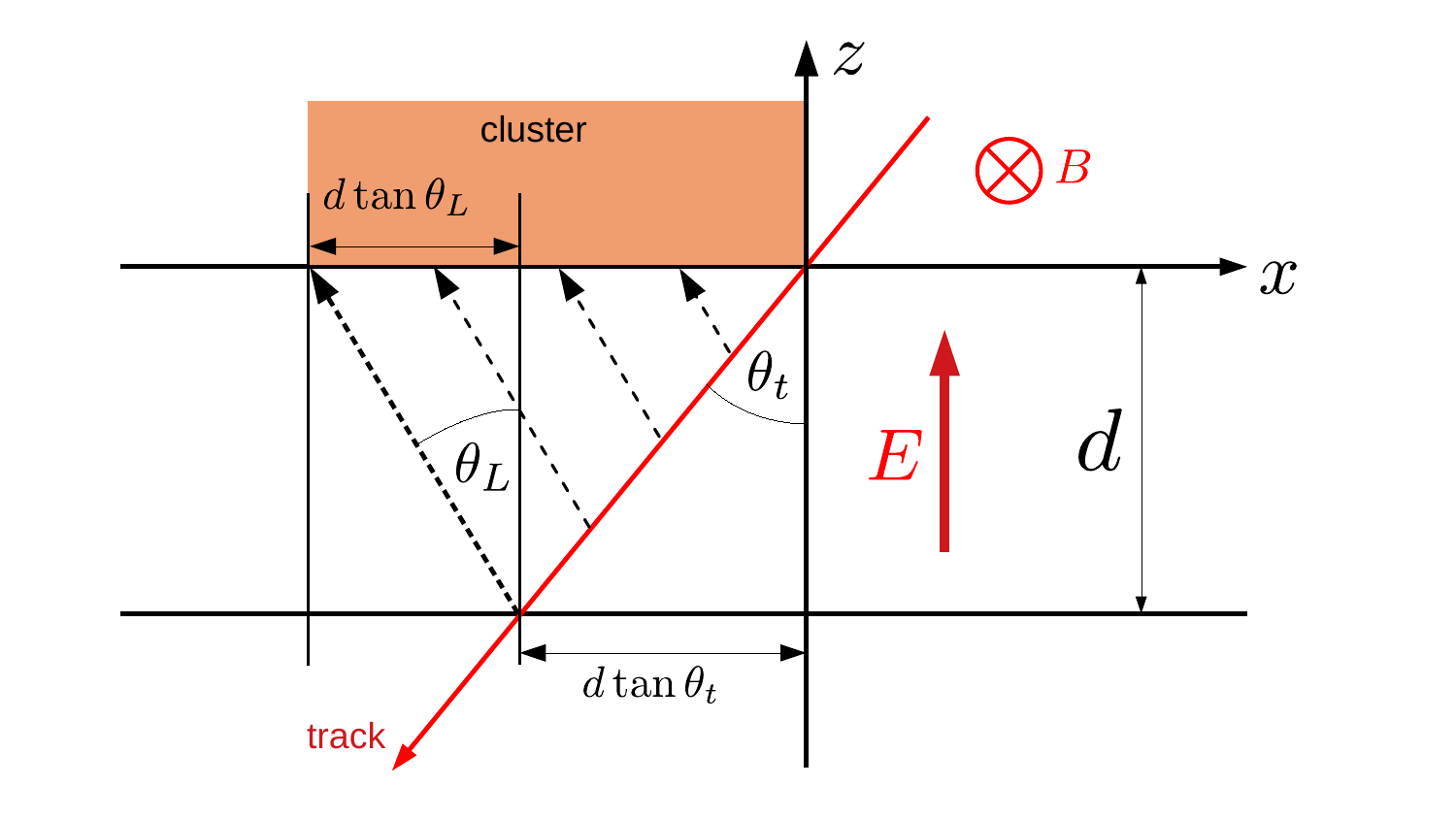}
\caption{Illustration of the shift due to the Lorentz force along the sensitive coordinate $x$ of a sensor of thickness $d$. The Lorentz angle is $\theta_{\mathrm{L}}$ and the $z$ axis is perpendicular to the sensor. The particle crosses the sensor with an incident angle $\theta_{\mathrm{t}}$. The dashed arrows represent the direction of drift of charge carriers produced in the silicon. The cluster is represented by the orange rectangle. The cluster size $d \tan\theta_{\mathrm{t}}$ is increased by $d \tan\theta_{\mathrm{L}}$ in the presence of the magnetic field.}
\label{fig:LA}
\end{figure}

The SST is operated in a homogeneous 3.8\unit{T} magnetic field $\vec{B}$ oriented parallel to the beam axis. 
In the detector modules, the electric field $\vec{E}$ generated by the bias voltage is oriented perpendicular to the sensor plane. In
the SST disks the $\vec{E}$ and $\vec{B}$ fields are approximately parallel, whereas in the inner and outer barrels, the electrical and 
magnetic fields are perpendicular to each other. This causes the charge carriers produced in the $n$-doped silicon bulk
of both the inner and the outer barrel detectors to experience a
Lorentz force in addition to their drift to the readout strips
under the influence of the electric field. As illustrated in Fig.~\ref{fig:LA}, because of the deflection induced by the Lorentz force, the cluster measured on the sensor surface is distorted in length and displaced. In this sketch, the $z$ axis is perpendicular to the sensor, $d$ is the sensor thickness, and $\theta_{\mathrm{t}}$ is the incident angle of the track on the sensor in the projection perpendicular to the strips. 
The charge carriers inside the sensor are deflected by an angle $\theta_{\mathrm{L}}$, called the Lorentz angle,  with respect to the electric field direction. This leads to a shift of the cluster position on the sensor surface measured in 
the local coordinate $x$ (perpendicular to the strip direction, parallel to the sensor). The size of a cluster ($d \tan\theta_{\mathrm{t}}$) is increased by $d \tan\theta_{\mathrm{L}}$ in this direction.

\begin{figure}[tb]
\centering
\includegraphics[width=0.5\textwidth]{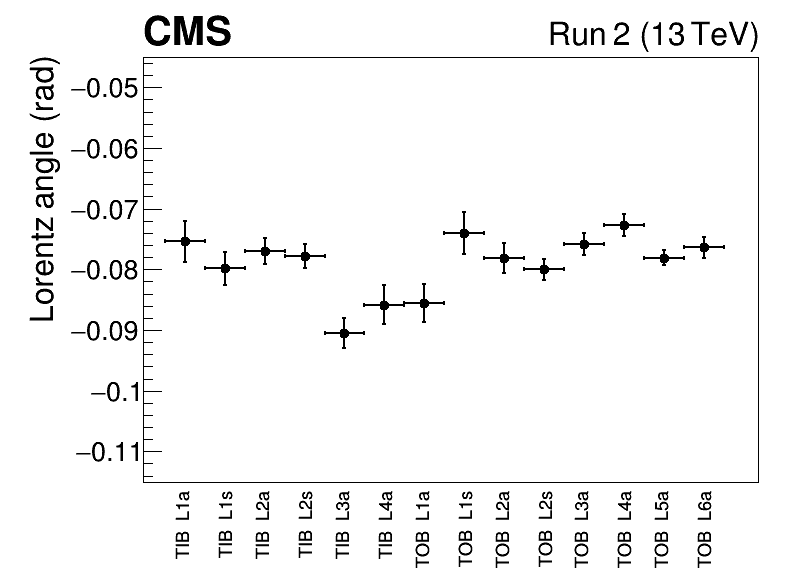}
\caption{Lorentz angle measured at the end of Run~2 for the different SST layers of the TIB and TOB. 
The measurement is displayed separately for modules with strips oriented along the $z$ direction, labeled a, and modules with the strips rotated by an angle of 100~mrad with respect to the $z$ direction, labeled s.}
\label{fig:LA_meas}
\end{figure}

To measure the Lorentz angle, the minimum of the distribution of the cluster size as a function of the particle incident angle with respect to the module surface is used~\cite{Bartsch_2003}. Without a magnetic field, when the Lorentz force and the
Lorentz angle are zero, this minimum is
observed for perpendicular incidence. With a magnetic field, the minimum occurs when the incident angle is equal to the Lorentz angle, where
the charge carrier drift is parallel to the particle trajectory. The results of the Lorentz angle measurement performed at the end of Run~2 for the different layers within the barrel are presented in Fig.~\ref{fig:LA_meas}.

\begin{figure}[tb]
\centering
\includegraphics[width=0.45\textwidth]{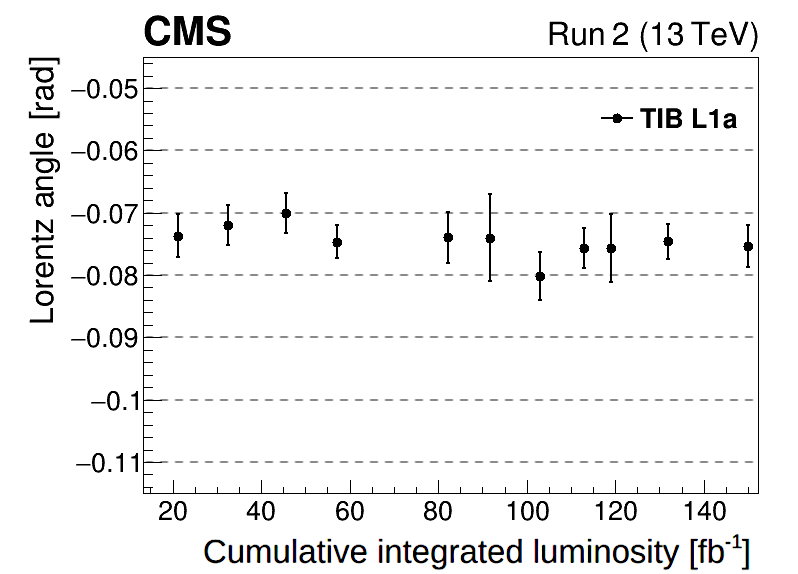}
\includegraphics[width=0.45\textwidth]{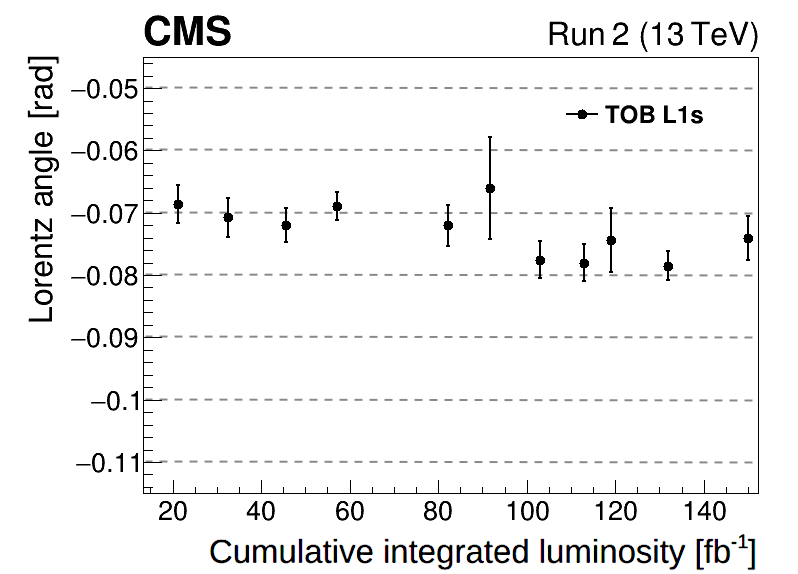}
\caption{Evolution of the Lorentz angle during Run~2, for modules with strips oriented along the $z$ direction belonging to the first layer of the TIB, L1a (left), and for modules with the strips rotated by an angle of 100~mrad with respect to the $z$ direction belonging to the TOB, L1s (right). The dashed lines are drawn to guide the eye.}
\label{fig:LA_trend}
\end{figure}

Once $\theta_{\mathrm{L}}$ is known, a correction to the reconstructed cluster position can be applied to derive the actual position on the sensor surface of the charged particle 
hitting this sensor. This correction is used in the offline
reconstruction and as input for the offline tracker alignment
procedure~\cite{TRK-20-001}. In addition, the Lorentz angle is
monitored, because changes could arise from degradation of the sensor
performance due to radiation damage. Figure~\ref{fig:LA_trend} shows measurements of the Lorentz angle for modules with strips oriented along the $z$ direction, belonging to the first layer of the TIB, and for  modules with the strips rotated by an angle of 100~mrad with respect to the $z$ direction, belonging to the TOB, as a function of integrated luminosity during Run~2. No significant degradation with accumulated luminosity is observed.

\subsection{Hit reconstruction efficiency}\label{sec:hit_efficiency}
\subsubsection{Measurement of the hit efficiency}
The efficient detection of a hit, defined as the response of the sensor and electronics to the passage of a charged particle, is important in the process of efficiently reconstructing the trajectory of the particle. The hit efficiency is defined as the ratio of detected hits to the number of expected hits belonging to a track. Regular measurements of the hit efficiency are made using collision and cosmic data.

Only high-quality tracks are selected in this analysis. This track
selection, defined in Ref.~\cite{Chatrchyan:2014fea}, is based on a
standard set of selection criteria, in particular on the chi-squared normalized to the degrees of freedom and on the compatibility of the tracks with originating from the interaction region.
To avoid inactive regions, such as the bonding region for modules with two sensors, trajectories 
passing near the edges of sensors or close to their readout electronics are 
excluded from consideration. A module is considered as efficient if the distance between the trajectory crossing point and the hit position is less than 15 strips 
(depending on the sensor type, this corresponds to distances of about 1 to $3$\unit{mm}).

\begin{figure}[th]
\centering
\includegraphics[width=0.45\textwidth]{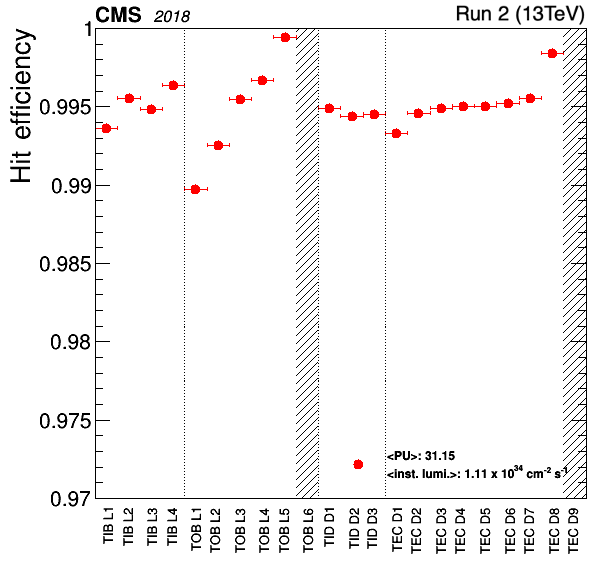}
\caption{Hit efficiency for the various layers of the SST at the end of Run~2. Known faulty modules are masked. 
The two gray bands correspond to the outermost layers where the efficiency cannot be measured.}
\label{fig:Hit_eff_layer}
\end{figure}

If the trajectory starts or ends in a module, it is not used for computing the efficiency of this module. 
Thus, the measurement in the first layer of the inner barrel relies on tracks with inner pixel detector hits, and no measurement is possible in the last layers, \ie, the last layer of the outer barrel and the last disk of the two endcaps.
To avoid biases, known bad modules (Section~\ref{sec:bad-components}) are excluded from the measurement. Likewise, modules with low efficiency, defined as those with efficiency 10\% below the layer average, are not included in the computation 
of the overall average efficiency. For layers with stereo modules, both sensors are taken into account.

The average hit efficiency at the end of Run~2 is above $99\%$, as shown in Fig.~\ref{fig:Hit_eff_layer} for typical beam conditions with an average number of interactions of 31 per bunch-crossing and an instantaneous luminosity of $1.1\ten{34}\percms$.

\subsubsection{Effect of the APV25 preamplifier saturation on the hit efficiency}

\begin{figure}[!th]
\centering
\includegraphics[width=0.45\textwidth]{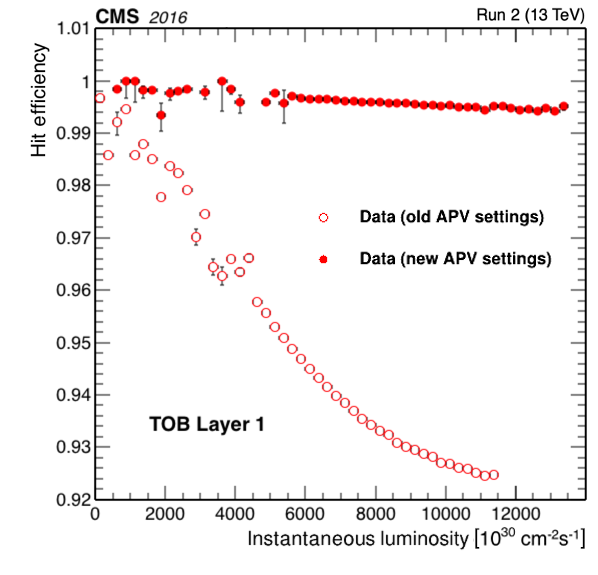}
\caption{Hit efficiency as a function of the instantaneous
  luminosity for the modules in the TOB layer~1 for runs before (open circles) and after (filled
  circles) the change
  of the APV25 preamplifier discharge speed. Error bars are statistical only.}
\label{fig:hit_eff_tob_1}
\end{figure}

The effect on the single-hit reconstruction efficiency from the APV25
preamplifier saturation is presented in Fig.~\ref{fig:hit_eff_tob_1},
again for modules in TOB layer~1 (as shown also in Fig.~\ref{fig:preampsat-sovernshape}). In runs affected by the preamplifier
saturation, a notable drop of the efficiency is already seen at
luminosities of a few $10^{33}\percms$, with
the effect reaching more than 7\% at
$1\times10^{34}\percms$. After the VFP
parameter change, only a very slight decrease of the efficiency is observed, even at instantaneous luminosities above the design value of $1\times10^{34}\percms$.

\subsubsection{Highly ionizing particles as the main source of the hit inefficiencies}
\begin{figure}[!th]
\centering
\includegraphics[width=0.45\textwidth]{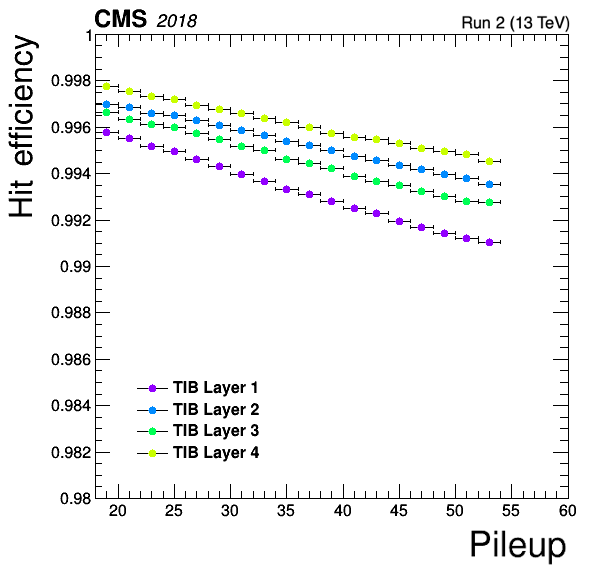}
\includegraphics[width=0.45\textwidth]{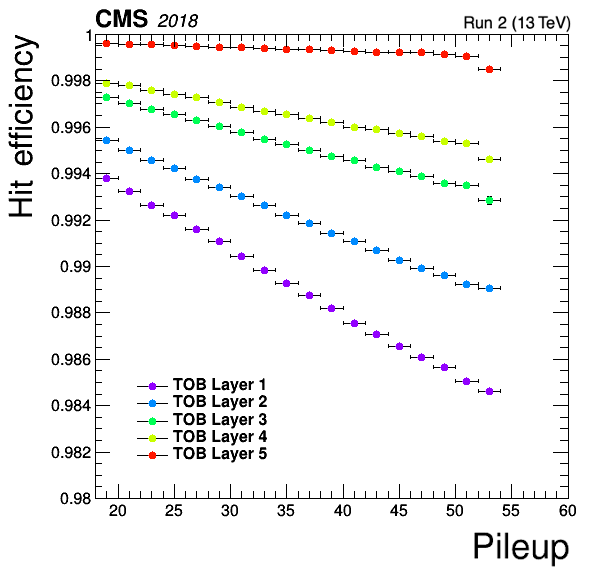}
\caption{Hit efficiency for different layers in the TIB (left)
  and TOB (right) as a function of the average pileup multiplicity.}
\label{fig:effvspu}
\end{figure}

Although the hit efficiency after eight years of operation is confirmed to remain very high, there is
typically a $1\%$ inefficiency, as shown in Fig.~\ref{fig:Hit_eff_layer}. 
To identify the origin, a measurement of this efficiency as a 
function of the number of pileup interactions has been carried out.  This indicates an almost linear dependency of the hit efficiency on the number of overlapping $\Pp\Pp$ interactions, as shown in Fig.~\ref{fig:effvspu}.

Highly ionizing particles (HIPs), generated from nuclear interactions in the SST sensors, give rise to large energy deposits (equivalent to several hundred MIPs) within the silicon sensors. These HIP events, though rare, are the source of a temporary saturation of the APV25 chip, leading to a deadtime of about five bunch crossings during the recovery process. This leads to a loss of efficiency, proportional to the interaction rate.

The effect of HIPs on the APV25 has been studied in Ref.~\cite{Adam:909141}. 
This study was performed during the design of the SST and is summarized in this section.

\begin{figure}[th]
  \centering
\includegraphics[width=0.5\textwidth]{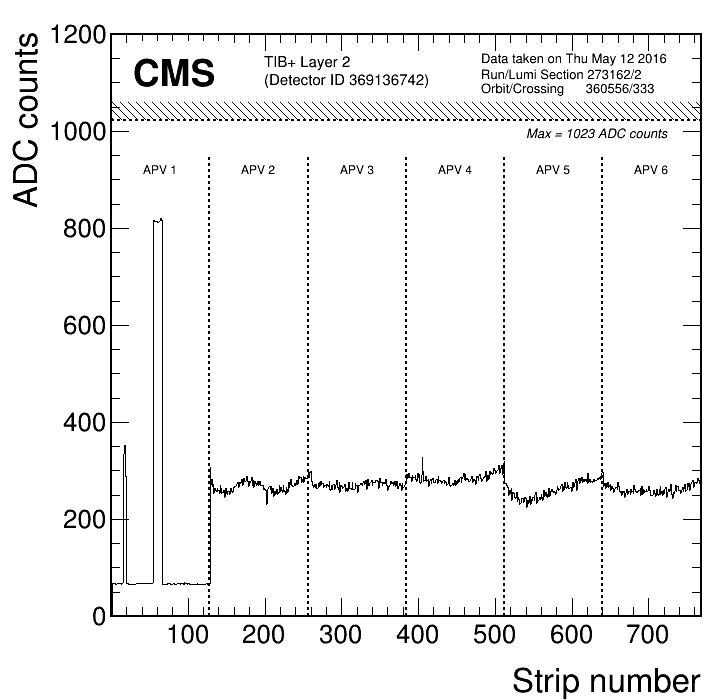}
\caption{ADC counts of the six APV25 chips in a TIB module during a collision in 2016. The large charge (at $\approx 800$ ADC counts located around strip number $\approx 60$) together with a drop of the baseline in the first chip is the signature of a HIP event.}
\label{fig:HIPevent}
\end{figure}

The HIP events are identified by the presence of a large signal in the affected APV25 chips, whereby the outputs of all other channels connected to the same APV25 are driven down to a level well below their pedestals. To illustrate this behavior, an example of a HIP event identified during a $\Pp\Pp$ collision is shown in Fig.~\ref{fig:HIPevent}. The typical signature of these events is a large charge deposited in the sensor, together with a dramatic drop of the baseline of the corresponding APV25, and accompanied by smaller-than-usual channel-to-channel fluctuations of this baseline.

\begin{figure}[th]
  \centering
  \includegraphics[width=0.45\textwidth]{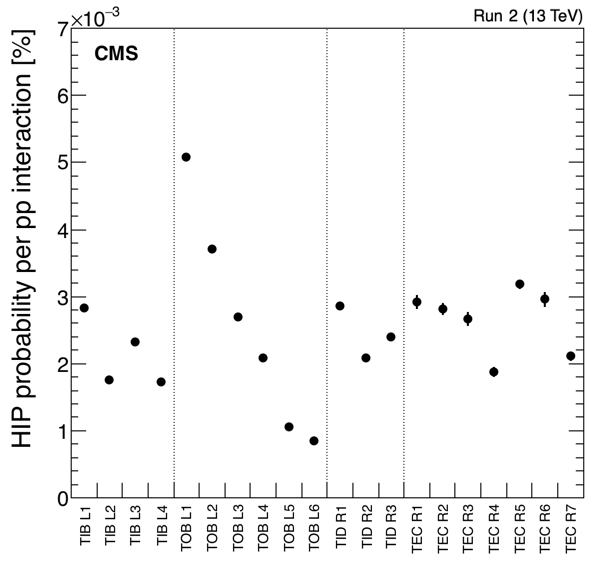}
\includegraphics[width=0.45\textwidth]{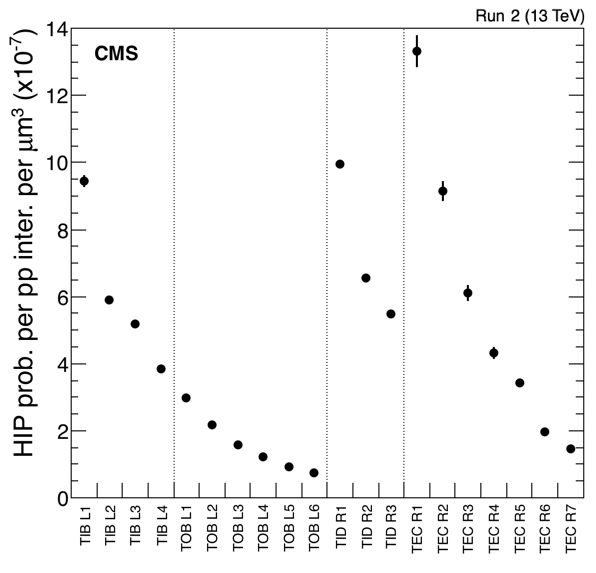}
\caption{Average probability of HIP event occurrence per $\Pp\Pp$
  interaction (left) and normalized to unit volume (right) for all
  layers of the silicon strip tracker. In the endcaps, the
  probability is reported per ring.}
\label{fig:hitproba}
\end{figure}

The probability of occurrence of such events within the SST was measured in several special NZS runs in 2018. The selection of the HIP events is based on the typical signature described above. The average probability for a HIP event to occur per $\Pp\Pp$ interaction is shown in Fig.~\ref{fig:hitproba}. 

\begin{figure}[t]
\centering
\includegraphics[width=0.45\textwidth]{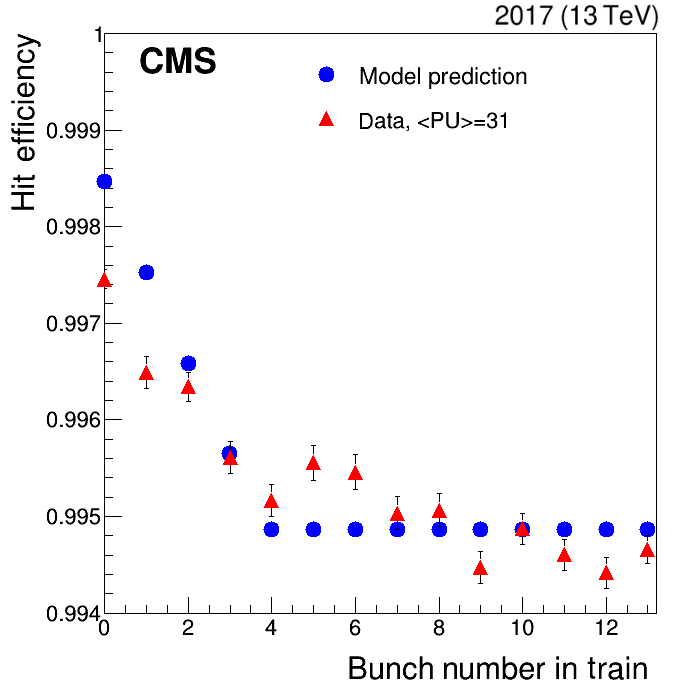}
\caption{Evolution of the hit efficiency as a function of bunch number within the train for TIB layer 1 modules, obtained from a representative selection of CMS 2017 data with an average pileup of 31. Bunches are separated by 25\unit{ns}. The data are represented by red triangles and the model prediction is shown with blue circles.}
\label{fig:effbunch}
\end{figure}

A model of the resulting inefficiency was developed, taking into account the HIP probability and the LHC bunch structure with trains of bunches of protons separated by 25\unit{ns}. The trains are separated by a gap of 3\mus. The deadtime, during which the chip is assumed to be fully inefficient after the occurence of a HIP, is used as a parameter. 
The prediction of this model for the first layer of the TIB is compared with the measured efficiency in Fig.~\ref{fig:effbunch}, using data from 2017 with an average pileup of 31, which is representative for the pileup during LHC Run~2. The agreement between this model and the data indicates that the time dependence within a 
train is well described, that HIP events are the dominant source of the hit inefficiency, and that HIP events lead to a typical deadtime of five bunch crossings, compatible with the measurements from previous dedicated beam tests, as presented in Ref.~\cite{Bainbridge:2002bda}.

\subsection{Single-hit resolution} \label{sec:hit-resolution}
The procedure to measure the spatial resolution of hits within the SST is explained in Ref.~\cite{Adam:1161239}, which also documents the measurements performed during LHC Run~1. The single-hit resolution measured during LHC Run~2 is presented in this section.

In the SST, silicon sensors slightly overlap with their neighbors to ensure a hermetic tracking coverage within the entire $\eta$ range, so that a particle can cross two sensors in the same layer. 
To compute the hit resolution, hits from tracks passing through regions where modules overlap within a layer are considered.
Tracking is redone without the hits in the layer, and the reconstructed track is used to predict the position of the impact point in each of the two overlapping modules A and B of the layer. 
For each module, the residual, \ie, the difference between the measured and predicted hit position, is determined. Then the so-called double-difference is calculated as:
\begin{equation}
(\text{hit}_A - \text{prediction}_A) - (\text{hit}_B-\text{prediction}_B),
\end{equation}
where $\text{hit}_{A,B}$ refers to the position of a hit in module $A$ or $B$, and $\text{prediction}_{A,B}$ refers to the position of a predicted impact point of the track in module $A$ or $B$.

The advantage of this method is that most uncertainties in the track propagation are cancelled in the difference $\text{prediction}_A - \text{prediction}_B$. For a perfectly aligned detector, the difference of residuals is expected to be zero on average. The width of this difference is a measure of the SST hit resolution, which depends on the thickness, the orientation, and the pitch of the sensor as well as the size of the clusters.

\begin{figure}[t]
\centering
\includegraphics[width=0.45\textwidth]{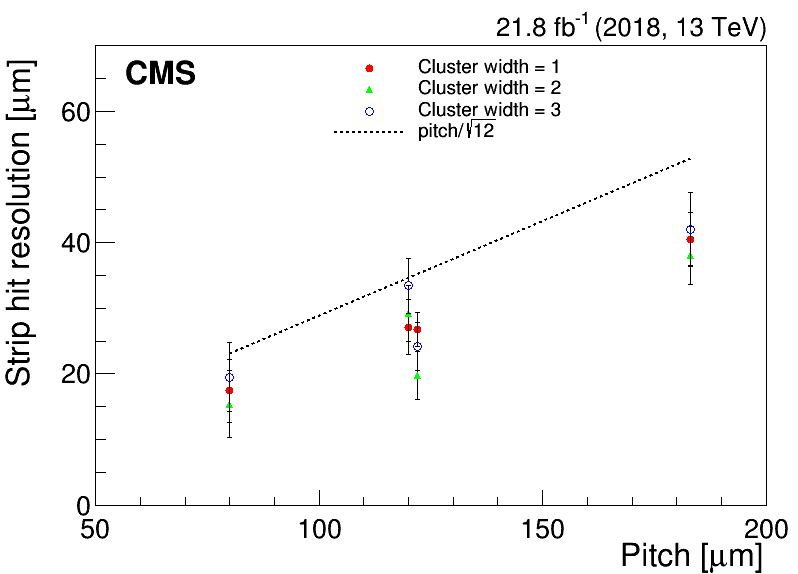}
\includegraphics[width=0.45\textwidth]{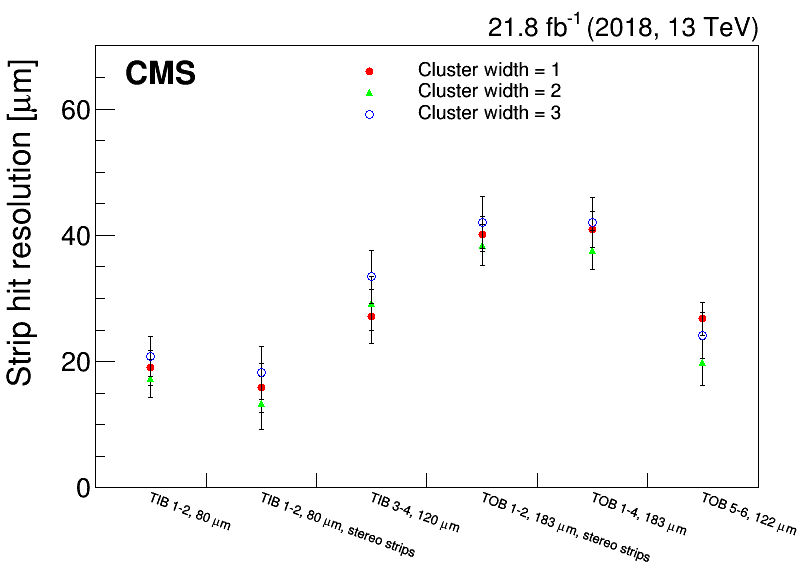}
\caption{Single-hit resolution as a function of the strip pitch (left) and for different detector regions (right). 
In the left plot, the expected resolution for binary readout (pitch/$\sqrt{12}$) is also shown for comparison.}
\label{fig:resoPitchGeom}
\end{figure}

The SST hit resolution measurements performed during the last year of Run 2 are shown in Fig. 41~\ref{fig:resoPitchGeom} functions of these parameters. In a detector without analog charge measurement, the typical resolution is limited to $p/\sqrt{12}$ where $p$ is the strip pitch. This binary resolution limit is displayed in the figure, highlighting the improvement achieved through the measurement of the charge sharing, which is made possible by the analog readout of the charge from each channel. 
By measuring the fraction of charge collected by adjacent strips, the hit position can be interpolated more precisely, leading to a resolution significantly better than the binary limit.

With typical values between 20 and 40\mum, the SST hit resolution matches the expectation from the detector design
\cite{CMS:1994hea}, without any sign of degradation compared to previous measurements~\cite{Adam:1161239}.

\subsection{Particle identification by ionization energy loss}\label{sec:dedx}

Although the primary function of the strip
tracker is to provide precise hit information for track reconstruction, and hence precise momentum determination, the wide linear
range of the analog output also provides a measurement of the ionizing energy
loss of the incident particles, which can be used for particle identification. 

The mean ionization energy loss per unit length $\ddinline Ex$ of a particle crossing a layer of material is given by the 
Bethe--Bloch formula~\cite{PDG2024}. In a restricted range of momentum $p$ of the incident particle of mass $m$ ($0.2<\beta\gamma <0.9$, where $\beta$ is the velocity and $\gamma$ the Lorentz factor), 
the Bethe--Bloch formula can be linearized in $m^2/p^2$~\cite{Quertenmont:2010ota}, where K and C are constants and can be extracted from a fit of the mean energy loss measurement:
\begin{equation}\label{eq:simplededx}
\left\langle{\frac{\textnormal{d}E}{\textnormal{d}x}}\right\rangle=K\frac{m^2}{p^2}+C.
\end{equation}

\begin{figure}[th]
\centering
\includegraphics[width=.5\textwidth]{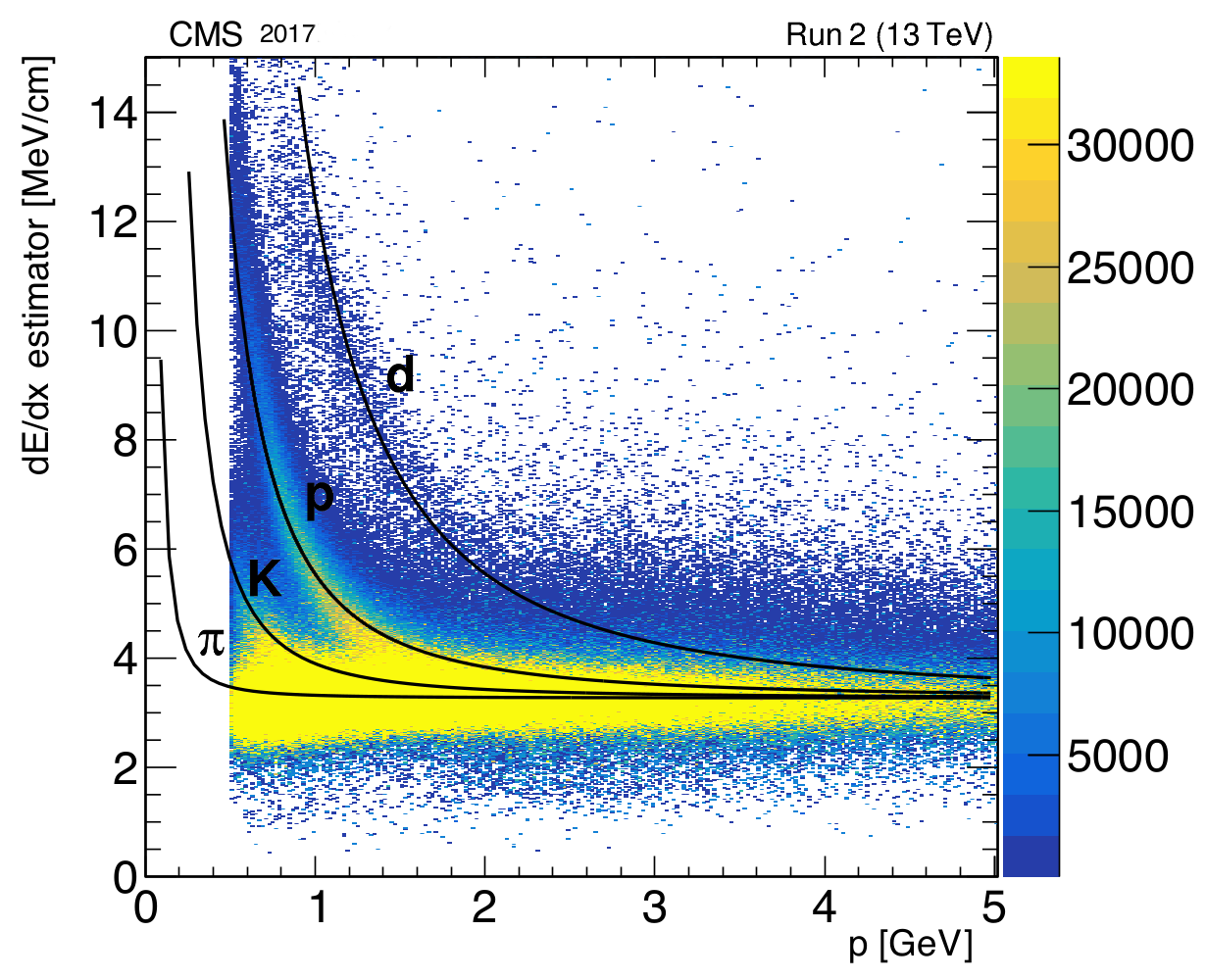}
\caption{Energy loss measurement in the SST  during LHC Run~2. Expected losses for pion, kaon, proton, and deuteron particles are also shown (black lines). Tracks with momentum below $0.5\GeV$ are not included in this plot.}
\label{fig:dedx}
\end{figure}

In the SST, the mean ionization energy loss per unit length is
computed by measuring the cluster charge generated in all the sensors
along the trajectory of a particle normalized to the path length
through each sensor. Because the SST sensors are thin, the
fluctuation of the energy loss within a sensor follows a Landau-like
distribution. The most probable energy loss
is estimated by combining the measurements along the particle
trajectory. Several estimators providing the most probable energy loss
have been evaluated. The harmonic-2 estimator, which is the harmonic mean
of power $-2$ defined as:
\begin{equation}\label{eq:harm}
 I_{\mathrm{h}}={\left( \frac{1}{N}\sum_{i=1}^{N}{\left(\ddinline Ex \right)}_i^{-2}\right)}^{-1/2} ,
\end{equation}
where $N$ is the number of measurements,  is chosen for its stronger discrimination power~\cite{Quertenmont:2010ota}.
The distribution of the most probable energy loss obtained from this estimator as a function of the particle momentum is shown in Fig.~\ref{fig:dedx}.
The region corresponding to the energy loss distribution for protons
was fitted using the function from Eq.~(\ref{eq:simplededx}) with $K$ and $C$
as fit parameters. The black lines in Fig.~\ref{fig:dedx} represent
the calculated energy loss for deuterons, kaons, and pions. However,
the region for pions at rising energy loss values is not visible due
to the cutoff at low momenta.
Particle identification using the energy loss measurement within the 
SST is used in searches for long-lived charged particles in $\Pp\Pp$ collisions~\cite{Khachatryan:2219676}.

\section{Radiation effects}  \label{sec:rad-mon}

Because of the continuous exposure to particles arising from the high-energy 
$\Pp\Pp$ collisions of the LHC, the detector modules of the SST and
their individual components are exposed to large levels of radiation over
their lifetime. The effect of the heavy ion data-taking periods can be neglected because of the very low integrated luminosity.
Since the start of collision data taking, two
effects have been regularly monitored: radiation damage to the optical link
system and radiation damage to the silicon sensors of the SST
modules. In both cases the dominant damage mechanism is displacement
damage in the crystal lattice through hadronic interactions. The
damage mechanism is commonly expressed through the NIEL (non ionizing
energy loss) concept~\cite{Moll:1999kv} in which radiation damage from
different particle types is scaled to the equivalent damage from
neutrons. As a result, fluences for silicon sensors are
expressed in units of 1\MeV neutron equivalent per cm$^2$ (1\MeV $n_{\text{eq}}/\text{cm}^2$). For radiation damage in the InGaAsP laser diodes in use in the SST, equivalent damage factors
for different particle types, that allow scaling to a common reference, is reported in the literature, \eg, in Ref.~\cite{Gill:2005ui}.

Simulated fluences in this chapter are obtained from the FLUKA
radiation simulation framework~\cite{Ferrari:898301,Bohlen:2109973}, 
within which the geometry and material distribution of the full CMS experiment have been
simulated. 

\subsection{Optical link radiation damage monitoring}

The optical link system is expected to degrade with increasing
radiation exposure.
The two main
radiation-induced effects expected are a decrease in link gain and an
increase in laser threshold current. The laser bias setting is
retuned during gain scans at regular
intervals to account for the change in threshold current (Section~\ref{sec:LLD_tuning}). The
threshold current is thus measured regularly and its change as a
function of time and luminosity can be studied.

\begin{figure}[th]
\centering
\includegraphics[width=0.95\textwidth]{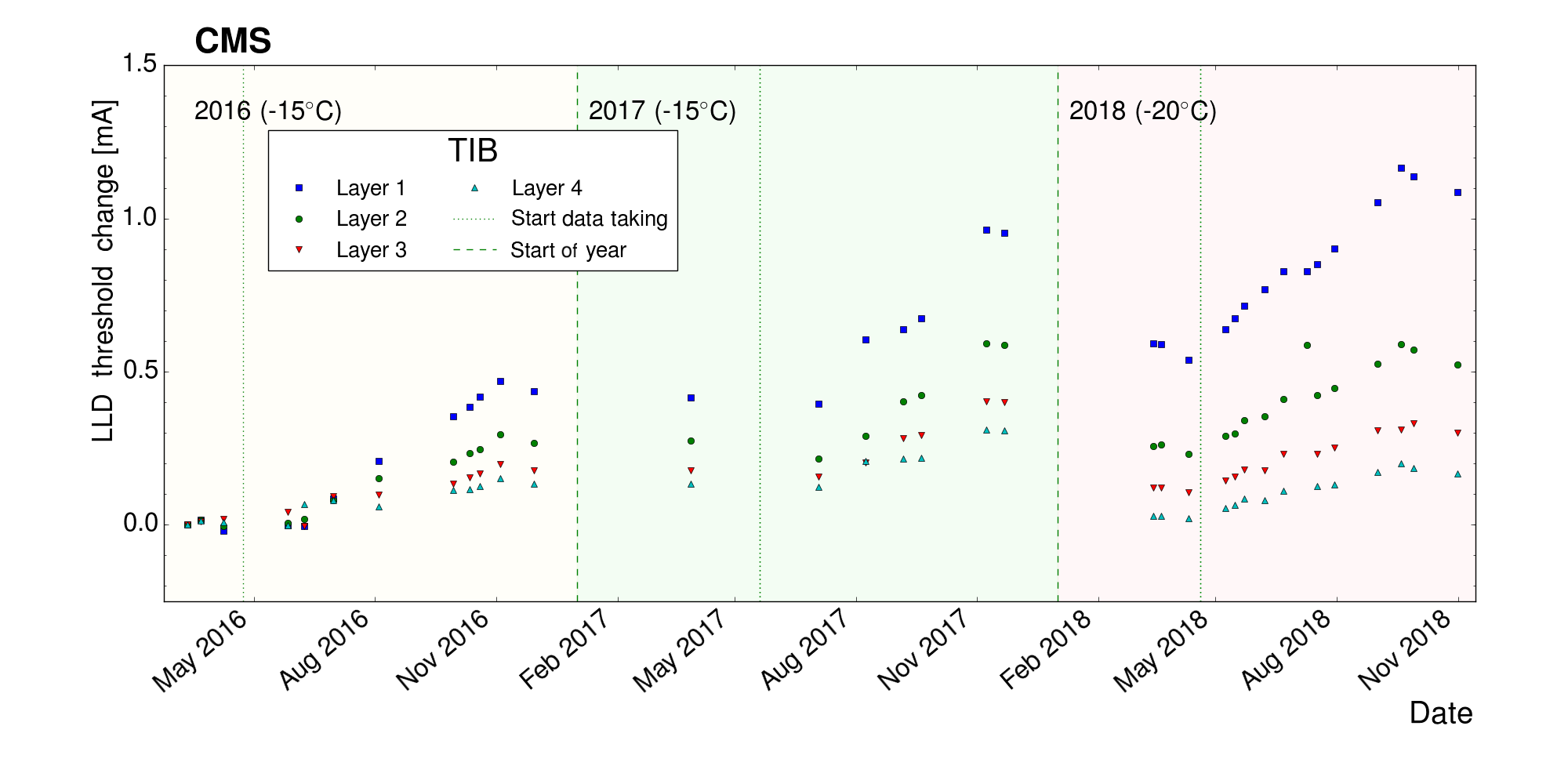}
\includegraphics[width=0.95\textwidth]{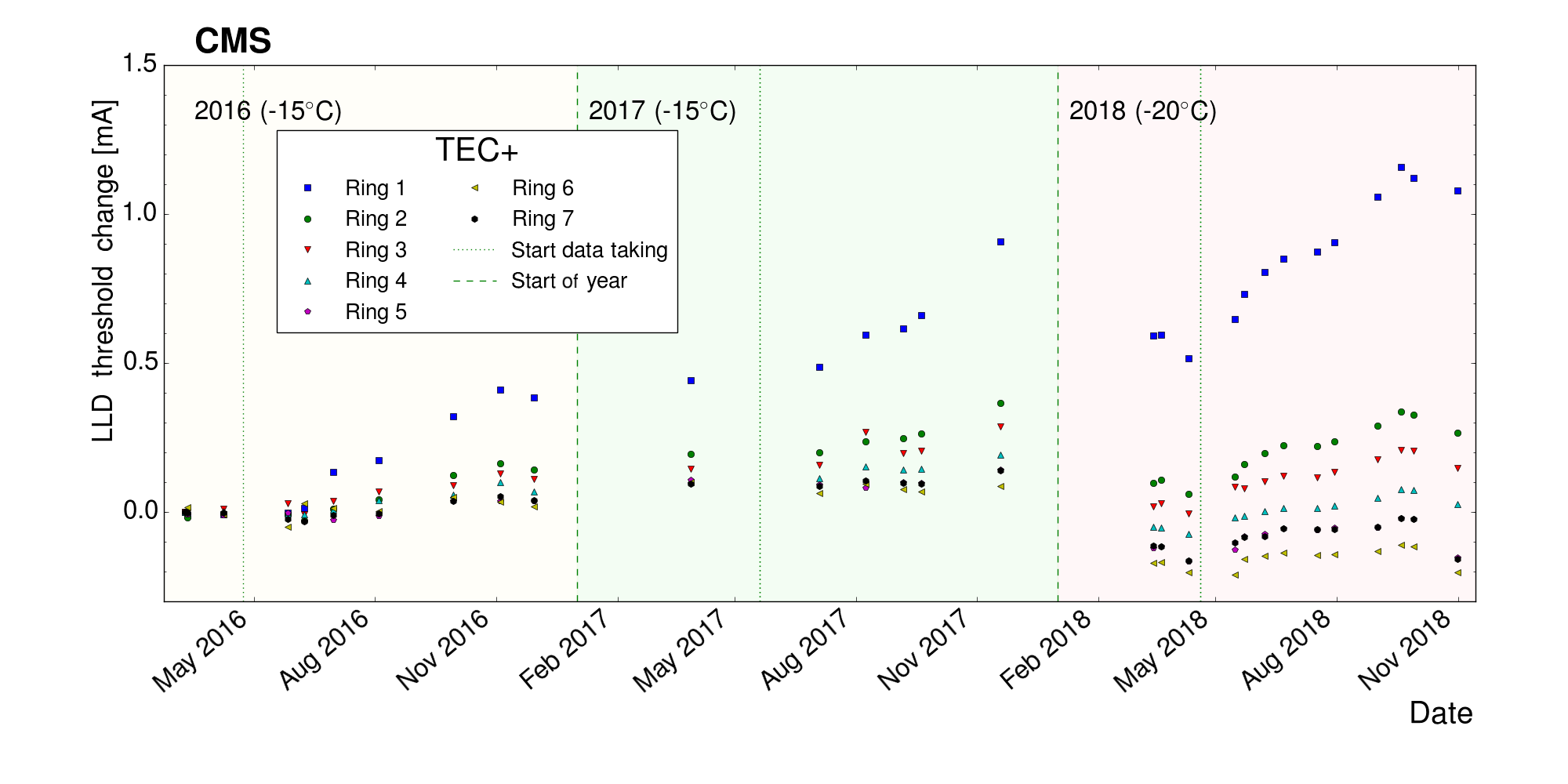}
\caption{Laser driver threshold increase versus time for laser drivers
  in TIB~(upper) and TEC$+$~(lower). A point in early 2016, to which
  subsequent runs are compared, is defined for both readout
  partitions. }
\label{fig:threshold_increase_notscaled}
\end{figure}

The radiation level in the SST varies strongly as a function of the
radial distance from the beam line, but has only a weak dependence on the $z$ position. Accordingly
results are combined for modules at equal radius in layers~(TIB/TOB)
and rings~(TID/TEC). Measurements of the laser driver thresholds are summarized in
Fig.~\ref{fig:threshold_increase_notscaled}. The threshold current is shown as a function
of time to better show annealing effects during periods where no integrated
luminosity is accumulated. A reference point in early 2016, to which subsequent runs are compared, is defined 
for each readout partition. The threshold
current change compared with the selected reference point is shown for
the TIB and TEC$+$. Both detector parts show an increase of the
threshold current during running periods, and the
increase is larger for modules located closer to the beam line.
The increase in threshold current closely follows the increase in
integrated luminosity. Annealing of the current increase (Section~\ref{sec:detector-status}) is visible in
periods with no beam, most prominently during the 
interruptions of the LHC data taking of 2--3 months, typically starting at the end of a given calendar 
year, where the SST was kept at room temperature typically for at
least a few days.  The step in the distribution at the beginning of
2018 corresponds to the reduction of the operating temperature of the
SST to $-20^{\circ}$C. Such a change arises from the dependence of the
threshold current of the laser on the temperature. The current, for
identical radiation exposure and annealing, varies as
\begin{equation}
  I_{\text{th}} = I_{\text{th}}(0) \exp\left( \Delta T / T_0 \right), 
\end{equation}
where $I_{\text{th}}(0)$ is the threshold current at a reference
temperature, $\Delta T$ is the temperature difference between the
laser and the reference value and $T_0$ is the characteristic temperature of the laser
diodes, measured to be around
65\unit{K}~\cite{Macias:CMS-TK-TR-0036}. The characteristic temperature is used to indicate the temperature sensitivity of the laser diode.

The threshold current increased by 1\unit{mA} from the beginning of the
2016 run to the end of the 2017 run~(91.4\fbinv) and by 0.75\unit{mA} over the 2018
running period~(67.9\fbinv), which does not reflect the ratio of luminosities in
the two periods. This is believed to be an effect of the different
operating temperatures, namely $-15^\circ$C in 2016--2017 and
$-20^\circ$C in 2018, which reduced the beneficial annealing effect in
2018.

The integrated luminosity over these three years was about 160\fbinv. The
absolute threshold current was between 3 and 4\unit{mA} in 2016. The maximum possible threshold current is 22.5\unit{mA}, which suggests
that there should be ample margin for operation up to 500\fbinv, the expected integrated luminosity prior to the HL-LHC upgrades and replacement of the existing system.

From irradiation tests on similar devices~\cite{Gill:623625}, a threshold increase
of less than 6\unit{mA} is expected over the full lifetime of the SST.

\subsection{Silicon sensor radiation damage monitoring}

The two main effects of radiation on the p-in-n silicon sensors are an
increasing leakage current under HV bias, also leading to increased
noise, as well as a change of effective bulk doping concentration
$\neff$ of the initially n-type bulk, that will eventually
cause type inversion to p-type and hence a change of polarity of the
bulk material. The depletion voltage decreases before type inversion
and increases again afterwards.

Measurements of both the leakage current and the depletion voltage are
described below. Both quantities are compared with the
predictions from simulations and to preinstallation results.
A more detailed discussion of the effects of bulk damage in
silicon detectors is reported elsewhere, \eg, in Ref.~\cite{Moll:1999kv}.

Simulations of the evolution of the leakage current and depletion voltage
have been performed assuming an integrated luminosity of 500\fbinv and a
detector operating temperature of $-25^{\circ}$C until the start of the LHC LS3.

\subsubsection{Leakage current evolution}\label{sec:leakage_current}

\begin{figure}[th]
\centering
\includegraphics[width=0.55\textwidth]{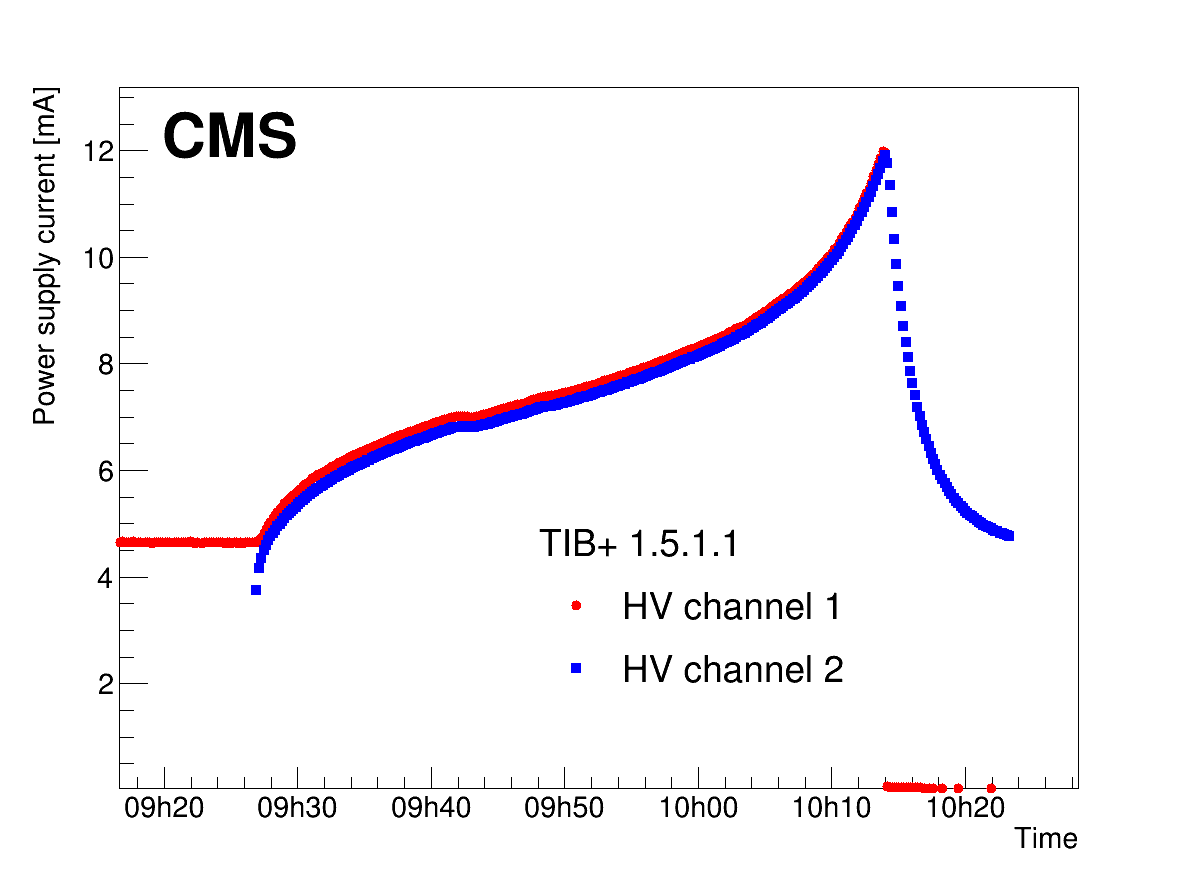}
\caption{Thermal runaway observed in one power group of the TIB during the 2017 running.
  A power group consists of two high-voltage channels.
  The maximum high-voltage current of each channel
  is 12~mA. Once one of the two channels reaches this limit it
  is switched off (red dots). The modules connected to the other channel cool down as a
  result of this and the current decreases (blue squares). }
\label{fig:thermal_runaway}
\end{figure}

In radiation-damaged silicon sensors, the leakage current increases linearly with the
fluence~\cite{Moll:1999kv}. This increase
is commonly expressed by the ``current-related
  damage rate'' or $\alpha$-parameter, which is defined as the current
increase, scaled to $+20^{\circ}$C, per sensor volume and 1\MeV $n_{\text{eq}}/\text{cm}^2$.
The leakage current increase $\Delta \ileak$ can then be written as
\begin{equation}\label{eq:alpha_param}
  \Delta \ileak = \alpha \Phi_{\text{eq}} V,
\end{equation}
where $\Phi_{\text{eq}}$ is the particle fluence in units of
1\MeV $n_{\text{eq}}/\text{cm}^2$ and $V$ is the volume
of the sensor under consideration.  The leakage current of a
radiation-damaged sensor is strongly temperature dependent with the leakage current roughly doubling for every $7^{\circ}$C of temperature increase.  The leakage current
at a reference temperature $T_{\text{ref}}$ can
be expressed as~\cite{Moll:1999kv}
\begin{equation}\label{eq:ileak_scaling}
  \ileak(T_{\text{ref}}) = \ileak(T_{\text{sil}}) \left( \frac{T_{\text{ref}}}{T_{\text{sil}}}\right)^2 \exp\left[ -\frac{E_{\mathrm{g}}}{2k_{\mathrm{B}}}\left( \frac{1}{T_{\text{ref}}} - \frac{1}{T_{\text{sil}}} \right) \right],
\end{equation}
where $E_{\mathrm{g}}$ is the bandgap energy of silicon (1.11\unit{eV}),
$k_{\mathrm{B}}$ is Boltzmann's constant, and $T_{\text{sil}}$ is the
temperature of the silicon sensor.  The temperature of the sensor
depends on the available cooling capacity for the module in question, which here 
is defined as the increase in temperature due
to an increase in power on the sensor. In the SST this is measured
using the DCUs on the individual modules during a bias voltage
scan. The increase in the total sensor power $P$, given by
$V_{\text{bias}} \ileak$, can be correlated with
the increase in sensor temperature to extract the thermal contact
coefficient $ \ddinline TP$. With sufficient cooling power available, the
additional heat generated by the increased current will not result in
a (significant) increase in the temperature of the silicon sensor. In
regions of the detector without adequate cooling, the increase is
sizable and can lead to thermal
runaway~\cite{Kohriki:1995ph}.
Thermal runaway occurs when an increase in leakage current results in
additional power dissipation in the sensor which in turn results in a
rise in temperature (self-heating) and thus a further increase in
leakage current, and this feedback loop continues uncontrollably.
Figure~\ref{fig:thermal_runaway} shows an example of this in a power
group in a stereo layer of the TIB during the 2017 running.

Separate HV channels provide bias to the two sides of each stereo module in this power group. Initially one of the two HV
channels is off. Switching on this channel causes the temperature of
both sets of modules to increase and the leakage currents 
increase accordingly. The increase continues until one of the HV channels
reaches the maximum power supply current of 12\unit{mA}, at which point the
voltage is abruptly ramped down. The current for the second channel
remains just below the 12\unit{mA} trip point and decreases rapidly once
the temperature decreases. It was possible to operate both HV channels
of this power group by lowering the bias voltage and thus reducing the
power dissipated on the sensors. A few more power groups experienced
thermal runaway during the late stage of operation at
$-15^{\circ}$C{}; no thermal runaway has been observed when
operating at $-20^{\circ}$C{} during Run~2. 

\begin{figure}[tbp]
\centering
\includegraphics[width=0.6\textwidth]{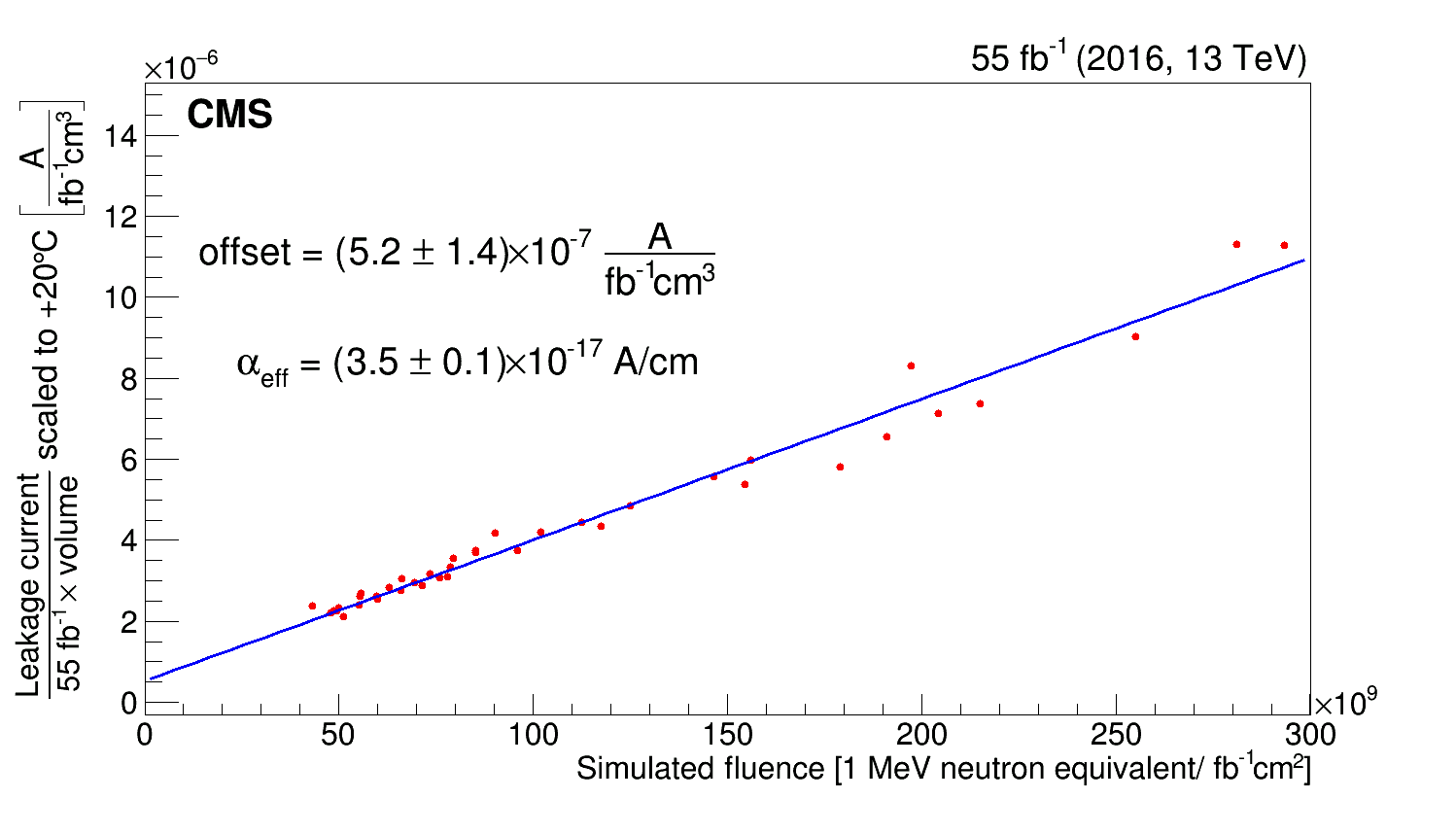}
\caption{Leakage current per unit volume and integrated luminosity,
  scaled to $+20^\circ$C, measured after an integrated luminosity of
  55.4\fbinv as a function of the simulated fluence for modules at
  different radii. A linear fit to the data is performed and used to extract the
  effective current related damage rate, $\alpha_{\text{eff}}$.
   }
\label{fig:alpha_eff_20C}
\end{figure}

The leakage current is measured with two complementary approaches, one
measuring the total current for a group of modules in the power supply
units, the other measuring the leakage current of the individual
modules using the DCU.  The behavior for
different parts of the strip tracker is investigated. The leakage current is extracted 
for a particular integrated luminosity and then scaled to $+20^{\circ}$C using Eq.~(\ref{eq:ileak_scaling}). 
These measurements are then plotted as functions of the fluence that the
individual parts of the detector have received up to this moment, where the fluence generally decreases with increasing distance from the beam
line. Results are reported in Fig.~\ref{fig:alpha_eff_20C}, where the
leakage current per unit volume, scaled to $+20^\circ$C, is shown as
a function of the fluence. A linear fit is performed to the data to
extract the $\alpha$-parameter. It should be noted that the amount of
annealing for the individual sensors is not always the same because of
the strong temperature differences in the SST
(Section~\ref{sec:detector-status}), hence resulting in an ``effective'' $\alpha$-parameter being derived
rather than the one defined in Eq.~(\ref{eq:alpha_param}), which does not include annealing effects.
Nevertheless, the fit describes the data reasonably well and a value of
$(3.5 \pm 0.1)\times10^{-17} \unit{A/cm}$ is obtained, which is in good
agreement with the value of $(3.79 \pm 0.27) \times
10^{-17} \unit{A/cm}$ found in Ref.~\cite{Chatrchyan:2008aa}, 
in measurements on irradiated SST sensors.

Simulation is used to predict the future variation of the 
leakage current. Such
simulations require the fluence as a function of the position inside the
detector, the sensor temperature, and the available cooling power for
each module as input parameters.
The temperature change due to changes in operating conditions
(\eg, change of coolant temperature) and self-heating can then be predicted.
The self-heating is included
iteratively in the simulation until either thermal runaway occurs or
the procedure converges (that is, the increase in leakage current due to self-heating
from the previous iteration is less than 1\unit{nA}).

At sufficiently high temperature, the radiation damage to the sensors will 
undergo annealing, which leads to a reduction of the leakage current.
The simulation must include simultaneous annealing and radiation damage effects. 
This is done by calculating the increase in leakage current
on a given day, based on the integrated luminosity or radiation dose on
this day. This damage is then annealed based on the sensor temperature
over the following days. The leakage current on any given day is thus
a superposition of the initial leakage current measured before 
radiation exposure and the increments from each successive day.

The predicted leakage current and sensor temperature from the
simulation are compared with measurements at specific times.

\begin{figure}[tbp]
\centering
\includegraphics[width=0.49\textwidth]{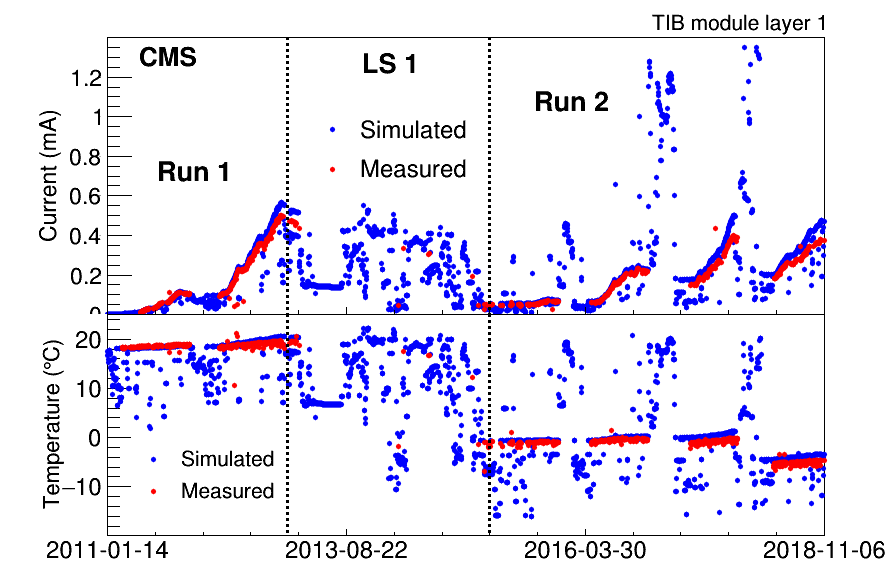}
\includegraphics[width=0.49\textwidth]{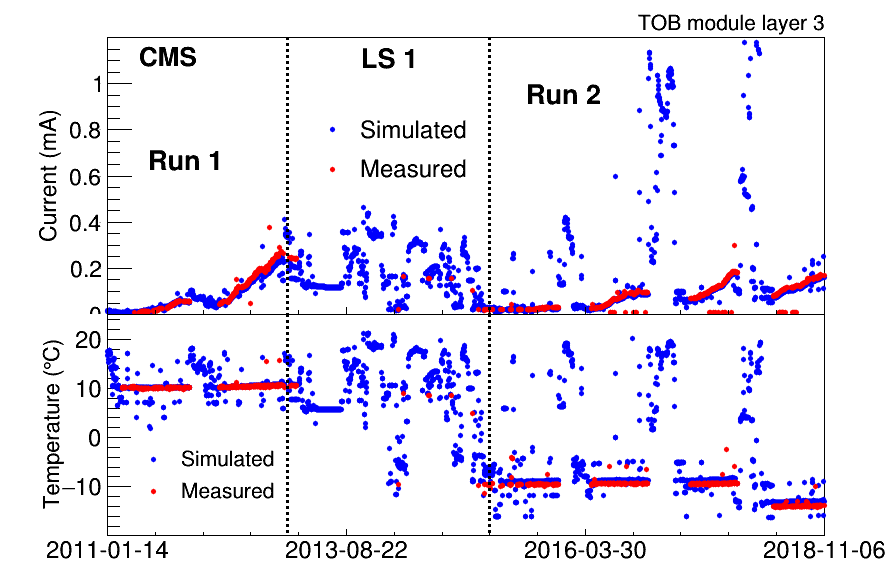}
\caption{Leakage current and temperature measured by the DCU compared with simulation for a
  single module in the TIB layer 1 (left) and a module in TOB layer 3
  (right).}
\label{fig:raddam_single_module_evolution}
\end{figure}

Figure~\ref{fig:raddam_single_module_evolution} shows two examples of
the variation of leakage current and temperature in both data and
simulation as a function of time. The simulation tracks the
measurements reasonably well. For both modules the temperatures during
the Run~1 period agree well between simulation and measurement. The
leakage current is well described for the TIB module during Run~1, with
the simulation slightly overestimating the leakage current towards the
end. For the TOB module the simulations similarly describe
the data well with a slight underestimation towards the end of
Run~1. During LS1 and two periods of Run~2, the
simulation has many more points than are present in the data. This
is because during periods when the detector is off there
are no per-module measurements of temperature and leakage current,
but the temperature can still be estimated rather precisely from
nearby temperature sensors that are continuously read out.
These temperature measurements are good proxies of the module temperature
when the LV of the system is off and the system is thus in thermal
equilibrium. The temperature measurements can also be used in the
simulation of annealing effects in periods without DCU
readings. The peaks visible during Run~2 correspond to periods where
the SST cooling plants were switched off. The high simulated current
corresponds to the expected value under these conditions; in reality the
sensors were never put under bias.

\begin{figure}[tbp]
\centering
\includegraphics[width=0.49\textwidth]{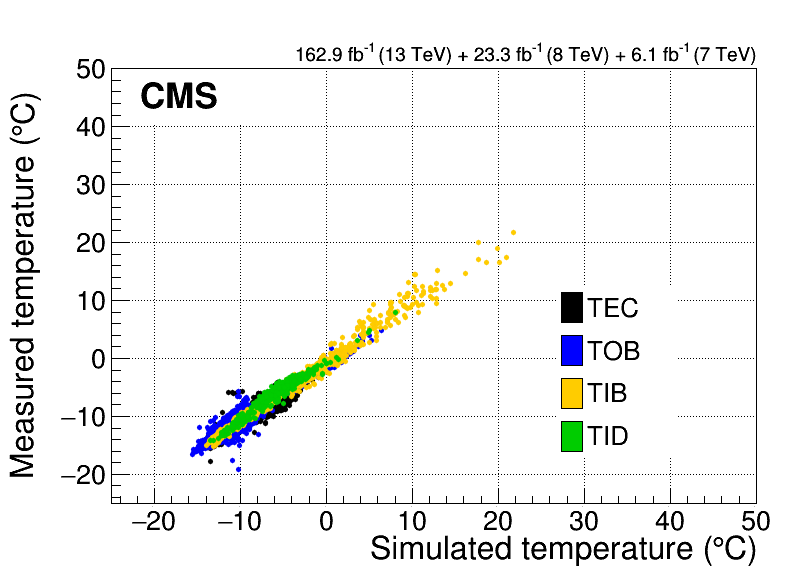}
\caption{Silicon temperature as measured by the DCUs of the
  individual modules after
  192.3\fbinv of integrated luminosity, compared with simulation. }
\label{fig:tsil_dcu_vs_simulation}
\end{figure}

\begin{figure}[tbp]
\centering
\includegraphics[width=0.49\textwidth]{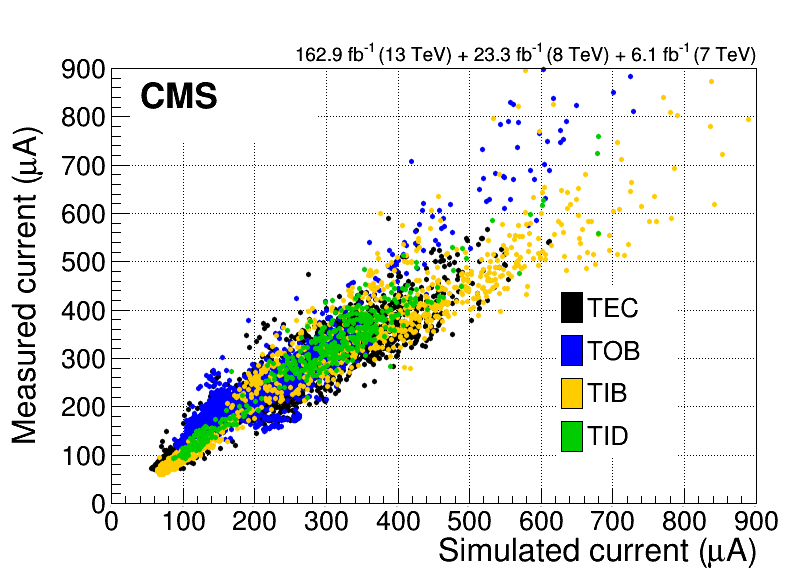}
\includegraphics[width=0.49\textwidth]{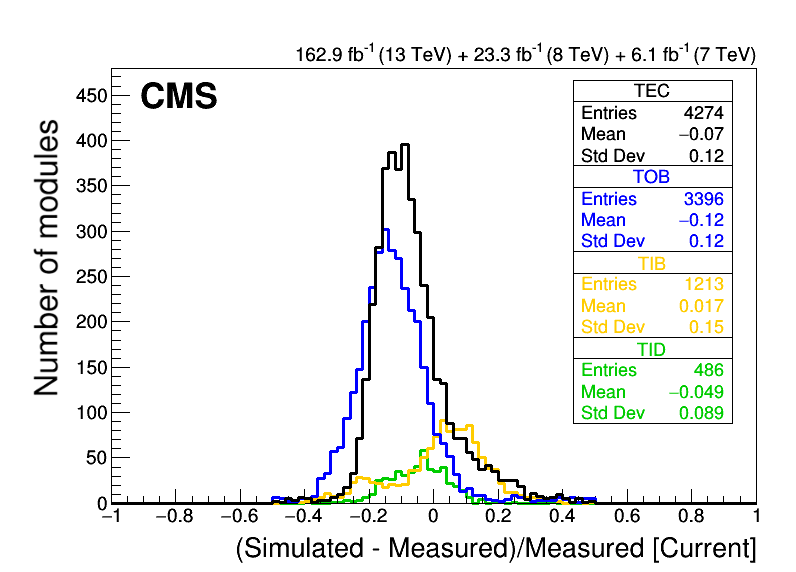}
\caption{Left: leakage current as measured by the DCUs of the individual
  modules after 192.3\fbinv
  of integrated luminosity, compared with simulation. Right: relative difference
  between predicted and measured leakage current after 192.3\fbinv.}
\label{fig:ileak_dcu_vs_simulation}
\end{figure}

The measured and simulated sensor temperatures after exposure to
192.3\fbinv of integrated luminosity are shown for all modules of the
SST in Fig.~\ref{fig:tsil_dcu_vs_simulation}. The simulated values
agree well with those measured. 
The corresponding plot for the leakage
current is shown in Fig.~\ref{fig:ileak_dcu_vs_simulation}. Again, 
most modules are well described by the simulation. The simulation slightly 
underestimates the measured values in the TID, TOB, and TEC, but
slightly overestimates them in the TIB.

\begin{figure}[tbp]
\centering
\includegraphics[width=0.49\textwidth]{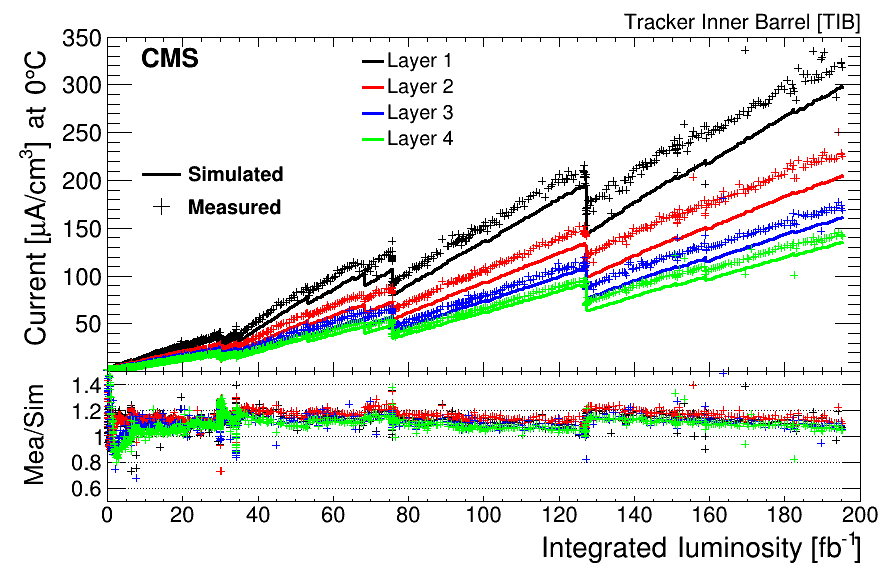}
\includegraphics[width=0.49\textwidth]{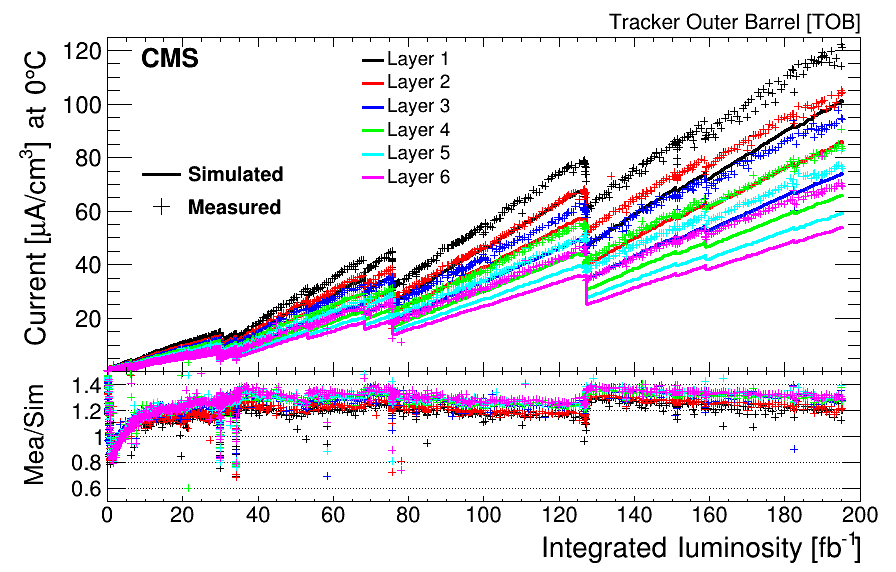}
\caption{Leakage current for each layer scaled to unit volume and
  0$^\circ$C{} as a function of the total delivered integrated
  luminosity in the TIB (left) and TOB (right). The lower part of each
  plot shows the ratio of the measured and simulated current.}
\label{fig:ileak-per-layer-scaled}
\end{figure}

The average leakage current for each layer, normalized to unit volume and
scaled to a common temperature (0$^\circ$C{}), is shown as a function
of the integrated luminosity in Fig.~\ref{fig:ileak-per-layer-scaled}
for the TIB (left) and the TOB (right). In general, the simulation
reproduces the features of the data well, but underestimates the leakage current by about
20\% consistently for all layers, even given the variation in their
radial position. This discrepancy is not yet understood;
possible factors are an imperfect modelling of the radiation
environment in FLUKA, and uncertainties in the cooling contact, which
can lead to an incorrect estimate for the self-heating.

\subsubsection{Depletion voltage evolution} \label{sec:depletion_voltage}

\begin{figure}[tbp]
  \centering
    \includegraphics[width=0.4\textwidth]{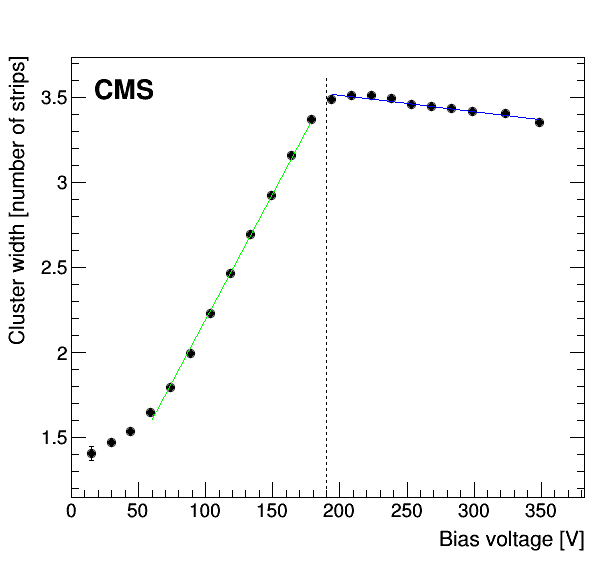}
  \caption{Example of the determination of the full-depletion voltage
    using the intersection of two linear fits to the cluster size data of one module.
    The dashed line indicates the derived full-depletion voltage.
  }
  \label{fig:example_bias_scans}
\end{figure}

The full-depletion voltage $\vdep$ of a silicon sensor is
the applied bias voltage at which the active silicon volume is
completely depleted of free charge carriers.
After installation, when direct measurements on the silicon sensors
are no longer possible, the full-depletion voltage can be determined
from the signal created by MIP-like particles. The variation of 
cluster charge and cluster size with
bias voltage is investigated. This is called a `bias scan' in the following.
At the full-depletion voltage there
is a noticeable change in behavior, an example of which is shown in
Fig.~\ref{fig:example_bias_scans} where the mean cluster size for
on-track clusters is presented as a function of the bias voltage.

The cluster size increases below the full depletion because of the incomplete
collection of the charge which results in lower strip charges, that
have a higher probability to be below the zero-suppression threshold,
especially in the tails of a cluster. Above the full depletion voltage
the cluster decreases due to the increased charge collection speed
which reduces the charge sharing between neighboring strips.

The
kink in the measured data indicates that the full-depletion voltage has been reached. 
To determine the precise full-depletion voltage while accounting
for the finite step size of the scan, the measurements are fitted with two
linear functions, one before the kink and one after it. The
abscissa of the intersection point of the two functions is the full
depletion voltage.

The effective doping concentration $\neff$ of silicon sensors changes with
radiation dose due to the creation of point and cluster defects in the
lattice structure~\cite{Moll:1999kv}. Defects can be caused by energy deposition from
both charged particles (mostly charged pions and protons) and neutral particles
(mostly neutrons).  The defects lead to (incomplete) donor removal and
the creation of stable acceptor levels. For the n-type sensors of
the SST this initially leads to a reduction of the effective doping
concentration to the point of type inversion, after which the bulk becomes increasingly p-type effectively. The full-depletion voltage $\vdep$ can
be related to the effective doping concentration $\neff$ and to the resistivity $\rho$ via
\begin{equation}\label{eq:neff_vdep}
  \vdep = \frac{\left \vert \neff\right\vert d^2 q_0}{2\epsilon\epsilon_0}\quad\text{and}\quad \vdep = \frac{d^2}{2\epsilon\epsilon_0\mu\rho},
\end{equation}
where $d$ is the sensor thickness, $q_0$ is the unit charge, $\mu$ is
the electron mobility, and $\epsilon$ and $\epsilon_0$ are the permittivity
of silicon and the vacuum, respectively. The left formula is valid
for the radiation fluences in the SST for which the effect of charge
trapping can be neglected. The initial resistivity of the thin and
thick sensor types is chosen to have similar initial
depletion voltages. The values are in the range of 
$\rho = 1.55$--$3.25$\unit{k$\Omega$ cm} for the 320\mum thick sensors and 
$\rho = 4$--$8$\unit{k$\Omega$ cm} for the 500\mum thick sensors. Because of the higher
resistivity of their bulk material, thick sensors are expected to reach
type inversion at lower fluences than the thin sensors.

\begin{figure}[tbp]
  \centering
  \includegraphics[width=0.49\textwidth]{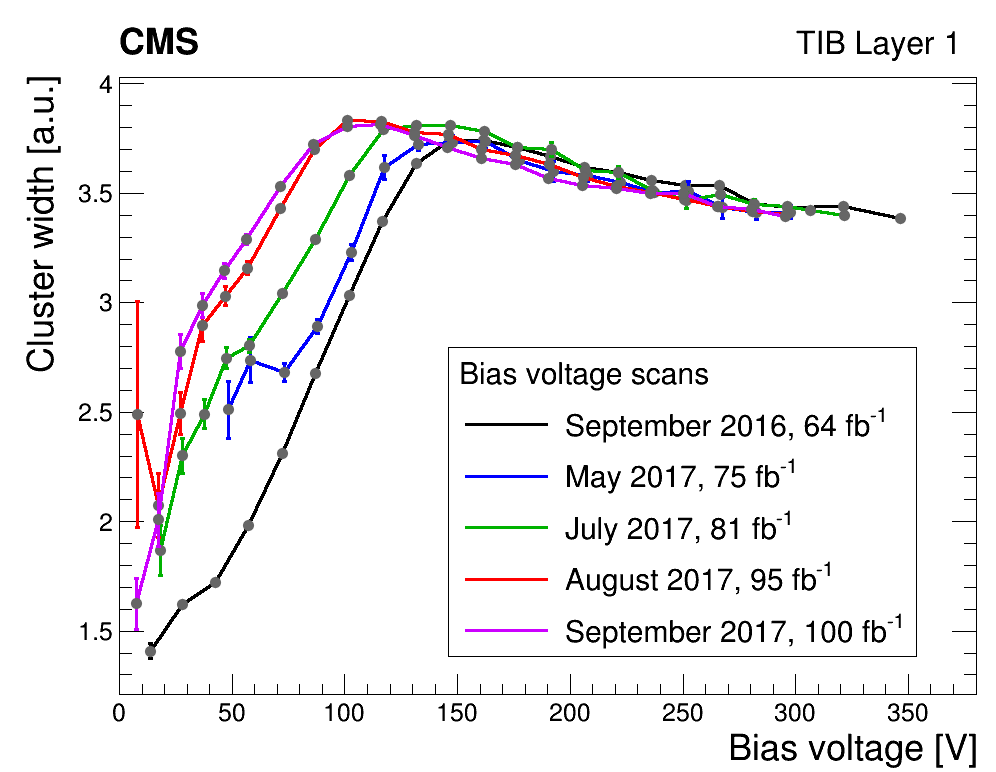}
  \includegraphics[width=0.49\textwidth]{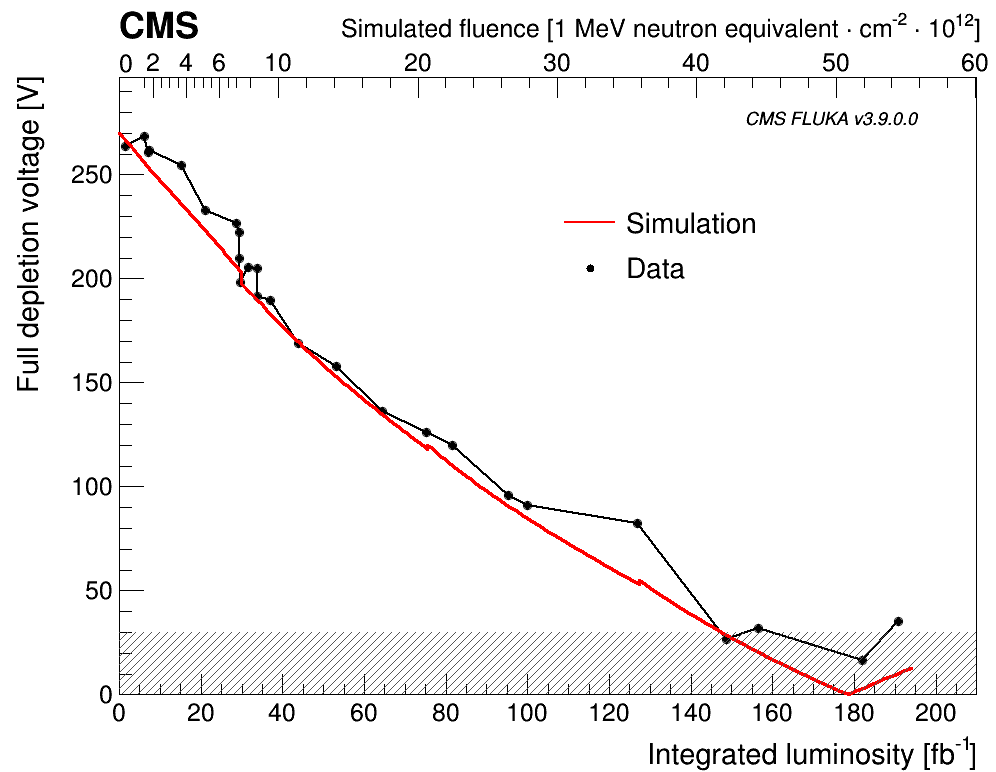}
  \caption{Left: mean cluster width as a function of the sensor bias
    voltage for one module in the TIB layer~1 for bias scans taken at
    various integrated luminosities. The curves have been normalized to
    the same cluster width at the highest bias voltage of the scans. Right:
    evolution of the full-depletion voltage for one TIB layer 1 module
    as a function of the integrated luminosity (lower $x$ axis) and fluence
    (upper $x$ axis) using the data from bias scans. The red line shows the
    predicted full-depletion voltage. The gray hashed area indicates
    the region where depletion voltage estimates from data have a
    large uncertainty. }
  \label{fig:example_bias_scan_single_mod}
\end{figure}

The evolution of the full-depletion voltage is monitored continuously
during the operation of the SST by means of regular bias scans.
In Fig.~\ref{fig:example_bias_scan_single_mod}~(left) an example of
the scan data used to extract the full-depletion voltage at various
times is shown. The evolution of the full-depletion voltage as
determined from these data is shown and compared with the expectations
from simulation in Fig.~\ref{fig:example_bias_scan_single_mod}~(right). In general, the simulation agrees
well with the data. A small step downwards is visible around 30\fbinv in both
data and simulation, with different magnitude. This corresponds to
reverse annealing that happened during LS1 where the SST was at room
temperature for extended periods of time. Comparing the size of this
step in many modules suggests that the amount of reverse annealing is
underestimated in the simulation.

\begin{figure}[th]
\centering
\includegraphics[width=0.49\textwidth]{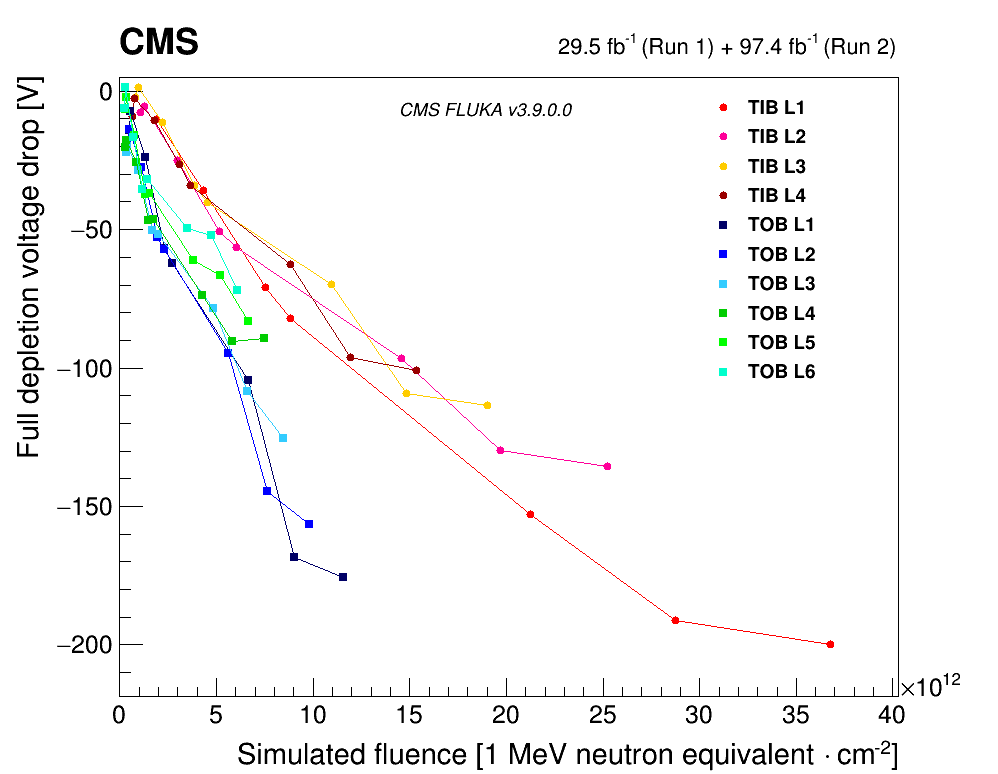}
\includegraphics[width=0.4825\textwidth]{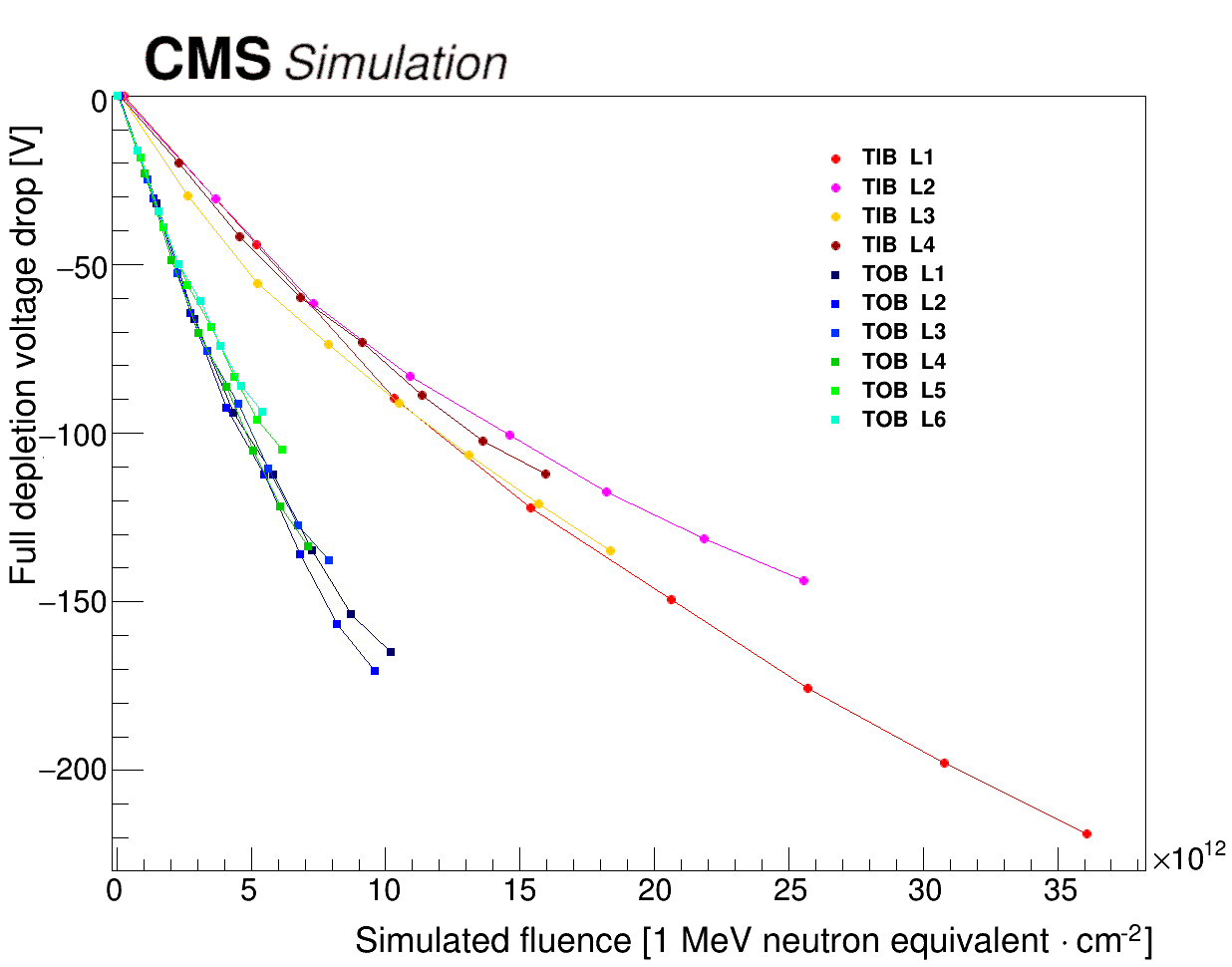}
\caption{ Left: change of the full-depletion voltage in different tracker
  layers as a function of the particle fluence in data. Scans at
  different integrated luminosities are shown. The luminosity is converted to
  particle fluence for each position in the detector using a FLUKA
  model. Right: simulated change of the full-depletion voltage for single modules in
  different tracker layers as a function of the particle fluence.  }
\label{fig:vdep_change}
\end{figure}

The average change of the full-depletion voltage for each layer in the TIB and
TOB as a function of the fluence is shown in Fig.~\ref{fig:vdep_change}
for both data~(left) and simulation~(right). For data the integrated
luminosity is converted to the expected fluence for each module 
using a FLUKA simulation. As expected, different behavior
for the thin and thick sensors is clearly visible in the data and
reproduced in simulation.  Furthermore, the change of
$\vdep$ is similar for different sensor geometries. The 
range of simulated fluence stops just before the sensors
reach type inversion; here the determination of
$\vdep$ becomes unreliable. At the end of Run~2 many sensors of the SST are at or close to type inversion.

\subsubsection{Radiation damage projections} \label{sec:raddam_projections}

Simulation is used to predict the status of the SST at its
end of life. This is assumed to be at the end of Run~3, after an
integrated luminosity of 500\fbinv while operating with a cooling fluid temperature of $-25^{\circ}$C.

The SST was warm for about 160~days during LS2, over
periods where cold operation was not possible. These were during the removal and
reinstallation of the pixel detector, as well as during removal,
reinstallation, and bake out of the central beam pipe, when there was
insufficient humidity control, \eg, in the bulkhead.
The simulation here assumes a total of 120 days at room temperature
during LS2. Although this is slightly lower than the actual, the equilibrium
temperature used in the simulation is $+18^{\circ}$C, whereas the actual
temperature of the detector was around $+16^{\circ}$C during warm
periods. During future year-end technical stops the temperature is assumed to
be maintained near or below $0^{\circ}$C{}, by means of the primary
detector cooling or with the thermal screen cooling, as described
in Section~\ref{sec:system_description}.

\begin{figure}[tp]
\centering
\includegraphics[width=0.95\textwidth]{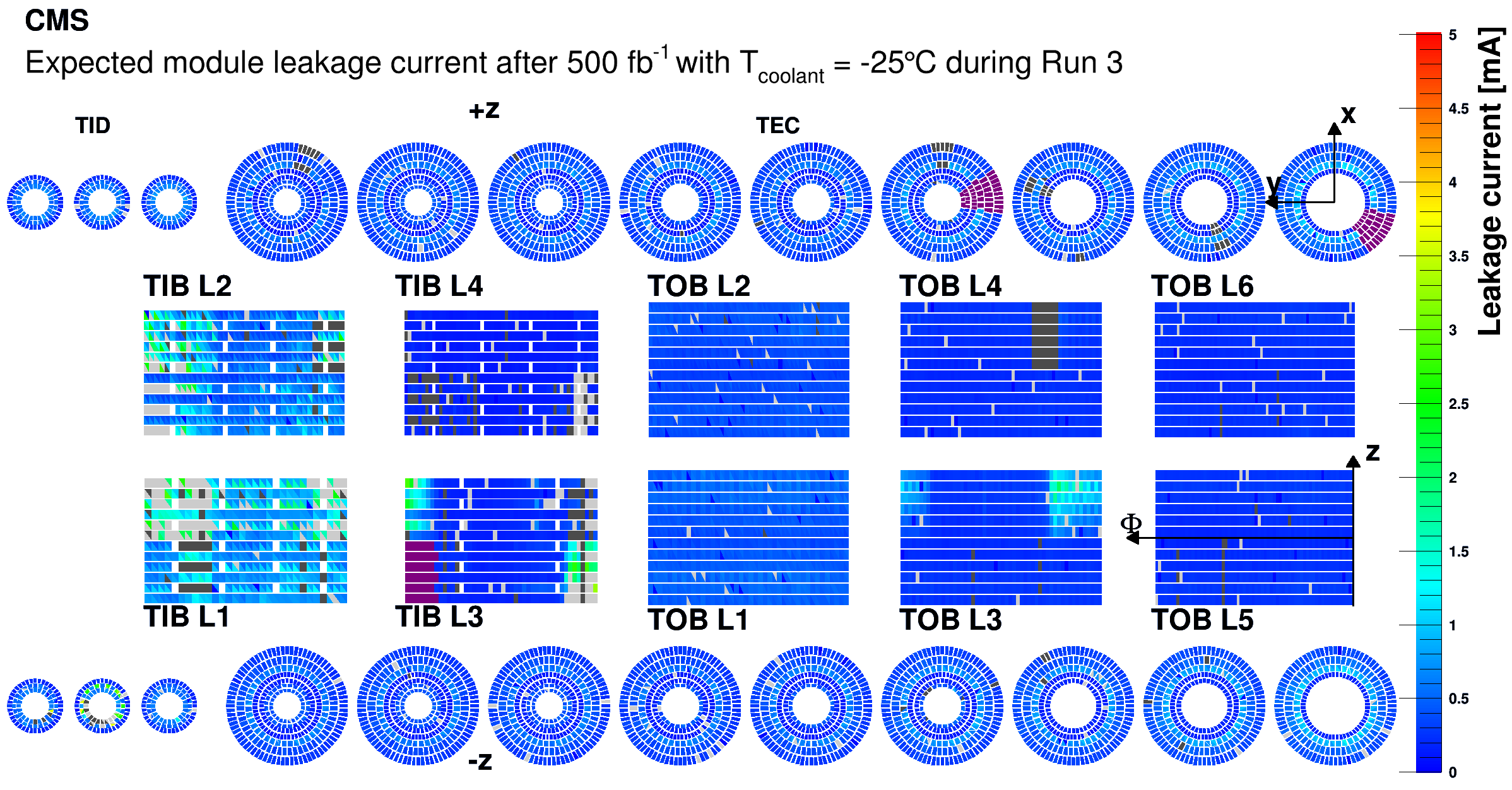}
\caption{
  Tracker map of predicted leakage current at the end of life of the SST, where each silicon module is represented by a
  rectangle in the barrel and a trapezoid in the endcap; in stereo
  modules each submodule constitutes one half of this area.
  The color scale shows the projected sensor leakage current for a total integrated luminosity of
  500\fbinv with the SST operated at $-25^{\circ}$C during all of
  Run~3. Modules in dark gray were already inoperable at the end of
  Run~2. Modules in light gray are expected to become inoperable during Run~3 because their projected leakage current will be too
  high. Regions in purple are control rings with DCU readout issues,
  implying no input data for the simulation.
  }
\label{fig:ileak-tkmap-500fb-25C}
\end{figure}

The predicted leakage current after 500\fbinv of integrated luminosity
is shown in Fig.~\ref{fig:ileak-tkmap-500fb-25C}. The gray areas
of the detector in TIB layers~1, 2, and 3, and in the $-z$ disk~2 of
the TID are regions where either the maximum power supply current of
12\unit{mA} per HV channel is reached or where one or more modules connected to the
same HV channel have experienced thermal runaway. The purple regions
are those that lacked appropriate input parameters for the simulation
(TEC$+$ disks~6 and 9, and TIB layer~3, TOB layer~4) as DCU readout on these control
rings was not possible. The modules that are expected to become
inoperable from excessive leakage current correspond very closely to
the regions with elevated temperatures from Fig.~\ref{fig:tkmap-tsil}.
The total number of inactive modules
is smaller than that anticipated
in~Fig.~2.24 of Ref.~\cite{thetrackergroupofthecmscollaboration2020cms}, suggesting a less severe
degradation of the system: fewer bad modules are
expected especially in TIB layer~2 and TOB layer~4. 
Some groups of modules in the TIB layers 1, 2, and 3 show elevated
leakage currents of around 2--3\unit{mA} per module and these could also
potentially become inoperable. Because of their limited number, the
overall assessment does not change.

\begin{figure}[bp]
\centering
\includegraphics[width=0.65\textwidth]{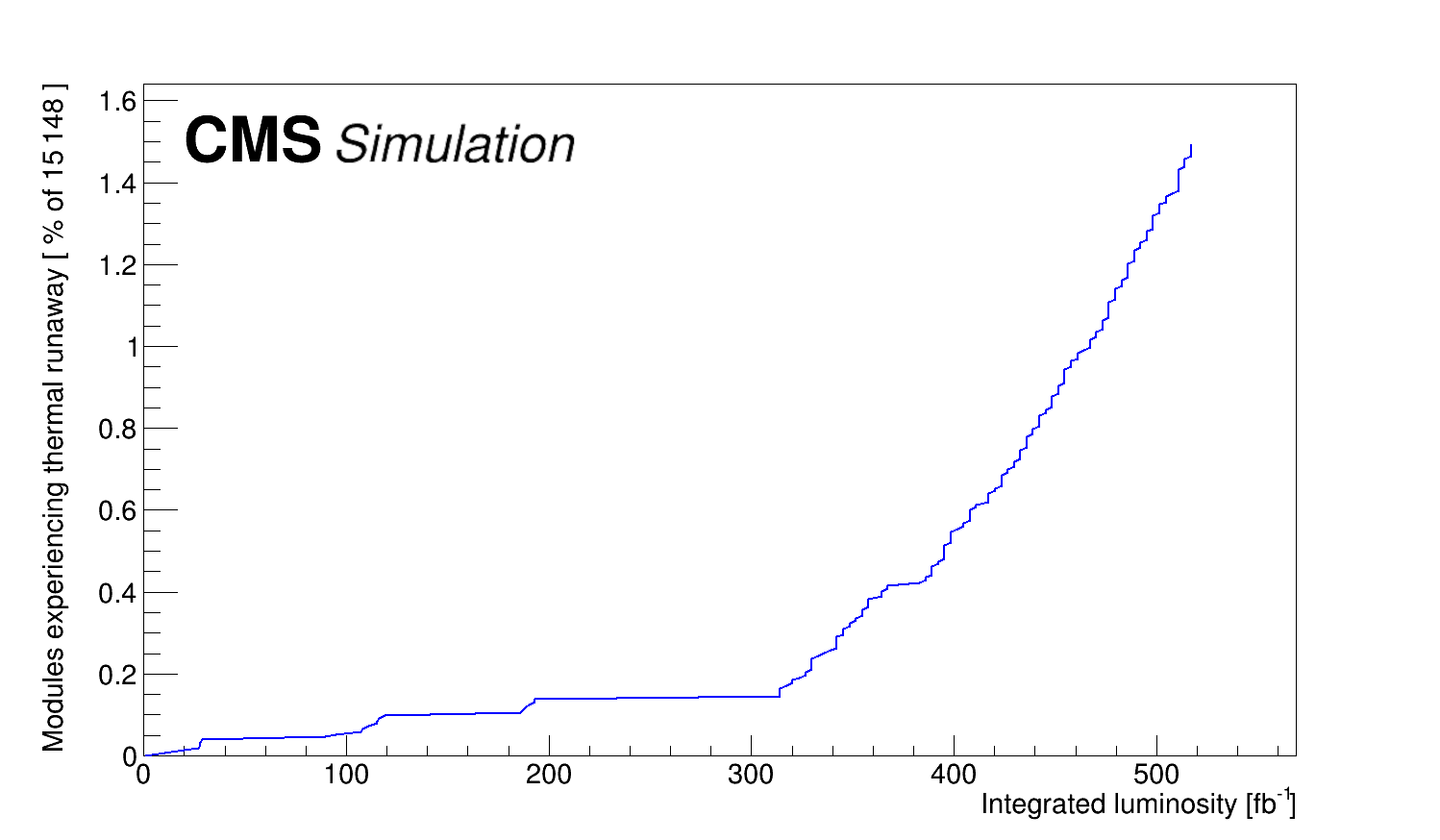}
\caption{Fraction of modules affected by thermal runaway as a function of
  the integrated luminosity.   }
\label{fig:thermal_runaway_vs_lumi}
\end{figure}

In Fig.~\ref{fig:thermal_runaway_vs_lumi}, the percentage of modules
experiencing thermal runaway is shown as a function of the integrated
luminosity. The number of such modules increases rapidly above
300\fbinv and reaches about 1.5\% at 500\fbinv. A module
is considered to reach thermal runaway if during the iterative
simulation the self-heating contribution continues to increase.

\begin{figure}[t!bh]
  \centering
  \includegraphics[width=0.95\textwidth]{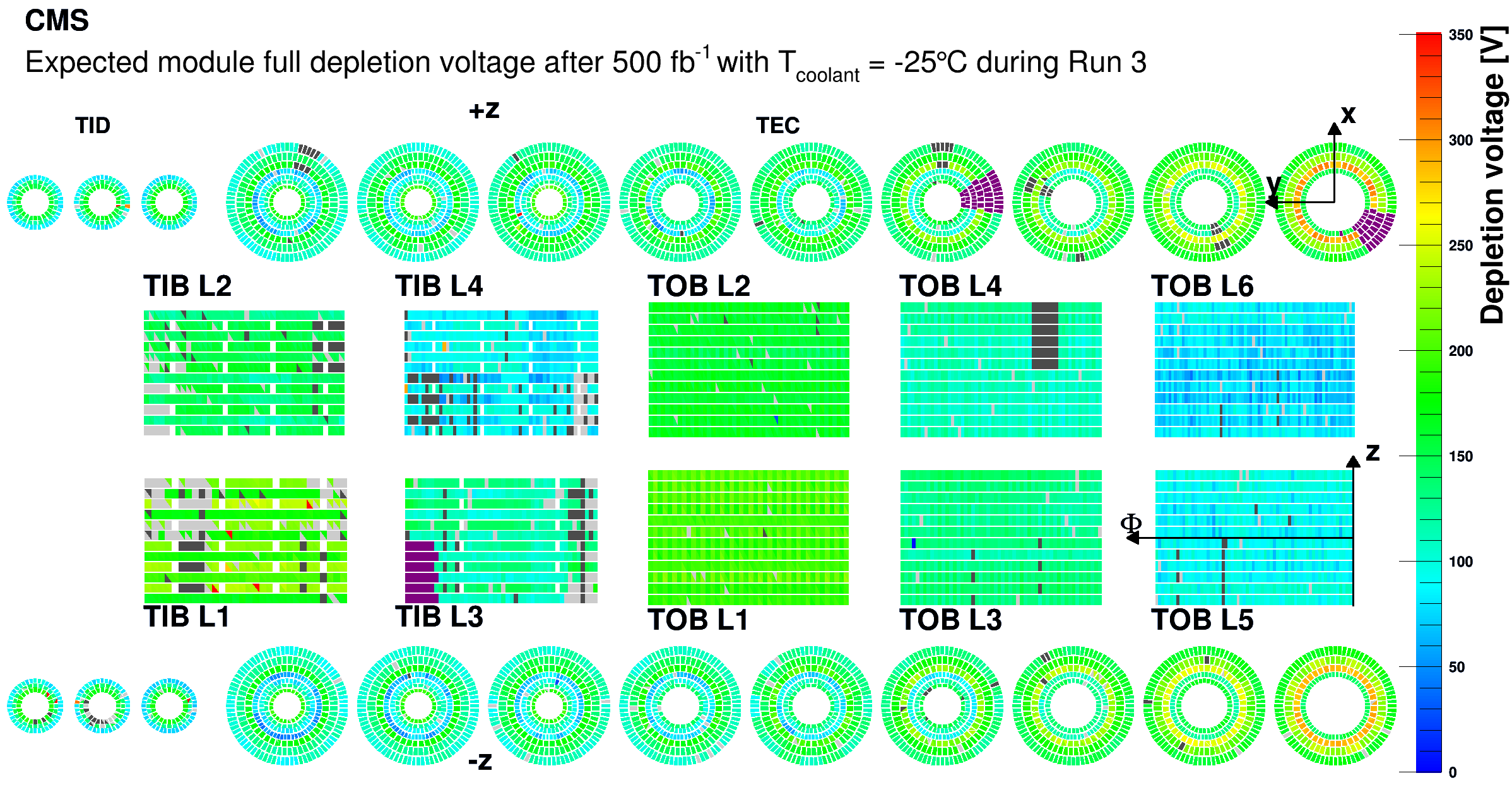}
  \caption{Tracker map of predicted full-depletion voltage at the end of life of the SST, where each silicon module is represented by a
    rectangle in the barrel and a trapezoid in the endcap; in stereo
    modules each submodule constitutes one half of this area.
    The color scale shows the projected full-depletion voltage for a total integrated luminosity of
    500\fbinv with the SST operated at $-25^{\circ}$C during all of
    Run~3. Modules in dark gray were already inoperable at the end of
    Run~2. Modules in light gray are expected to become inoperable during Run~3 because their projected leakage current will be too
    high and are identical to the modules depicted in light grey in Fig.~\ref{fig:ileak-tkmap-500fb-25C}. Regions in purple are control rings with DCU readout issues,
    implying no input data for the simulation.}
  \label{fig:vdep_projection_tkmap}
\end{figure}

Predictions are made for the full-depletion voltage at the end of life of the SST. 
The predicted full-depletion voltage at 500\fbinv is shown in
Fig.~\ref{fig:vdep_projection_tkmap}.

The highest full-depletion voltage is expected to
be around 275\unit{V}, suggesting that there is ample margin
with respect to the maximum bias voltage of 600\unit{V} that can be
delivered by the power supply system. Modules that
suffer most from reverse annealing due to high operating temperatures
are not included here since they are expected to either reach the power
supply current limit or experience thermal runaway before a possible
limit in full-depletion voltage becomes relevant.

\section{Summary and outlook}\label{sec:outlook}

In this paper, the calibration, operation, and performance
of the CMS silicon strip tracker (SST) were presented. The SST has successfully 
delivered measurements for the re\-con\-struc\-tion of\linebreak charged particle
trajectories since the start of LHC operation.

First the calibration of the SST was discussed. It was demonstrated that the system behavior is well understood
over 10 years of operation, which includes different operating
temperatures and ever increasing instantaneous luminosities over the course of the data taking.

Next, the ingredients necessary for a proper simulation of the SST were
considered. Measurements required as input to the simulation, obtained
in special runs, were presented.

The performance with proton-proton collisions was then extensively discussed. The single-hit occupancy of the system is in line with expectations even while
running at twice the design instantaneous luminosity. The dynamic
identification of bad components, crucial for a proper handling of
missing hits in the subsequent reconstruction of particle tracks, was
introduced and shown to work well. Saturation effects
in the APV25 preamplifier, which resulted in a reduced track reconstruction efficiency, were identified during 2015--2016. This problem was understood and resolved in mid-2016. The
signal-to-noise performance of the system was excellent
even after almost 200\fbinv{} of integrated luminosity. The outlook
toward the expected end-of-life integrated luminosity is in line with
expectations. As a result, the system shows excellent single-hit
reconstruction efficiency because of the analog readout that enables
a hit resolution significantly better than the binary limit for the
most relevant cluster sizes. The evolution of the Lorentz angle as a function of the integrated luminosity was shown, as was a measurement of
the specific energy loss of particles, made possible by the analog
readout, which is useful for more sophisticated analyses of collision
events.

Finally, radiation effects in the SST were examined, with focus on the effects in the laser diodes of the optical link system and the effects
on the properties of the silicon sensors. The threshold current
increase in the laser diodes 
was shown to behave as expected, leaving ample margin for the
future operation of the SST. For the bulk damage of the silicon sensor, the
resulting effects on the leakage current and full-depletion voltage
were shown and compared with simulations. The evolution of both leakage current
and depletion voltage is well described by models, and the outlook to the
expected end of life of the SST shows that the vast majority of the
modules should continue to function well. Those modules that are
expected to become inoperable are in regions of the SST affected by
cooling-related problems encountered during early operation; the
extent of problems is expected to be lower than anticipated
from earlier projections. The small fraction of about 1.5\% of affected
modules is not expected to severely affect the SST's ability to
provide high-quality measurements for track reconstruction until the
start of LS3, when the tracking system will be replaced for the High-Luminosity LHC
era.

\begin{acknowledgments}
  We congratulate our colleagues in the CERN accelerator departments for the excellent performance of the LHC and thank the technical and administrative staffs at CERN and at other CMS institutes for their contributions to the success of the CMS effort. In addition, we gratefully acknowledge the computing centers and personnel of the Worldwide LHC Computing Grid and other centers for delivering so effectively the computing infrastructure essential to our analyses. Finally, we acknowledge the enduring support for the construction and operation of the LHC, the CMS detector, and the supporting computing infrastructure provided by the following funding agencies: SC (Armenia), BMBWF and FWF (Austria); FNRS and FWO (Belgium); CNPq, CAPES, FAPERJ, FAPERGS, and FAPESP (Brazil); MES and BNSF (Bulgaria); CERN; CAS, MoST, and NSFC (China); MINCIENCIAS (Colombia); MSES and CSF (Croatia); RIF (Cyprus); SENESCYT (Ecuador); ERC PRG, RVTT3 and MoER TK202 (Estonia); Academy of Finland, MEC, and HIP (Finland); CEA and CNRS/IN2P3 (France); SRNSF (Georgia); BMBF, DFG, and HGF (Germany); GSRI (Greece); NKFIH (Hungary); DAE and DST (India); IPM (Iran); SFI (Ireland); INFN (Italy); MSIP and NRF (Republic of Korea); MES (Latvia); LMTLT (Lithuania); MOE and UM (Malaysia); BUAP, CINVESTAV, CONACYT, LNS, SEP, and UASLP-FAI (Mexico); MOS (Montenegro); MBIE (New Zealand); PAEC (Pakistan); MES and NSC (Poland); FCT (Portugal); MESTD (Serbia); MICIU/AEI and PCTI (Spain); MOSTR (Sri Lanka); Swiss Funding Agencies (Switzerland); MST (Taipei); MHESI and NSTDA (Thailand); TUBITAK and TENMAK (Turkey); NASU (Ukraine); STFC (United Kingdom); DOE and NSF (USA).
  
  \hyphenation{Rachada-pisek} Individuals have received support from the Marie-Curie program and the European Research Council and Horizon 2020 Grant, contract Nos.\ 675440, 724704, 752730, 758316, 765710, 824093, 101115353, 101002207, and COST Action CA16108 (European Union); the Leventis Foundation; the Alfred P.\ Sloan Foundation; the Alexander von Humboldt Foundation; the Science Committee, project no. 22rl-037 (Armenia); the Fonds pour la Formation \`a la Recherche dans l'Industrie et dans l'Agriculture (FRIA-Belgium); the Beijing Municipal Science \& Technology Commission, No. Z191100007219010 and Fundamental Research Funds for the Central Universities (China); the Ministry of Education, Youth and Sports (MEYS) of the Czech Republic; the Shota Rustaveli National Science Foundation, grant FR-22-985 (Georgia); the Deutsche Forschungsgemeinschaft (DFG), among others, under Germany's Excellence Strategy -- EXC 2121 ``Quantum Universe" -- 390833306, and under project number 400140256 - GRK2497; the Hellenic Foundation for Research and Innovation (HFRI), Project Number 2288 (Greece); the Hungarian Academy of Sciences, the New National Excellence Program - \'UNKP, the NKFIH research grants K 131991, K 133046, K 138136, K 143460, K 143477, K 146913, K 146914, K 147048, 2020-2.2.1-ED-2021-00181, TKP2021-NKTA-64, and 2021-4.1.2-NEMZ\_KI-2024-00036 (Hungary); the Council of Science and Industrial Research, India; ICSC -- National Research Center for High Performance Computing, Big Data and Quantum Computing and FAIR -- Future Artificial Intelligence Research, funded by the NextGenerationEU program (Italy); the Latvian Council of Science; the Ministry of Education and Science, project no. 2022/WK/14, and the National Science Center, contracts Opus 2021/41/B/ST2/01369 and 2021/43/B/ST2/01552 (Poland); the Funda\c{c}\~ao para a Ci\^encia e a Tecnologia, grant CEECIND/01334/2018 (Portugal); the National Priorities Research Program by Qatar National Research Fund; MICIU/AEI/10.13039/501100011033, ERDF/EU, "European Union NextGenerationEU/PRTR", and Programa Severo Ochoa del Principado de Asturias (Spain); the Chulalongkorn Academic into Its 2nd Century Project Advancement Project, and the National Science, Research and Innovation Fund via the Program Management Unit for Human Resources \& Institutional Development, Research and Innovation, grant B39G670016 (Thailand); the Kavli Foundation; the Nvidia Corporation; the SuperMicro Corporation; the Welch Foundation, contract C-1845; and the Weston Havens Foundation (USA).
\end{acknowledgments}

\bibliography{auto_generated} 

\appendix
\numberwithin{table}{section}

\section{Glossary of special terms and acronyms}
\label{chap:glossary}

\begin{footnotesize}%

\begin{tabbing}
\hspace*{2.2cm}\=\hspace*{5cm} \kill
ADC:\>Analog-to-digital converter \\
AOH:\>Analog Optohybrid, used in the auxiliary electronics of the \SST{}\\
APSP:\> Analog pulse shape processor\\
APV:\> Analog pipeline voltage\\
APV25:\> Readout chip used in the \SST{}\\
APVMUX:\> Auxiliary chip to multiplex the outputs of two APV25 chips\\
ASIC:\> Application-specific integrated circuit\\
BX:\> Bunch crossing\\
CCU:\> Communication and Control Unit, used in the auxiliary electronics of the \SST{} \\
DAQ:\> Data acquisition\\
DCU:\> Detector Control Unit, used in the auxiliary electronics of the \SST{}\\
DOH:\> Digital Optohybrid, used in the auxiliary electronics of the \SST{}\\
ENC:\> Equivalent noise charge\\
FEC:\> Front-End Controller, used in the DAQ system of the \SST{}\\
FED:\> Front-End Driver, used in the DAQ system of the \SST{}\\
FEH:\> Front-end hybrid\\
FET:\> Field-effect transistor\\
HIP:\> Highly ionizing particle\\
HL-LHC:\> High-Luminosity LHC\\
HV:\> High voltage\\
I$^2$C:\> Inter-integrated circuit\\
$\ileak$:\> Silicon sensor leakage current\\ %
$\lambda_0$:\> Interaction length\\
L1A:\> CMS level-1 trigger accept signal\\ %
LHC:\> Large Hadron Collider\\
LLD:\> Linear Laser Driver, used in the auxiliary electronics of the \SST{}\\
LS1:\> Long Shutdown 1 (2013--2014)\\
LS2:\> Long Shutdown 2 (2019--2021)\\
LS3:\> Long Shutdown 3 (2026--2030)\\
LV:\> Low voltage\\
MC:\> Monte Carlo\\
mFEC:\> Mezzanine card on the FEC, used in the DAQ system of the \SST{},\\
     \> for the communication to an individual control ring\\
MIP:\> Minimum-ionizing particle\\
MPV:\> Most probable value, of a Landau distribution\\
$\neff$:\> Effective doping concentration of a silicon sensor\\
NIEL:\> Non-ionizing energy loss\\
NZS:\> Data with no zero suppression applied\\
PLL:\> Phase-locked loop\\
PSU:\> Power supply unit\\
PU:\> Pileup\\
RMS:\> Root-mean-square\\
Run 1:\> First data-taking period at the LHC (2009--2012)\\
Run 2:\> Second data-taking period at the LHC (2015--2018)\\
Run 3:\> Third data-taking period at the LHC (2022--2026)\\
\sovern{}:\> Signal-to-noise ratio\\
SST:\> Silicon strip tracker of the CMS experiment\\
TIB:\> Tracker Inner Barrel, a subdetector of the SST\\
TID:\> Tracker Inner Disk, a subdetector of the SST\\
TEC:\> Tracker Endcap, a subdetector of the SST\\
TOB:\> Tracker Outer Barrel, a subdetector of the SST\\
TTS:\> Trigger throttling system\\
$\vdep$:\> Silicon sensor full-depletion voltage\\
VFP:\> Preamplifier feedback voltage bias, setting to control the discharge speed\\
\> of the APV25 preamplifier\\
$X_0$:\> Radiation length\\
ZS:\> Zero-suppressed\\

\end{tabbing}

\end{footnotesize}

\cleardoublepage \section{The CMS Collaboration \label{app:collab}}\begin{sloppypar}\hyphenpenalty=5000\widowpenalty=500\clubpenalty=5000\input{TRK-20-002-public-authorlist.tex}\end{sloppypar}
\end{document}

%% file: TRK-20-002-public-authorlist.tex
\cmsinstitute{Yerevan Physics Institute, Yerevan, Armenia}
{\tolerance=6000
A.~Hayrapetyan, A.~Tumasyan\cmsAuthorMark{1}\cmsorcid{0009-0000-0684-6742}
\par}
\cmsinstitute{Institut f\"{u}r Hochenergiephysik, Vienna, Austria}
{\tolerance=6000
W.~Adam\cmsorcid{0000-0001-9099-4341}, J.W.~Andrejkovic, T.~Bergauer\cmsorcid{0000-0002-5786-0293}, S.~Chatterjee\cmsorcid{0000-0003-2660-0349}, K.~Damanakis\cmsorcid{0000-0001-5389-2872}, M.~Dragicevic\cmsorcid{0000-0003-1967-6783}, M.~Friedl, E.~Fr\"{u}hwirth, P.S.~Hussain\cmsorcid{0000-0002-4825-5278}, M.~Jeitler\cmsAuthorMark{2}\cmsorcid{0000-0002-5141-9560}, N.~Krammer\cmsorcid{0000-0002-0548-0985}, A.~Li\cmsorcid{0000-0002-4547-116X}, D.~Liko\cmsorcid{0000-0002-3380-473X}, I.~Mikulec\cmsorcid{0000-0003-0385-2746}, J.~Schieck\cmsAuthorMark{2}\cmsorcid{0000-0002-1058-8093}, R.~Sch\"{o}fbeck\cmsorcid{0000-0002-2332-8784}, D.~Schwarz\cmsorcid{0000-0002-3821-7331}, M.~Sonawane\cmsorcid{0000-0003-0510-7010}, H.~Steininger, S.~Templ\cmsorcid{0000-0003-3137-5692}, W.~Waltenberger\cmsorcid{0000-0002-6215-7228}, C.-E.~Wulz\cmsAuthorMark{2}\cmsorcid{0000-0001-9226-5812}
% ...
\par}
\vskip\cmsinstskip
\dag:~Deceased\\
$^{1}$Also at Yerevan State University, Yerevan, Armenia\\
$^{2}$Also at TU Wien, Vienna, Austria\\
% ...
